\documentclass[twocolumn,aps,prb,superscriptaddress,nofootinbib]{revtex4-1}
\usepackage{graphicx}
\usepackage{times}
\usepackage{bm}
\usepackage[linktocpage=true,colorlinks=true,urlcolor=blue]{hyperref}
\usepackage{pgfplots}
\usepackage{amsfonts}
\usepackage{amsmath}
\usepackage{braket}
\usepackage{relsize}
\usetikzlibrary{arrows}
\usepackage{enumerate}
\usetikzlibrary[patterns] % ConTEXt when using Tik Z
\usetikzlibrary[snakes] % ConTEXt when using Tik Z
\usetikzlibrary[shapes] % ConTEXt when using Tik Z
\usepackage{pgfplots}
\usepackage{newfloat}
\DeclareFloatingEnvironment[name={Supplementary Figure}]{suppfigure}

\graphicspath{ {Figures/} }

\begin{document}

\title{Effective theory approach to the Schr\"{o}dinger-Poisson problem in semiconductor Majorana devices}
\author{Benjamin D. Woods}
\author{Tudor D. Stanescu}
\affiliation{Department of Physics and Astronomy, West Virginia University, Morgantown, West Virginia 26506, USA}
\author{Sankar Das Sarma}
\affiliation{Condensed Matter Theory Center and Joint Quantum Institute, Department of Physics, University of Maryland, College Park, Maryland, 20742-4111, USA}

\begin{abstract}
We propose a method for solving the Schr\"{o}dinger-Poisson problem that can be efficiently implemented in realistic 3D tight-binding models of semiconductor-based Majorana devices. The method is based on two key ideas:  (i) For a given geometry, the Poisson problem is only solved once (for each local orbital) and the results are stored as an interaction tensor; using this Green's function scheme, the Poisson component of the iteration procedure is reduced to a few simple summations. (ii) The 3D problem is mapped into an effective multi-orbital 1D problem with molecular orbitals calculated  self-consistently as the transverse modes of an infinite wire with the same electrostatic potential as the local
electrostatic potential of the finite 3D device. These two ideas considerably simplify the numerical complexity of the full 3D Schr\"{o}dinger-Poisson problem for the nanowire, enabling a tractable effective theory with predictive power.
To demonstrate the capabilities of our approach, we calculate the response of the system to an external magnetic field, the dependence of the effective chemical potential on the work function difference, and  the dependence of the effective semiconductor-superconductor coupling  on the applied gate potential. We find that, within a wide range of parameters, different low-energy bands 
are characterized by similar effective couplings, which results in induced gap features characterized by a single energy scale. We also find that electrostatic effects are responsible for a partial suppression of the Majorana energy splitting oscillations. Finally, we show that a position-dependent work function difference can produce a non-homogeneous effective potential that is not affected by the screening due to the superconductor and is only partially suppressed by the charge inside the wire. In turn, this potential can induce trivial low-energy states that mimic the phenomenology of Majorana zero modes. Thus any position-dependent work function difference (even at the 1$\%$ level) along the nanowire must be avoided through carefully engineered semiconductor-superconductor interfaces.
\newline
\newline
DOI: \href{https://journals.aps.org/prb/abstract/10.1103/PhysRevB.98.035428}{10.1103/PhysRevB.98.035428} 
\end{abstract}

\maketitle

\section{Introduction}
Motivated by the theoretical model proposed by Kitaev\cite{Kitaev2001} and the concrete predictions\cite{Sau2010a,Alicea2010,Oreg2010,Lutchyn2010,Sau2010} about the existence  of  zero-energy  Majorana  modes  in proximity-coupled semiconductor-superconductor (SM-SC)  hybrid  structures,  a systematic experimental search for Majorana zero modes\cite{Majorana1937,Read2000} (MZMs) has gained momentum in the past few years.\cite{Mourik2012,Deng2012,Das2012,Rokhinson2012,Churchill2013,Finck2013,Chang2015,Albrecht2016,Deng2016,Chen2017}
 Recent  improvements in materials science and nanofabrication\cite{Zhang2017,Nichele2017,Zhang2018} have led to the observation of stable zero-energy subgap states that  manifest the predicted $2e^2/h$ quantization of the zero bias tunneling differential conductance at low temperatures.\cite{Sengupta2001,Law2009,Flensberg2010} The signatures observed experimentally provide strong indication that MZMs localized at the ends of  proximitized  semiconductor nanowires may have been realized in the laboratory. However, based on the existing evidence one cannot rule out the possibility that these experimental signatures are, in fact, generated by non-topological Andreev bound states (ABSs), which are ubiquitous in the presence of non-uniform system parameters (e.g., variations of the electrochemical potential) or when the wire is coupled to a quantum dot.\cite{Kells2012,Prada2012,SanJose2013,Ojanen2013,Stanescu2014a,Lee2014,Cayao2015,Klinovaja2015,Stanescu2016,Liu2017a,Moore2018} 
In particular, the possible presence of partially separated ABSs (ps-ABSs) consisting of pairs of Majorana bound states separated by a distance comparable to or larger than the characteristic Majorana  length-scale (but  less  than  the length of the wire) should raise serious concern, as one cannot distinguish between these trivial low-energy modes and genuine non-Abelian MZMs using {\em any} type of local measurement at the end of the wire.\cite{Moore2018} Considering this rather disturbing state of affairs, in conjunction with the promising proposals\cite{Alicea2011,Sau2011,Aasen2016,Karzig2017,Litinski2017} for testing the predicted non-Abelian properties of the MZMs and building topological qubits, which will require exquisite control of the hybrid system, it becomes clear that a major theoretical task is to develop a more detailed modeling of semiconductor-superconductor Majorana devices. The minimal model used extensively so far in the Majorana nanowire literature is simply insufficient for describing the SM-SC structure at the level of essential details necessary to distinguish between MZMs and ABSs, as well as in the elucidation of other basic properties of the hybrid device.

A key component of this task is to account for the electrostatic effects that  are naturally induced by the presence of  the superconductor-semiconductor interface and external potential gates. Understanding these effects is critical in the context of two important aspects of the modeling of hybrid devices. On the one hand, they control three basic system parameters: the chemical potential, the Rashba spin-orbit coupling, and the induced superconducting pair potential. Typically, these parameters are treated as independent phenomenological parameters. In fact, they are all controlled by the effective electrostatic potential inside the wire generated by the work function difference at the SM-SC interface and by the applied gate potential in the presence of  a low (but non-vanishing) electron density. The work function difference and the gate potential determine the number of charge carriers in the wire (hence the value of the chemical potential relative to the bottom of the conduction band). In addition, the transverse profile of the effective potential is directly linked to the Rashba  spin-orbit coupling and determines the amplitudes of the wave functions at the SM-SC interface, which, in turn, control the strength of the proximity coupling to the superconductor. Understanding the dependence of these system properties on control parameters such as external gate potentials and applied magnetic fields is important for correctly interpreting the experimental data and optimizing the Majorana devices. On the other hand, electrostatic effects are critical ingredients of existing and proposed Majorana devices, ranging from the controllable tunnel barrier in a charge transport measurement, to the electrostatic confinement in two-dimensional SM-SC structures,\cite{Suominen2017,Hell2017,Hell2017a} and electrostatic operations in Majorana nanowire-based topological circuits,\cite{Alicea2011,Sau2011,Aasen2016,Karzig2017,Litinski2017} while being a major potential source of unwanted inhomogeneity in the active segments of these devices (i.e. those that host the non-Abelian MZMs). From this perspective, understanding in detail the electrostatic effects in semiconductor Majorana devices represents a requirement. Clearly, the minimal model in which all of these crucial parameters (e.g. chemical potential, spin-orbit coupling, proximity-induced pair potential) are assumed to be independent adjustable parameters, is highly inaccurate (and perhaps even incorrect) and non-predictive, since these parameters cannot be freely tuned in any experimental hybrid system by controlling the electrostatic environment (i.e. various gate voltages).
 
In general, accounting for electrostatic effects requires solving a Schr\"{o}dinger-Poisson problem self-consistently. The Schr\"{o}dinger-Poisson problem is important in understanding the properties of low-dimensional semiconductor structures and, indeed, over the years many self-consistent treatments have been carried out in semiconductor inversion and accumulation layers \cite{Stern1972,Ando1976,DSarma1982}, semiconductor heterojunctions \cite{Stern1984} and quantum wells \cite{Bastard1983,Bauer1986}, semiconductor nanowires \cite{Lai1986,Laux1986}, and semiconductor quantum dots \cite{Stopa1996}.  Most of these self-consistent theories are carried out within the continuum effective mass approximation (sometimes with additional approximations to simplify the numerics) where the self-consistency is limited to the electrons in the semiconductor itself, thus motivating our work.  In general, these theories capture the electronic structure of the low-dimensional semiconductor systems extremely well \cite{Ando1982,Bastard1988}, and have become a standard tool in the semiconductor industry.  Our goal here is to develop a similar self-consistent tool in  hybrid structures with SM-SC interfaces,  whereas by contrast the standard low-dimensional semiconductor systems have typically SM-SM (or SM-insulator, as in Si MOSFETs) interfaces.  The presence of superconductivity, spin-orbit coupling, and magnetic field makes our problem much richer (and more difficult technically) than the above-mentioned pure semiconductor low-dimensional systems.
 
The most relevant components that determine the electrostatic effects in SM-SC structures are the applied gate potentials, the work function difference at the SM-SC interface, and the screening  due to the presence of the superconductor and the finite charge in the wire.  The topological superconducting phase and the emerging MZMs have been been found to be relatively stable against disorder and  weak interaction \cite{Gangadharaiah2011,Stoudenmire2011,Sela2011,Lutchyn2011,Lobos2012,Crepin2014,Manolescu2014}. Considering the properties of the semiconductor materials used in the fabrication of  Majorana devices and the strong screening by the superconductor, it is reasonable to assume that the main effects of electron-electron interaction are faithfully captured at the mean-field level, i.e. within the Hartree approximation.
Exchange-correlation effects may have some small quantitative effects, but given that the typical semiconductor materials used in Majorana nanowires (e.g. InSb and InAs) have very small electron effective masses and rather large lattice dielectric constants, we expect such exchange-correlation corrections to be rather negligible since the relevant dimensionless interaction coupling constant (the so-called $r_s$ value) is very small. Therefore the task at hand is to find a self-consistent solution of a three-dimensional (3D) Schr\"{o}dinger-Poisson problem associated with a given semiconductor-superconductor Majorana device. This task, however, poses a significant challenge due to the enormous number of relevant degrees of freedom that have to be taken into account.  A possible path would be the brute force approach to the 3D Schr\"{o}dinger-Poisson problem. This could be helpful in the engineering process of a specific device, but has two major disadvantages: it is an extremely costly numerical scheme and it provides virtually no additional understanding of the relevant physics and has limited predictive power. An additional (and rather serious) empirical problem associated with a brute-force 3D Schr\"{o}dinger-Poisson approach is that the relevant experimental parameters are simply not known at the level of accuracy necessary for such a method to provide reliable results at the $1\!-\!100~\mu$eV energy scale operational for the MZM problem of interest here.
 
In this work, we propose and develop an alternative approach involving an effective theory of the 3D Schr\"{o}dinger-Poisson problem that can be efficiently implemented numerically and can provide insight into the low-energy physics of the SM-SC device, particularly in terms of the dependence of key low-energy features on the SM-SC materials parameters and the applied gate voltages. Our method is based on two key ideas. (i) We split the actual 3D problem into a 2D problem corresponding to an infinite (uniform) wire and an effective multi orbital 1D problem with ``molecular'' orbitals calculated (self-consistently) using the infinite 2D system. (ii) For a given geometry, the Poisson problem is solved once for each lattice site and the results are stored; using this Green's function scheme, the Poisson component of the iteration procedure becomes trivial. More specifically, we first consider an infinite nanowire-superconductor system in the presence of an external gate potential that is translation invariant (along the wire) and calculate the transverse profiles of  the wave functions associated with each confinement-induced band by solving self-consistently the corresponding 2D Schr\"{o}dinger-Poisson problem. Next, we construct an effective multi-orbital 1D model of the 3D device by dividing the system into $N$ ``slices'' and associating to each ``slice'' molecular orbitals given by the transverse profiles of  the confinement-induced bands corresponding to an infinite wire with the same electrostatic potential as the local electrostatic potential of the ``slice,'' which is obtained by solving a 3D Laplace equation. Of course, including all the bands would simply imply a change of basis. The point is that the subspace spanned by a relatively small number of bands calculated self-consistently by solving the (2D) infinite wire problem provides a good approximation for the low-energy Hilbert space of the 3D system. The projection reduces the numerical complexity of the problem enormously, since it eliminates a large number of (irrelevant) high-energy degrees of freedom that have to be considered when using the brute force approach to the full 3D problem.
We note that both the 2D problem and the 1D effective model are solved self-consistently. The first self-consistency condition ensures that the calculated transverse profiles  (hence, the ``molecular'' orbitals) accurately include interaction effects (at the Hartree level of Coulomb energetics), while the second condition ensures that the charge is correctly distributed along the wire (within the same approximation).   
This effective approach is both computationally efficient and physically substantive, as demonstrated explicitly in the current work, being characterized by numerical tractability and  predictive power.

This work focuses on a method to effectively solve the Schr\"{o}dinger-Poisson problem in semiconductor Majorana devices, elucidating the implicit approximations as well as additional possible simplifications and refinements of the proposed approach. In addition, we provide specific examples of how one  can use this method to address important questions regarding the low-energy physics of proximity-coupled SM-SC structures.  We first consider the case of an infinite semiconductor wire in the presence of an external gate potential and a work function difference at the interface between the wire and the superconductor. We calculate the response to an external magnetic field and compare the predictions based on first order perturbation theory \cite{Vuik2016} with the fully self-consistent results. We also calculate the dependence of the ``effective chemical potential'' (in fact, the energies of the interacting semiconductor bands) on the work function difference and show that the corresponding linear coefficient is of order unity. By contrast, the dependence on the applied gate potential is strongly suppressed due to the screening provided by the superconductor. We also investigate the dependence of the band-dependent induced pair potential  on the work function difference and the applied potential and find that the low-energy bands are characterized by similar values of this parameter, in sharp contrast with predictions based on simple noninteracting models. This result is corroborated by a direct calculation of the induced gap as a function of the applied potential within a model that includes the parent superconductor explicitly.  

Next, we consider a finite wire and investigate the energy splitting oscillations of the Majorana modes arising from the overlap of the MZMs at the two wire ends of the wire. We find that interaction partially suppresses these oscillations \cite{Dominguez2016,Escribano2018}, which is an effect arising from the self-consistency in the  Schr\"{o}dinger-Poisson solution. We then consider a finite system with a nonuniform work function difference at the SM-SC interface. This non uniformity in the work function could arise, for example, from physical structural fluctuations at the interface, which are invariable at the few mono-layer level even in the best epitaxial interfaces. We find that small variations of the work function difference (of the order of $1\%-2\%$) can generate variations of the effective electrostatic potential larger than the induced gap. The screening by the superconductor plays no role in suppressing the emergence of this inhomogeneous  potential, while the screening by the charge inside the wire is only effective at high occupancies. This calculation provides concrete support to the possibility of long-range potential inhomogeneities in proximitized nanowires, which are predicted\cite{Stanescu2016,Liu2017a,Moore2018} to induce trivial low-energy states that mimic the (local) signatures of non-Abelian MZMs. We note that the typical absolute work function at the SM-SC interface is of the order of hundreds on meVs whereas the relevant low-energy energy scale (e.g. the induced gap in the nanowire) is only $\sim 100~\mu$eV, making the homogeneous control of the work function along the whole SM-SC interface a rather formidable materials science, fabrication, and engineering challenge, which must be solved for future progress in the field.  We mention as an aside that the work function inhomogeneity issue discovered in the current work is quite distinct from the short-range disorder problem associated with the SM-SC interface discussed earlier in the literature within the minimal model.\cite{Takei2013}  

The rest of the paper is organized as follows. In Sec. \ref{Th_m} we present our approach to the Schr\"{o}dinger-Poisson problem in proximitized semiconductor nanowires. We describe the Green's function scheme (Sec. \ref{Greens_Method}), its implementation in the case of infinite nanowires (Sec. \ref{Methods_infinite}), and the scheme for constructing and solving the effective 1D problem corresponding to finite systems  (Sec. \ref{Effective_method}). In Sec. \ref{Applic1}, we apply  our method to infinite Majorana nanowires and investigate the response to an external magnetic field  (Sec. \ref{Applic1A}), the dependence of the effective chemical potential on the work function difference at the SM-SC interface  (Sec. \ref{Applic1B}), and dependence of the proximity-induced pair potential on the relevant parameters  (Sec. \ref{effcoupl}). We also use our scheme to study the dependence of the induced gap on the applied gate potential for a system in the intermediate coupling regime  (Sec. \ref{indgap}). 
Section \ref{Applic2} is dedicated to finite hybrid structures of experimental relevance. We discuss the suppression of the Majorana splitting oscillations due to interaction (Sec. \ref{Moscill}) and the emergence of  inhomogeneous potentials in systems with a non-uniform work function difference (Sec. \ref{Applic2B}). The convergence of our effective theory scheme is discussed in Sec. \ref{coverg}. We conclude in Sec. \ref{Cnc} with a summary of the results and a discussion of the relevance of this work to future studies of Majorana systems.

\section{Theoretical methods} \label{Th_m}

In this section we describe our approach to the Schr\"{o}dinger-Poisson problem in proximitized semiconductor nanowires. We discuss (A) the Green's function scheme, (B) the infinite wire case, and (C) the effective 1D problem. We focus on the weak coupling regime, i.e. we assume that the low-energy wave functions have almost all their weight inside the semiconductor nanowire (with an exponentially-small tail penetrating inside the superconductor). The parent superconductor is treated as a ``boundary condition'' for the electrostatic potential. We show that the strong/intermediate coupling regime, which is expected to exhibit interesting new physics at low energies \cite{Stanescu2017a}, can also be addressed within our theoretical framework by explicitly including the superconductor in the model Hamiltonian. However, this approach is limited to simple effective models of the parent superconductor.  A more general theory of the strong/intermediate coupling regime will be discussed elsewhere. 

\subsection{The Green's function scheme} \label{Greens_Method}

Consider a $d$-dimensional semiconductor system described by a multi-orbital tight-binding Hamiltonian of the form
\begin{equation}
H = H_0 + H_{\rm int},           \label{Ham}
\end{equation}
where $H_0$ is a non-interacting Hamiltonian, which includes hopping terms, spin-orbit coupling, and external field contributions, and  $H_{\rm int}$ accounts for the electron-electron interaction. At the mean-field level, the interaction term has the form 
\begin{equation}
H_{\rm int} = \sum_{i,j}\sum_{\alpha,\beta}{U}_{ij}^{\alpha\beta} c_{i\alpha}^\dagger c_{j\beta},   \label{Eq2}
\end{equation}
where $i$ and $j$ label the lattice on which the tight-binding model is defined, $\alpha$ and $\beta$ are combined orbital and spin indices, and $U_{ij}^{\alpha\beta} = -e\langle i,\alpha|{U}|j,\beta\rangle$ are matrix elements of the Hartree potential ${U}({\bm r})$ with the basis states $|i,\alpha\rangle$ of the tight-binding model. The operator $c_{i\alpha}^\dagger$ creates an electron in a single particle state with orbital/spin index $\alpha$ centered at site $i$ (i.e. the state $|i,\alpha\rangle$). The Hartree (or Coulomb) potential satisfies the Poisson equation
\begin{equation}
{\nabla}^{2}{U}(\mathbf{r}) = - {\rho(\mathbf{r}) \over \epsilon}, \label{Poiss}
\end{equation}
where $\epsilon$ is the background dielectric constant of the semiconductor and  $\rho(\mathbf{r})$ is the charge density. In turn, the charge density can be expressed in terms of the eigenstates $\psi_n$ of Hamiltonian (\ref{Ham}) as a sum over the occupied states,
\begin{equation}
\rho({\bm r}) = -e\sum_n^{occ.}|\psi_n({\bm r})|^2.  \label{rho}
\end{equation} 
Equations (\ref{Ham}-\ref{rho}) define a Schr\"{o}dinger-Poisson problem that has to be solved self-consistently.  
The self-consistency arises from the fact that the eigenstates $\psi_n({\bm r})$, which define the charge density through  Eq. (\ref{rho}),  are in turn determined by the charge density through Eqs. (\ref{Eq2}) and (\ref{Poiss}).
We note that having a unique solution of the Poisson equation (\ref{Poiss}) requires specified boundary conditions. Also, in general, the  non-interacting Hamiltonian $H_0$ contains an external electrostatic potential generated, for example, by an applied gate voltage. Finding the spatial dependence of this external potential may require solving an additional Laplace equation, which involves knowledge of various geometrical and materials details characterizing each given device.
  It is convenient to solve the Poisson equation (\ref{Poiss}) with homogeneous boundary conditions and incorporate all non-homogeneous contributions (e.g., a non-vanishing gate voltage) into the boundary conditions of the Laplace equation. Note that the Laplace equation has to be solved once (for a given external potential configuration), while the Poisson equation has to be solved self-consistently, together with the Schr\"{o}dinger problem defined by Hamiltonian (\ref{Ham}),  within an iterative scheme, which can be computationally expensive. For example, having to solve the Poisson equation numerically at every iteration represents a serious practical obstacle when exploring the large parameter space that typically characterize the heterostructure model. In addition, numerical accuracy demands very precise solutions of the Poisson equation, making this the essential roadblock in the efficiency of the computational scheme.

To address this challenge, we reformulate the problem so that the Poisson component of each iteration becomes trivial. First, we write the eigenstates in terms of the localized basis states as 
\begin{equation}
\left|\psi_{n}\right> = \sum\limits_{j,\alpha} A_{nj\alpha} \left|j ,\alpha \right>.   \label{Eq5}
\end{equation}
Defining $\mathbb{A}^{\alpha\beta}_{nij} = A^{*}_{ni\alpha} A_{nj\beta}$, we can write the charge density in the form 
\begin{equation}
\begin{aligned}
\rho(\mathbf{r}) =&-e\sum_n^{occ.}\sum\limits_{j,\alpha} \mathbb{A}^{\alpha\alpha}_{njj} {\left| \varphi_{j\alpha}(\mathbf{r})\right|}^{2} \\
&- e\sum_{n}^{occ.} \sum_{(i,\alpha)\neq(j,\beta)}\mathbb{A}^{\alpha\beta}_{nij}~ \varphi^{*}_{i\alpha}(\mathbf{r}) \varphi_{j\beta}(\mathbf{r}),   \label{rhoA}
\end{aligned}
\end{equation}
where $\varphi_{j\alpha}({\bm r}) =\langle {\bm r}|j,\alpha\rangle$ are local orbitals. Note that the second term in Eq. (\ref{rhoA}) is due to orbital overlap and can be neglected in single-band models (see below). 

Next, we introduce the Green's function  $G_{nm}^{\alpha\beta}$  defined by the equation
\begin{equation}
{\nabla}^{2}G_{ij}^{\alpha\beta}(\mathbf{r}) =   \frac{e}{\epsilon}\varphi^{*}_{i\alpha}(\mathbf{r}) \varphi_{j\beta}(\mathbf{r})  \label{Gij}
\end{equation}
with homogeneous boundary conditions. Note that $G_{ii}^{\alpha\alpha}(\mathbf{r})$ represents the electrostatic potential generated by an electron occupying the orbital $\alpha$ at site $i$. Finally, we define the following ``interaction tensor'':
\begin{equation}
\nu^{\alpha\beta\gamma\delta}_{ijkl} = -e\int  \varphi_{i\alpha}^{*}(\mathbf{r})\varphi_{j\beta}(\mathbf{r}) G_{kl}^{\gamma\delta}(\mathbf{r}) d^{3}r.     \label{nuij}
\end{equation}
The element $\nu^{\alpha\alpha\gamma\gamma}_{iikk}$ represents the interaction energy between two electrons occupying the orbitals $\alpha$ at site $i$ and $\gamma$ at site $k$, respectively. Note that, in general, the Green's function defined by Eq. (\ref{Gij}) and the interaction tensor defined by Eq. (\ref{nuij}) are complex quantities. Using these quantities, we can write the matrix elements of the Hartree potential in the form
\begin{eqnarray}
U_{ij}^{\alpha\beta} &=&\sum_n^{occ.}\sum_{k,l}\sum_{\gamma,\delta}  \nu^{\alpha\beta\gamma\delta}_{ijkl} ~\mathbb{A}^{\gamma\delta}_{nkl}.   \label{UnuA}
\end{eqnarray}
Our strategy is to solve Eq. (\ref{Gij})  for every lattice site in the system, which can be done numerically or, in some cases, even analytically (see, for example,  Appendix \ref{App1}), perform the integration in Eq. (\ref{nuij}), and store the interaction tensor.  The Poisson component of the iterative scheme reduces to tensor contraction in Eq. (\ref{UnuA}). We note that, in practice, many elements of the interaction tensor are small and can be safely neglected. Also, the calculation of the interaction tensor using Eq. (\ref{nuij}) requires knowledge of the basis states $\varphi_{i\alpha}(\mathbf{r})$, which can be found using ab-intio techniques. In the applications discussed in this work we only consider single-orbital models and 
we assume that $\varphi_{i}(\mathbf{r})$ has spherical symmetry and is strongly localized near site $i$ (i.e. we neglect the overlap with neighboring orbitals). 

The general scheme described above simplifies significantly in the case of single-orbital tight-binding models.  Since the only internal degree of freedom is spin, we have $\alpha \rightarrow \sigma$, where $\sigma=\pm$ is the spin index. Furthermore, the spatial profile of the local orbital is spin-independent, so that we have $\langle{\bm r}|i,\sigma\rangle = \varphi_i({\bm r})|\sigma\rangle$, and we neglect the overlap between neighboring orbitals, $ \varphi_i({\bm r}) \varphi_j({\bm r}) =\delta_{ij}|\varphi_i({\bm r})|^2$. With these simplifications the relevant Green's function that has to be calculated (for each lattice site $i$) becomes
\begin{equation}
{\nabla}^{2}G_{i}(\mathbf{r}) = \frac{e}{\epsilon}  {\left| \varphi_{i}(\mathbf{r}) \right|}^{2} \label{Geq1}
\end{equation}
and the interaction tensor (\ref{nuij}) reduces to an interaction matrix,
\begin{equation}
\nu_{ij} =  -e \int G_{i}(\mathbf{r}) {\left| \varphi_{j}(\mathbf{r}) \right|}^{2} d^{3}r. \label{Int1}
\end{equation}
Note that $\nu_{ij}$ is simply the effective Coulomb interaction energy between two electrons at sites $i$ and $j$, respectively. Finally, the interaction term from Hamiltonian (\ref{Ham}) becomes local and can be expressed in terms of the matrix elements of the Hartree potential as 
\begin{eqnarray}
H_{int} &=&\sum_{i,\sigma}U_i~c_{i\sigma}^\dagger c_{i\sigma}, \nonumber \\
U_i &=& \sum_n^{occ.}\sum_{j\sigma} \nu_{ij} ~|A_{nj\sigma}|^2. \label{U1}
\end{eqnarray}

The usefulness of this method becomes clear if we consider exploring a large parameter space within a given device geometry. As long as the geometry of the system remains fixed, we can change various system parameters, such as back gate potentials, magnetic fields, and spin-orbit couplings,  while using the same interaction matrix, which is determined once at the beginning of the calculation.
Moreover, since finite element computational methods can automatically handle unconventional and complicated device geometries, this method can be applied to devices having arbitrary shape, with any number of gates, different dielectric materials, and arbitrary spatial dimension $d$. Thus, the method described above is of wide applicability to actual systems of experimental relevance.

A generalization of this method that explicitly incorporates the parent superconductor is straightforward. The Hamiltonian of the hybrid system has the generic form $H=H_0+H_{\rm int} + H_{\rm sc} + H_{\rm sm-sc}$, where the first two terms correspond to the Hamiltonian in Eq. (\ref{Ham}), while $H_{\rm sc}$ models the parent superconductor. To preserve the numerical efficiency of the method, the modeling of the superconductor has to be simple, e.g., $H_{\rm sc}$ can be a single-band tight-binding model with superconducting correlations incorporated at the mean-field level through an on-site pairing potential $\Delta_0$. The last term in the Hamiltonian describes the coupling between the semiconductor wire and the superconductor. We note that within this generalization there are no constraints regarding the coupling strength at the SM-SC interface, i.e., the method can be applied to both weak-coupled and intermediate/strong-coupled hybrid systems. In the generalized scheme, the charge density inside the SM wire is calculated using the particle components of the eigenstates of the total Hamiltonian. Explicitly, Eq. (\ref{rho}) is replaced by 
\begin{equation}
\rho({\bm r}) = -e \sum_{n, \sigma}^{occ.} |u_{n\sigma}(\bm r)|^2, \label{rhobis}
\end{equation}
where ${\bm r}$ is a position vector inside the SM wire and $u_{n\sigma}$ are the particle components (corresponding to the spin projection 
$\sigma$) of the spinor $\psi_{n} =(u_{n\uparrow}, u_{n\downarrow}, v_{n\uparrow}, v_{n\downarrow})^T$ representing an eigenstate of the 
full Hamiltonian. Note that, in general, the eigenstates extent into both the SM wire and the parent SC, but only the components inside the wire 
contribute to the charge density $\rho({\bm r})$. Finally, Eq. (\ref{Eq5}) becomes
\begin{equation}
\left|u_{n}\right> = \sum\limits_{j,\alpha} A_{n\!j\alpha} \left|j ,\alpha \right>,
\end{equation}
where we have incorporated the spin into the state label, $(n,\sigma)\rightarrow n$, to simplify the notation. The remaining steps can be implemented as described above. In particular, the Green's function (\ref{Gij})  that provides the solution of the Poisson component of the Schr\"{o}dinger-Poisson problem remains unchanged. Note that the key difference between the basic scheme and this generalization is that the explicit treatment of the parent SC allows one to account for the fact that the low-energy states have spectral weight inside both the SM wire and the SC. This is particularly important in the intermediate/strong coupling regime. The generalized scheme should be used to calculate key effective model parameters, such as the induced gap, the Rashba spin-orbit coupling, and the $g$ factor. We emphasize that (i) these are {\em not} independent parameters, but should be determined self-consistently as functions of the electrostatic parameters of the system (e.g., work function difference and gate potentials) and the coupling strength across the SM-SC interface, and (ii) these parameters can be strongly position-dependent, particularly near the ends of the wire or inside tunnel barrier regions, hence it is important to determine them using a 3D Schr\"{o}dinger-Poisson scheme (see below, Sec. \ref{Effective_method}).

\subsection{Schr\"{o}dinger-Poisson scheme for infinite nanowires} \label{Methods_infinite} %%%%%%%%%%%%%%%%%%%%%%%%%%%%

While using the Green's function method makes a 3D Schr\"{o}dinger-Poisson problem significantly more manageable (this technique being clearly preferable to the pure brute force self-consistent approach), a direct 3D calculation may still be prohibitively costly due to the large number of (relevant) degrees of freedom. To overcome this challenge, we split the 3D problem into a 2D problem corresponding to an infinite uniform wire and an effective 1D problem associated with the actual finite structure. In this section we describe the self-consistent procedure for solving the 2D Schr\"{o}dinger-Poisson problem using the general framework discussed above. 

%%%%%%%%%%%%%%%%%%%%%%%%%%%%%%%%
%%%%%%%%%%%%%%%%%%%%%%%%%%%%
\begin{figure}[t]
\begin{center}
\includegraphics[width=0.36\textwidth]{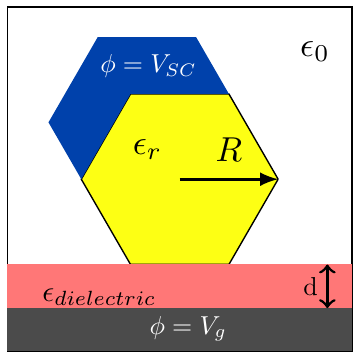}
\end{center}
\caption{(Color online) Typical transverse profile of a  Majorana SM-SC heterostructure. The SM nanowire (yellow) is partially covered by an s-wave SC (blue)  and placed on an insulating substrate (light red). A back gate (black) creates a controllable electrostatic potential.}
\label{FIG1}
\vspace{-3mm}
\end{figure}
%%%%%%%%%%%%%%%%%%%%%%	
%%%%%%%%%%%%%%%%%%%%%%%%%%%%%%%%

Consider an infinite quasi-1D semiconductor (SM) nanowire proximity coupled to an s-wave superconductor (SC). The axis of the wire is oriented along the $x$-direction, while the finite cross section has a geometry similar to that shown in Fig. \ref{FIG1}, which is the typical experimental setup for Majorana nanowires. The semiconductor nanowire (e.g., InSb or InAs) is partially covered by an s-wave superconductor (e.g., Al or NbTiN)  and placed on an insulating substrate. A controllable back gate allows one to change the electrostatic potential across the wire. For clarity and to avoid cumbersome notations, we restrict ourselves to single-orbital tight-binding models and we neglect the overlap between neighboring orbitals, which allows us to use the simplified version of the Green's function scheme described above. However, we emphasize that the approach is generic and can be directly generalized to the multi-orbital case.
The non-interacting part of the Hamiltonian describing the nanowire has the form
  \begin{equation}
\begin{aligned}
H_{0}^{\infty} =&  \sum_{i,j,k,\sigma} \left[t_{ij} + \left({{\hbar}^2 {k}^2 \over 2{m}^*}  + V_i + E_{0}\right)\delta_{i,j}\right]
{c}^\dagger_{ik\sigma} c_{jk\sigma} \\
& +\sum\limits_{i,k,\sigma,\sigma^{'}} \Gamma~ {c}^\dagger_{ik\sigma}(\sigma_{x})_{\sigma\sigma^{'}} c_{ik\sigma^{'}} \\
& + \sum\limits_{i,k,\sigma,\sigma^{'}}  {\alpha_R k} \left[  {c}^\dagger_{ik\sigma}(\sigma_{y})_{\sigma\sigma^{'}}   c_{ik\sigma^{'}} + H.C. \right],  \label{HamInf}
\end{aligned}
\end{equation}
where $i, j\in{\mathcal L}$ are position labels in the transverse ($y$-$z$) plane (i.e. normal to the nanowire direction taken to be the $x$-direction throughout) and ${c}^\dagger_{ik\sigma}$ creates an electron at position $i$ with longitudinal wave vector $k$ and spin $\sigma$.  Note that the lattice ${\mathcal L}$ is only defined inside the SM nanowire.
In Eq. (\ref{HamInf}), $t_{ij}$ are matrix elements for hopping across the wire, $\hbar^2 {k}^2 \over 2{m}^*$ (with $m^*$ being the effective mass) is the longitudinal component of the kinetic energy, $V_i$ represents the external potential at site $i$ arising from the back gate and the work function difference at the SC-SM interface, and $E_0$ is a reference energy (determined by the value of the SM band gap and the possible presence of dopants) that controls the minimum of the (noninteracting) spectrum for an isolated SM wire. In the last two terms,  $\Gamma$ represents the (half) Zeeman splitting due to a magnetic field applied parallel to the wire,  $\alpha_R$ is the Rashba spin-orbit coupling, and $\sigma_\mu$ (with $\mu=x,y,z$) are Pauli matrices associated with the spin degree of freedom. Note that the (infinite) wire has translational invariance in the $x$ direction and, therefore, $k\equiv k_x$ is a good quantum number. Also, we assume that the SM-SC coupling is weak, which means that the SC can be treated as (i) a source of Cooper pairs for the wire (with pairing potential $\Delta$) and (ii) a boundary condition for the electrostatic problem. The weak-coupling assumption, used extensively in the Majorana nanowire literature, enables one to integrate out all the complications of the underlying superconductor in terms of a single pairing potential parameter characterizing the induced proximity effect.

The electrostatic potential $V_i$ has to be calculated by solving a Laplace equation with boundary conditions determined by the geometry of the problem and by two key parameters: the gate voltage $V_g$ and the work function difference at the interface, $V_{SC}$ (see Fig. \ref{FIG1}). We emphasize that, for a given SM model, the parameters $V_g$, $V_{SC}$, and $E_0$ completely determine the carrier concentration in the nanowire and the transverse profiles of the wave functions and effective electrostatic potential (which includes the interaction effects at the mean-field level). Hence the chemical potential of the wire (relative to, e.g., the bottom of the spectrum), the Rashba coefficient $\alpha_R$, and the induced pairing potential $\Delta$ are not independent parameters (as implicitly assumed in the extensively used minimal model), but rather functions of $V_g$, $V_{SC}$, and $E_0$, the actual independent parameters of the microscopic theory.

The interaction effects are incorporated at the mean field (Hartree) level by adding to Hamiltonian (\ref{HamInf}) the term 
\begin{equation}
H_{int} = \sum_{i,k,\sigma} U_i {c}^\dagger_{ik\sigma} c_{ik\sigma},  \label{Hint1}
\end{equation}
where $U_i$ are the matrix elements of the Hartree potential. These matrix elements are determined by the interaction matrix (\ref{Int1}) and  by eigenstates $\psi_{n k \sigma}(i) \equiv A_{n k i \sigma}$ of the full Hamiltonian $H^\infty=H_0^\infty+H_{int}$,  where $n$ labels the confinement-induced transverse modes. Explicitly, we can write the matrix elements of the Hartree potential in the form
\begin{equation}
U_i = \sum_{n,k,\sigma}^{occ.}\sum_{j} \nu_{ij} ~|A_{nkj\sigma}|^2.  \label{Uinf}
\end{equation}
 Solving the Schr\"{o}dinger-Poisson problem for the infinite wire implies solving Eq. (\ref{Geq1}) with homogeneous boundary conditions (once) for each lattice site $i$ corresponding to a transverse section of the wire, calculating and storing the interaction matrix $\nu_{ij}$ given by Eq. (\ref{Int1}), then solving self-consistently the Schr\"{o}dinger problem for  $H^\infty=H_0^\infty+H_{int}$ with the matrix elements of the Hartree potential being given by Eq. (\ref{Uinf}). 
 
%%%%%%%%%%%%%%%%%%%%%%%%%%%%%%%%
%%%%%%%%%%%%%%%%%%%%%%%%%%%%
\begin{figure}[t]
\begin{center}
\includegraphics[width=0.48\textwidth]{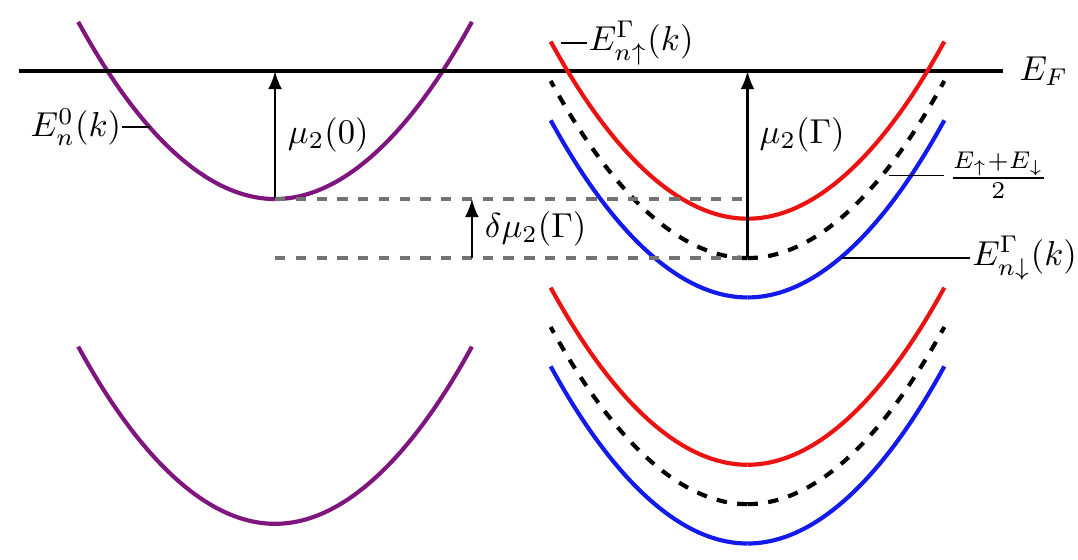}
\end{center}
\caption{(Color online) ({\em  Left}): Confinement-induced bands for an infinite wire with $\alpha_R = 0$ in the absence of an applied Zeeman field (i.e. for $\Gamma=0$). ({\em Right}): Applying a magnetic field ($\Gamma \neq 0$) splits the bands into pairs of spin sub-bands (solid lines). The dashed lines represent the (fictitious) energy dispersion corresponding to $\Gamma=0$ that is used in the definition of the effective chemical potential $\mu_n$. For each band $\mu_n$ is defined with respect to the minimum (at $k=0$) of the corresponding dashed line. The effective chemical potential of the second band,  $\mu_{2}$, is shown as an example.}
\label{FIG2}
\vspace{-3mm}
\end{figure}
%%%%%%%%%%%%%%%%%%%%%%	
%%%%%%%%%%%%%%%%%%%%%%%%%%%%%%%%

A few comments regarding the practical implementation of this scheme are warranted. First, we note that the basis states  $\varphi_i$ of the tight-binding model are typically unspecified. Moreover, we often deal with effective tight-binding models defined on a lattice having a unit cell much larger then the atomic unit cell of the semiconductor. Hence,  $\varphi_i$ should not necessarily be regarded as atomic-type orbitals. In such cases, a reasonable approximation that can be easily implemented numerically is based on the assumption that the charge  associated with $\varphi_i({\bm r})$ is uniformly distributed throughout the unit cell. Second, we note that, imposing only minor additional restrictions, we can find an analytic solution of Eq. (\ref{Geq1}). The main idea is to solve the Poisson problem in a cylindrical geometry, then use a conformal mapping to obtain the results for, e.g., a hexagonal wire. The details of this calculation are provided in Appendix \ref{App1}.

Finally, let us discuss qualitatively the effect of the (mean-field) electron-electron interaction on the energy spectrum of the infinite SM wire. A quantitative analysis will follow in Sec. \ref{Applic1}. The transverse confinement of the nanowire gives rise to confinement-induced one-dimensional sub-bands (henceforth referred to as ``bands''), as shown in Fig. \ref{FIG2}. Here, for simplicity, we take $\alpha_R=0$. We define the effective chemical potential measured relative to the bottom of a given band $n$ as $\mu_n = - E_n^0(0)$, where $E_n^0(0)$ is the n$^{\rm th}$ energy eigenvalue of $H^\infty$ corresponding to $\Gamma=0$ (i.e. no Zeeman field) and $k=0$. In the presence of a Zeeman field, the effective chemical potential is defined as $\mu_n(\Gamma) = -\frac{1}{2}[E_{n\uparrow}^\Gamma(0) + E_{n\downarrow}^\Gamma(0)]$, where $E_{n\sigma}^\Gamma(0)$ is the energy (at $k=0$) of the corresponding spin-split sub-band (see Fig. \ref{FIG2}).
Defining such a quantity can be useful in the context of Majorana physics, for example, when discussing the ``topological condition,''  $\Gamma > \sqrt{{\mu}_{n}^{2} + \Delta^{2} }$, where $n$ is the topmost occupied band. Note that $\mu_n$ is positive for occupied bands and negative for empty bands. Neglecting interactions results in a chemical potential that is independent of the applied Zeeman field, $\mu_n(\Gamma) = \mu_n(0)=const.$
However, due to interaction effects, the dependence of $\mu_n$ on control parameters such as the Zeeman field becomes nontrivial. Indeed, turning on $\Gamma$ splits each band into two spin sub-bands, as shown in Fig. \ref{FIG2}. With increasing $\Gamma$, the higher-energy spin sub-band ``loses'' occupied states, while its lower energy partner gains occupied states. The net gain (or loss) is, in general, nonzero, which implies that the occupation of each band will change and, consequently, the Hartree potential (\ref{Uinf}) will change. In turn, this shifts the effective chemical potential of each band by an amount $\delta\mu_n(\Gamma)$ that has to be determined self-consistently. 
We conclude that applying a magnetic field does not simply split the bands. Instead, due to interactions, the Zeeman effect has to be supplemented by band-dependent energy shifts that can only be determined by solving the Schr\"{o}dinger-Poisson problem self-consistently. Hence the effective chemical potential varies with $\Gamma$, leading to important consequences regarding the dependence of various low-energy features on the applied magnetic field.

%%%%%%%%%%%%%%%%%%%%%%%%%%%%%%%%
%%%%%%%%%%%%%%%%%%%%%%%%%%%%
\begin{figure}[t]
\begin{center}
\includegraphics[width=0.48\textwidth]{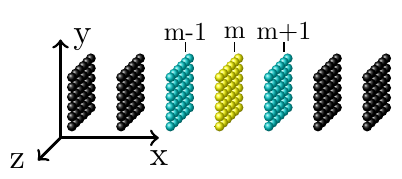}
\end{center}
\caption{(Color online) Schematic representation of the layers (slices) used for constructing the effective 1D model. A generic site of the 3D lattice is labeled $(i, m)$, where $m$ is the layer index and $i$ indicates the transverse position within the layer.}
\label{FIG3}
\vspace{-3mm}
\end{figure}
%%%%%%%%%%%%%%%%%%%%%%	
%%%%%%%%%%%%%%%%%%%%%%%%%%%%%%%%

\subsection{The effective 1D problem for finite systems}           \label{Effective_method} %%%%%%%%%%%%%%%%%%%%%%%%

Consider a finite nanowire oriented along the $x$ direction and having a certain transverse profile. We divide the wire into $N_{x}$ layers (or slices), each containing $N_{\bot}$ sites, as shown schematically in Fig. \ref{FIG3}. The corresponding 3D Hamiltonian has the form
\begin{equation}
\begin{aligned}
H_{3D} =& \sum\limits_{i,j,m,\sigma} t^{\bot}_{ij} c_{im\sigma}^{\dagger}c_{jm\sigma}
 + \sum\limits_{i,m,n,\sigma} t^{\parallel}_{mn} c_{im\sigma}^{\dagger}c_{in\sigma} \\
 & + \sum\limits_{i,m,\sigma} \left(V_{im}+U_{im}\right) n_{im\sigma} \\
  & + \sum\limits_{i,m,\sigma,\sigma^{\prime}} i\alpha_{R}  \left[c_{i(m+1)\sigma}^{\dagger} \left(\sigma_{y}\right)_{\sigma\sigma^{\prime}} c_{im\sigma^{\prime}}  + h.c. \right]\\
  & +\sum\limits_{i,m,\sigma,\sigma^{\prime}} \Gamma~ c_{im\sigma}^{\dagger} \left(\sigma_{x}\right)_{\sigma\sigma^{\prime}} c_{im\sigma^{\prime}}, \label{H3D}  
\end{aligned}
\end{equation}
where $c_{im\sigma}^{\dagger}$ creates an electron with spin $\sigma$ localized near the site $i$ of layer $m$,  $n_{im\sigma} = c_{im\sigma}^{\dagger}c_{im\sigma}$ is the number operator,
$t^{\bot}_{ij}$ and $t^{\parallel}_{mn}$ are intra- and inter-layer nearest neighbor hopping matrix elements, respectively, $\Gamma$ is the (half) Zeeman splitting, and  $\alpha_{R}$ is the Rashba spin-orbit coefficient.  The electrostatic effects are described by the external potential $V_{im}$ and by the mean field contribution $U_{im}$, which will be determined self consistently using the generic Green's function method discussed in Sec. \ref{Greens_Method} and the procedure described below.  Note that a reference energy $E_0$ that controls the minimum of the (noninteracting) spectrum for an isolated SM wire [see Eq. (\ref{HamInf})] can be incorporated into $V_{im}$.

First, for each layer $m$ we define the following auxiliary Hamiltonian: 
\begin{equation}
\begin{gathered}
H^{(m)}_{aux} = \sum\limits_{i,j,k,\sigma} \!\!\left[ t^{\bot}_{ij} + \left({\hbar^{2}k^{2} \over 2m^{*}} 
\!+\!V_{i}^{(m)} \!\!+\! U_{i}^{(m)} \right) \delta_{ij} \right] c_{ik\sigma}^{\dagger} c_{jk\sigma} \\
 +\sum\limits_{ik\sigma\sigma^{\prime}} \alpha_{R}k ~c_{ik\sigma}^{\dagger} \left(\sigma_{y}\right)_{\sigma\sigma^{\prime}} c_{ik\sigma^{\prime}}, \label {Haux}
 \end{gathered}
\end{equation} 
where $V_{i}^{(m)} = V_{im}$. The auxiliary model, which describes an infinite wire, is defined on a lattice with a transverse profile that matches the lattice of layer $m$, i.e. the local transverse profile of the original 3D system. Note that, Hamiltonian (\ref{Haux}) represents a specific case of the infinite wire problem considered in Sec. \ref{Methods_infinite} corresponding to an external potential  $V_{i}^{(m)} = V_{im}$ and no Zeeman field, i.e. $\Gamma=0$. In other words, the auxiliary Hamiltonian $H^{(m)}_{aux}$ describes an infinite system in the presence of a translation-invariant external potential that matches the local external potential of the actual 3D wire on layer $m$.

The $k$-independent transverse components of the single-particle eigenstates of the auxiliary Hamiltonian have the form
\begin{equation}
\left| \varphi_{\alpha} ^{m} \right > = \sum\limits_{j} S_{\alpha j}^{m} \left|j\right>,  \label{Smtx}
\end{equation}
where $\left|j\right>$ is the local orbital at site $j$ and $\alpha$ is a band index. Note that the label for the spin degree of freedom has been suppressed. By convention,  the Roman letters $i, j, \dots$ label (transverse) positions within the wire, as well as the corresponding local orbitals. On the other hand, the Greek letters $\alpha, \beta, \dots$ will be used to designate confinement-induced bands and the  ``molecular orbitals'' $\left| \varphi_{\alpha} ^{m} \right >$ associated with the transverse profile of the corresponding band. 

Next, we perform a change of basis in the tight-binding Hamiltonian $H_{3D}$, from the local orbitals $|j m\rangle$ to the molecular orbitals $\left| \varphi_{\alpha} ^{m} \right >$ given (for each layer) by the $k=0$ eigenstates of the auxiliary problem (\ref{Haux}). For convenience, we introduce the ``vector'' operator $\bar{c}$ with components $\bar{c}_\ell= c_{im\sigma}$ labeled by $\ell= \ell(m,i,\sigma)=2 (m-1) N_{\bot} +2i-1+\sigma$. Here, we have $1\leq m \leq N_x$, $1\leq i\leq N_{\bot}$, and $\sigma =  \uparrow  \equiv 0$ or $\sigma = \downarrow \equiv 1$, so that the total number of degrees of freedom (which gives the size of $\bar{c}$) is $2 N_x N_{\bot}$. Similarly, we label the molecular orbital basis with $\nu=\nu(m,\alpha,\sigma) = 2 (m-1) N_{\bot} +2\alpha-1+\sigma$.
Using these notation, we rewrite the Hamiltonian (\ref{H3D}) in a more compact (and generic) form as
\begin{equation}
H_{3D} = \sum \limits_{\ell\ell^\prime} \bar{c}_{\ell}^{\dagger} \left[ \bar{ t}_{\ell\ell^\prime}^{\bot} + \bar{t}_{\ell\ell^\prime}^{\parallel} + (\bar{V}_{\ell} + \bar{U}_{\ell})\delta_{\ell\ell^\prime} + \bar{\Gamma}_{\ell\ell^\prime} + \bar{\alpha}_{\ell\ell^\prime} \right] \bar{c}_{\ell^\prime}, \label{H3Dgen}
\end{equation}
where the nonzero matrix elements match the corresponding quantities from Eq. (\ref{H3D}). The structure of these matrices is discussed in Appendix \ref{App2}. Now let $\bar{S}$ be the transformation matrix that generates the desired change of basis. The element $\bar{ S} _{\nu \ell}$ of the transformation matrix corresponding to $\nu=\nu(m,\alpha, \sigma)$ and $\ell =\ell( m, j,\sigma)$ is given by the coefficient in Eq. (\ref{Smtx}), $\bar{ S} _{\nu \ell} = S_{\alpha j}^{m}$.
Inserting the identity $\sum\limits_{\nu} \bar{S}_{\ell\nu}^{ \dagger}\bar{ S} _{\nu \ell^\prime} = \delta_{\ell\ell^\prime}$ in Eq. (\ref{H3Dgen}) and defining the annihilation operator for the molecular orbital, $\tilde{c}_{\nu} = \sum_{\ell}  \bar{S}_{\nu \ell } \bar{c}_{\ell}$, leads to
\begin{equation}
\begin{aligned}
H_{3D} =& \sum \limits_{\nu\nu^\prime} \tilde{c}_{\nu}^{\dagger} 
\left[ \sum\limits_{\ell\ell^\prime} \bar{S}_{\nu\ell} \left[\bar{t}^{\bot}_{\ell\ell^\prime} + (\bar{V}_{\ell} + \bar{U}^{aux}_{\ell})\delta_{\ell\ell^\prime}\right] 
\bar{S}^{\dagger}_{\ell^\prime\nu^\prime} \right] \widetilde{c}_{\nu^\prime} \\
 +& \sum \limits_{\nu\nu^\prime} \tilde{c}_{\nu}^{\dagger} \left[\tilde{t}^{\parallel}_{\nu\nu^\prime} +\widetilde{\Delta U}_{\nu\nu^\prime} + \widetilde{\Gamma}_{\nu\nu^\prime} + \widetilde{\alpha}_{\nu\nu^\prime} \right] \widetilde{c}_{\nu^\prime}, \label{H3Dtrans}
\end{aligned}
\end{equation}
where  $\widetilde{D} = \bar{S} \bar{D} \bar{S}^{\dagger}$ for all matrices $\bar{D}$ from Eq. (\ref{H3Dgen}). The potential 
$\bar{U}^{aux}$ is the mean field contribution determined self-consistently by solving the auxiliary problem (\ref{Haux}) for each layer $m$, i.e. for $\ell=\ell(m,i,\sigma)$ we have $\bar{U}^{aux}_{\ell} = U_i^{(m)}$. The additional term $\widetilde{\Delta U} = \bar{U} - \bar{U}^{aux}$ represents the difference between the mean-field potential $\bar{U}$ calculated self-consistently for the original 3D problem and $\bar{U}^{aux}$.  Noticing that  the quantity between the square brackets in the first term of Eq. (\ref{H3Dtrans})  is nothing but an eigenvalue $\epsilon_\nu=\epsilon_\alpha^m$ of the auxiliary Hamiltonian (\ref{Haux}) for $k = 0$, we can write the 3D Hamiltonian in the form
\begin{equation}
H_{3D} = \sum \limits_{\nu\nu^\prime} \tilde{c}_{\nu}^{\dagger} \left[\epsilon_{\nu}\delta_{\nu\nu^\prime}+\widetilde{t}^{\parallel}_{\nu\nu^\prime} +\widetilde{\Delta U}_{\nu\nu^\prime} + \widetilde{\Gamma}_{\nu\nu^\prime} + \widetilde{\alpha}_{\nu\nu^\prime} \right] \tilde{c}_{\nu^\prime}   \label{H3Dx}
\end{equation}

So far, we have made no approximation; the physics described by Eq. (\ref{H3Dx}) is exactly the same as that described by the original Hamiltonian  (\ref{H3D}). However, the key point of this construction is that the low-energy sub-space of the original problem (which is the relevant sub-space for understanding Majorana physics) is well approximated by the low-energy subspace spanned by a relatively small number $n_o$ of molecular orbitals.  In other words, we can project the 3D Hamiltonian onto the low-energy sub-space spanned by the molecular orbitals $\left| \varphi_{\alpha} ^{m} \right >$ with $\alpha < n_o$. The projection generates the following effective 1D Hamiltonian
\begin{equation}
\begin{gathered}
H_{\rm eff} = \sum_{m,n,\sigma}\sum_{\alpha,\beta} ^\bullet \tilde{t}^{\parallel}_{m\alpha,n\beta} ~ {c}_{m\alpha\sigma}^{\dagger}{c}_{n\beta\sigma} +\sum_{m,\sigma}\sum_{\alpha} ^\bullet \epsilon_{\alpha}^m ~{n}_{m\alpha\sigma} \\
+\sum_{m,\sigma\sigma^\prime}\sum_{\alpha,\beta} ^\bullet \left[\widetilde{\Delta U}_{\alpha \beta}^{m}~\delta_{\sigma\sigma^\prime} + {\Gamma}\left(\sigma_{x}\right)_{\sigma\sigma^{\prime}}\delta_{\alpha\beta}\right]c_{m\alpha\sigma}^{\dagger}  c_{m \beta \sigma^{\prime}} \\
+\sum_{m,n,\sigma\sigma^\prime}\sum_{\alpha,\beta} ^\bullet i \alpha_{\alpha\beta}^{mn} (\sigma_y)_{\sigma\sigma^\prime} ~c_{m\alpha\sigma}^{\dagger}  c_{n \beta \sigma^{\prime}}     \label{Heff}
\end{gathered}
\end{equation}
where $m$ and $n$ label the sites of the (finite) 1D lattice, $\alpha$ and $\beta$ designate the molecular orbitals, and the summations marked by a $\bullet$ symbol are restricted to the lowest energy orbitals, i.e, $\alpha,\beta \leq n_o$.  The hopping matrix elements $\tilde{t}^{\parallel}_{m\alpha,n\beta}$ can be written in terms of the hopping matrix $[T^\parallel]_{im,jn} = {t}^{\parallel}_{m n}~\delta_{ij}$ between layers $m$ and $n$ as
\begin{equation}
\tilde{t}^{\parallel}_{m\alpha,n\beta} = \langle\varphi_\alpha^m|T^\parallel|\varphi_\beta^n\rangle.
\end{equation}
Starting with nearest-neighbor hopping  ${t}^{\parallel}_{m n}$ in Eq. (\ref{H3D}) results in an effective 1D model with  nearest-neighbor hopping  $\tilde{t}^{\parallel}_{m\alpha,n\beta}$. Note, however, that the hopping matrix elements of the effective Hamiltonian are, in general, orbital- and position-dependent. The position dependence and orbital mixing can be particularly strong at the ends of the wire or inside the transition regions between a segment of the wire that is covered by a superconductor and a segment that is not covered (e.g., a tunnel barrier region). This behavior is generated by the transverse profiles (i.e. molecular orbitals) being position-dependent inside the transition region. Similar considerations also apply to the spin-orbit coupling term. However, for a quantitative description of position-dependent spin-orbit coupling one should start with a more detailed model of the 3D wire, e.g., using an eight-band Kane-type Hamiltonian, rather than the simple phenomenological term discussed here. This is certainly doable, but unnecessary at this stage, as we focus on the basic ideas of the effective theory.
Nonetheless, it is important to emphasize that, based on the present analysis, we can conclude that accurate modeling of inhomogeneous regions such as, for example, the tunnel barrier region at the end of a proximitized wire, using effective 1D Hamiltonians should necessarily involve position-dependent hopping/spin-orbit coupling and orbital mixing terms, in addition to the potential barriers that are typically considered in the literature.  This physics of the position dependence is not accounted for in the usual minimal model of Majorana nanowires.

Calculating the matrix elements $\widetilde{\Delta U}_{\alpha \beta}^{m}$ of the mean-field potential is a straightforward extension of the Green's function method discussed in Sec. \ref{Greens_Method}. 
%However, the projection onto the low energy subspace would do us little good if for every iteration we had construct $\tilde{U}_{\alpha \beta}$ using the Green's functions, $G_{j}$, of every lattice site. Fortunatly, we can calculate some quanties using the Green's functions once before starting the self-consistent iteration scheme that make reconstruction of the Hamiltonian far more efficient.   
Let $\left|\psi_\lambda \right>$ be an eigenstate of the effective Hamiltonian (\ref{Heff}). Expanding it in terms of molecular orbitals, $|\varphi_\alpha^m\rangle$, then in terms of local orbitals, $|j m\rangle$, we have
\begin{equation}
\left|\psi_\lambda \right> = \sum_{m,\sigma}\sum_{\alpha}^\bullet A_{\lambda, m \alpha\sigma} \left|\varphi_{\alpha}^m\right> =\!\!
 \sum_{m, j, \sigma}\sum_{\alpha}^\bullet A_{\lambda, m \alpha\sigma}~ S_{\alpha j}^{m}~|j m\rangle.  
\end{equation}
The interaction matrix $\nu_{im, jn}$ of the original 3D problem is determined by solving equations (\ref{Geq1}) and (\ref{Int1}) for the corresponding system. This encodes the interaction energy between two electrons occupying the local orbitals $|im\rangle$ and $|j n\rangle$, respectively.
It is convenient to define the molecular orbital interaction tensor given by 
\begin{equation}
\widetilde{\nu}_{~m~\!n}^{\alpha \beta ~\!\gamma \delta} = \sum_{i, j} \mathbb{S}_{\alpha \beta, i m} ~\nu_{im,jn}~\mathbb{S}_{\gamma \delta, jn}, \label{lamdba}
\end{equation}
where $\mathbb{S}_{\alpha\beta, im} = [{S}^{*}]_{\alpha i}^m {S}_{\beta i}^m$. Note that $\widetilde{\nu}$ has the same structure as the interaction tensor (\ref{nuij}), with $i=j\rightarrow m$ and $k=l\rightarrow n$. In particular, the element $\widetilde{\nu}_{~m~n}^{\alpha \alpha ~\gamma \gamma}$ represents the interaction energy between two electrons occupying the molecular orbitals $\alpha$ on site $m$ and $\gamma$ on site $n$, respectively. Finally, 
using the results of Sec. \ref{Greens_Method}, one finds that the matrix elements of the mean-field potential are given by 
\begin{equation}
\widetilde{\Delta U}_{\alpha\beta}^m = \sum \limits_\lambda^{occ.} \sum_{n,\sigma}\sum\limits_{\gamma\delta}^\bullet \widetilde{\nu}_{~m~\!n}^{\alpha \beta ~\!\gamma \delta} ~\mathbb{A}_{\lambda, n\sigma}^{\gamma\delta} -\langle\varphi_\alpha^m|U^{(m)}|\varphi_\beta^m\rangle, \label{DUab}
\end{equation}
where $\mathbb{A}_{\lambda, n \sigma}^{\gamma\delta} = A^*_{\lambda,n\gamma\sigma}~A_{\lambda,n\delta\sigma}$ and we have subtracted the matrix elements of the mean-field potential associated with the auxiliary problem (\ref{Haux}).

We conclude this section with a summary of our approach to the  Schr\"{o}dinger-Poisson problem in semiconductor Majorana devices. 
Assume that a finite nanowire described by a 3D tight-binding model, e.g., the Hamiltonian given by  Eq. (\ref{H3D}), is weakly coupled to a superconductor. The first step is to calculate the external electrostatic potential  $V_{im}$ by solving a Laplace equation with appropriate boundary conditions. The result  will depend on the geometry of the system, as well as the applied gate potential $V_g$ (or, more generally, $V_{g1}, V_{g2}, \dots$ in a system with multiple gates) and the work function difference at the SM-SC interface, $V_{SC}$. Second, we divide the nanowire into $N_x$ layers and solve the auxiliary (infinite wire) problem (\ref{Haux}) for each layer, following the self-consistent procedure described in Sec. \ref{Methods_infinite}. The third and final step involves solving the effective 1D problem (\ref{Heff}) self-consistently using the matrix elements (\ref{DUab}) of the mean-field potential. We emphasize that the properties of the system in the superconducting state are obtained by solving the Bogoliubov de Gennes (BdG) problem defined by the Hamiltonian
\begin{equation}
H_{BdG} = H_{\rm eff} + \sum_{n,m,\alpha,\beta} \left[\langle\varphi_\alpha^n|\Delta_{\rm ind}|\varphi_\beta^m\rangle c_{n\alpha \uparrow}^\dagger c_{m\beta \downarrow}^\dagger + h.c.\right],
\end{equation} 
where $\Delta_{\rm ind}(i,j)$ is a proximity-induced anomalous term defined at the SM-SC interface. Note that in the presence of low-energy states (e.g., at finite magnetic fields) the charge density ia always calculated using Eq. (\ref{rhobis}) [instead of  Eq. (\ref{rho})], to account for particle-hole mixing. The essence of the approximation involved in this effective theory approach is the ansatz that the transverse profiles of the low-energy states at a given location along the wire are similar to the profiles of the low-energy confinement-induced bands of an infinite wire under the same electrostatic conditions. The theory includes mode mixing due to off diagonal terms in the effective Hamiltonian, which allows for corrections to these profiles. If one includes enough molecular orbitals into the basis of the effective model, the low-energy physics of the system is accurately described. One can systematically check if enough orbitals have been included by increasing $n_o$ and monitoring the convergence of the results.  Finally, we emphasize that both the auxiliary (infinite wire) problem and the effective 1D problem are solved self-consistently. Using the Green's function approach reduces the Poisson components of these problems to the summations in Eqns. (\ref{Uinf}) and (\ref{DUab}), respectively.

\section{Electrostatic effects in infinite wires} \label{Applic1}

In this section we illustrate the implementation of the general scheme described above focusing on the infinite wire case. We address three basic questions: (i) how are the spectral features (in particular the effective chemical potential) modified by the presence of an external Zeeman field, (ii) what is the dependence on the work function difference $V_{SC}$, and (iii)  how does the effective SM-SC coupling depend on the back gate voltage $V_g$? Throughout this section we consider an infinitely long wire of radius  $R=50~$nm  (see Fig. \ref{FIG1}) described by a Hamiltonian $H^\infty = H_0^\infty + H_{int}$ given by Eqns. (\ref{HamInf}) and (\ref{Hint1}). The parameters of the model correspond to an InSb nanowire and we have  $m^{*} = 0.014m_{0}$, where $m_{0}$ is the bare electron mass, the nearest-neighbor hopping matrix element $t_{ij} = -0.083~$eV, and the relative permittivity $\epsilon_{r} = 17.7$. The total number of lattice sites corresponding to the hexagonal cross section of the wire is  $N_\bot=1176$. In Sec. \ref{indgap} we use a smaller lattice spacing corresponding to $t_{ij} = -0.453~$eV and $N_\bot=2206$. For simplicity, we ignore the spin-orbit coupling (i.e., we set $\alpha_R = 0$), and use the analytical solution of the Green's function described in appendix \ref{App1},  except in Sec. \ref{indgap}, where we have a Rasba coefficient of $500~$meV$\cdot$\AA$~$ and we find the Green's function numerically. The self-consistent Schr\"{o}dinger-Poisson scheme that we use is discussed in Sec. \ref{Methods_infinite}.

%%%%%%%%%%%%%%%%%%%%%%%%%%%%%%%%
%%%%%%%%%%%%%%%%%%%%%%%%%%%%
\begin{figure}[t]
\begin{center}
\includegraphics[width=0.48\textwidth]{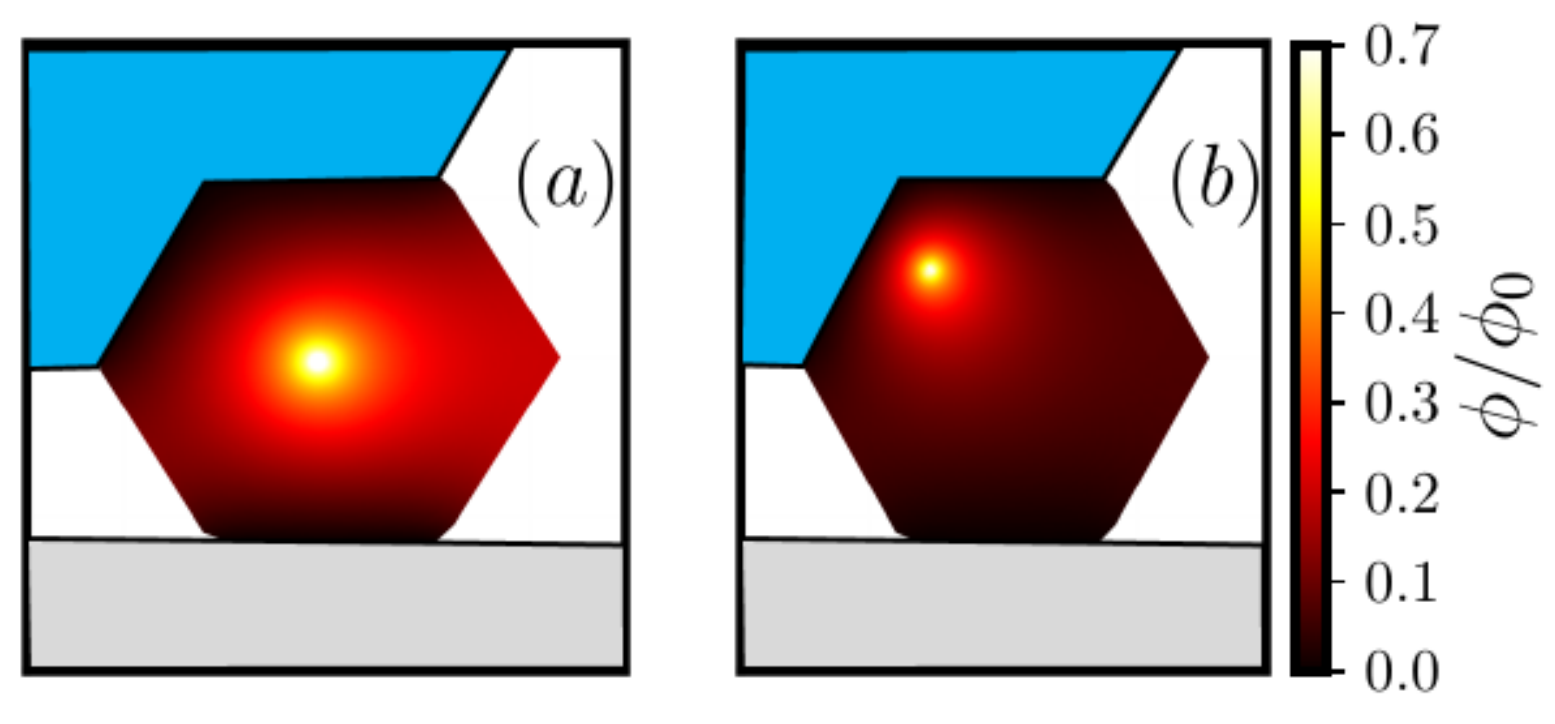}
\end{center}
\vspace{-3mm}
\caption{(Color online)  Normalized potential profiles corresponding to the Green's function $G_i({\bm r})$ generated by an infinite line charge placed inside the nanowire at a position given by the lattice site $i$. Placing the charge in the vicinity of the superconductor [panel (b)] results in a strongly screened potential.}
\label{FIG4}
\vspace{-3mm}
\end{figure}
%%%%%%%%%%%%%%%%%%%%%%	
%%%%%%%%%%%%%%%%%%%%%%%%%%%%%%%%

Before addressing the main questions, we make two general remarks. First, we note that the potential created by the charge inside the semiconductor is strongly screened by the superconductor and the back gate. To illustrate this point and to show the structure of the Green's function, we calculate  the potential profile created by an infinite line charge placed inside the nanowire at a position corresponding to the lattice site $i$, i.e., we calculate the Green's function $G_i({\bm r})$. The results are shown in  Fig.(\ref{FIG4}). Note that a charge placed near the middle of the wire, i.e. far from the SM-SC interface and the back gate [panel (a)],  generates a potential characterized by a spatial extent much larger than that of a potential created by a charge in the vicinity of the SM-SC interface [panel (b)]. This implies that the effect of Coulomb  interactions (at the mean-field level) is significantly reduced due to screening by the SC. The back gate has a similar effect. Consequently, the spatial profile of the (occupied) transverse modes is expected to determine the strength of interaction effects: the effects will be strong if the charge is located away from the SM-SC and back-gate interfaces  and weak if (most of) the charge is localized in the vicinity of an interface. 

%%%%%%%%%%%%%%%%%%%%%%%%%%%%%%%%
%%%%%%%%%%%%%%%%%%%%%%%%%%%%
\begin{figure}[t]
\begin{center}
\includegraphics[width=0.46\textwidth]{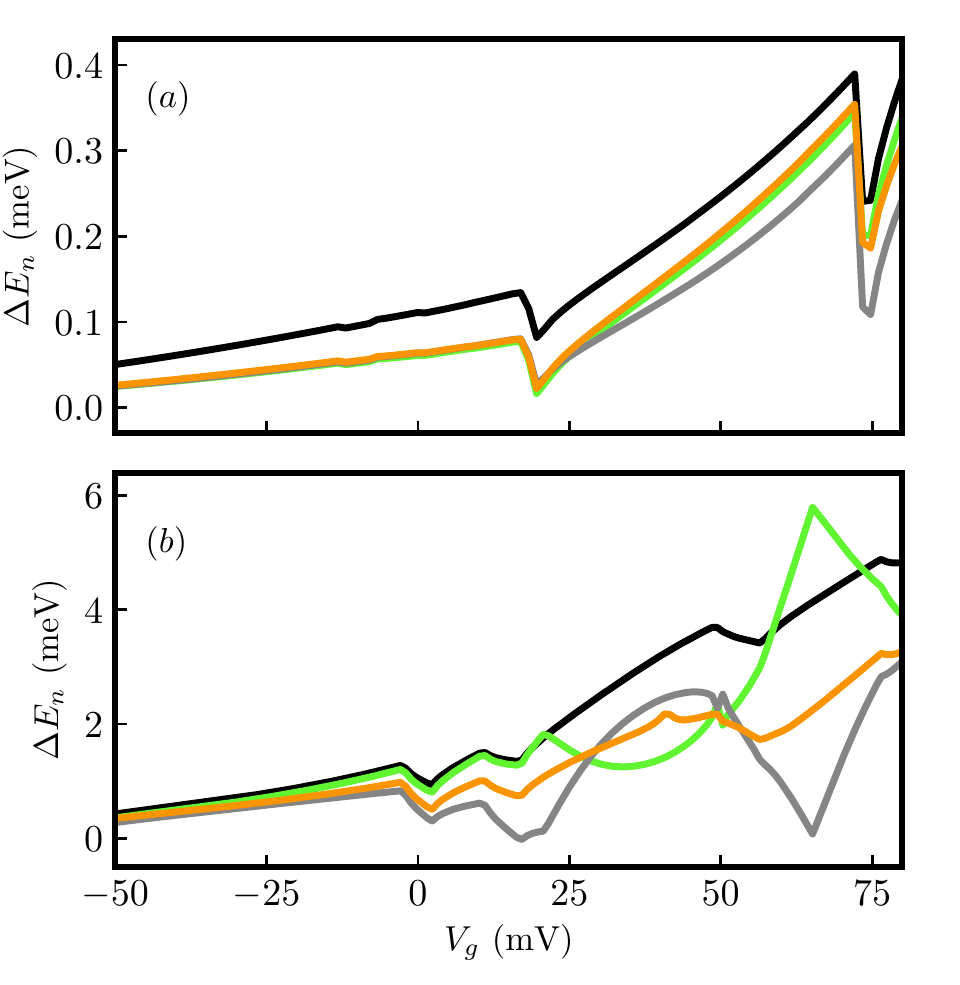}
\end{center}
\vspace{-3mm}
\caption{(Color online) Energy difference $\Delta E_n$ between eigenvalues calculated (i) fully self-consistently and (ii) using perturbation theory as function of the back gate voltage, $V_g$. Only the lowest four bands are shown. The parameters used in the calculations are: (a) $V_{SC} = 150~$mV,  $E_0 = 100~$meV, and (b) $V_{SC}$ = 50 mV,  $E_0 = -10~$meV. Note that the energy scales in the two panels differ by an order of magnitude. The color code for the bands is: black ($n=1$), green ($n=2$), gray ($n=3$), orange ($n=4$). We note that $\Delta E_n$ is a measure of how strongly the wave function profiles are affected by interactions.}     
\label{FIG5}
\vspace{-3mm}
\end{figure}
%%%%%%%%%%%%%%%%%%%%%%	
%%%%%%%%%%%%%%%%%%%%%%%%%%%%%%%%

Second, we would like to estimate the importance of self-consistency in solving the Schr\"{o}dinger-Poisson problem. Are fully self-consistent calculations really necessary? This is obviously important from a practical viewpoint since the self-consistent procedure is computationally costly even within our effective theory approach (and hopelessly complicated in a brute-force direct 3D approach).
To address this question, we compare fully self-consistent calculations with results obtained by treating electronic interactions within first order perturbation theory. We note that the first order perturbation theory relies on the assumption that the wave functions associated with different transverse modes are not affected by interactions (i.e., that they are solely determined by the external fields). Therefore, any discrepancy between the two methods is a result of the electronic interactions changing the wave function profiles. Details concerning
the perturbative calculations are given in Appendix \ref{App3}. A comparison between self-consistent calculations and perturbative results  for two different sets of parameters is shown in Fig. \ref{FIG5}. We plot the energy difference $\Delta E_n$ between the eigenstates calculated using the two methods (for the lowest four bands) as function of the applied gate voltage. To understand the behavior illustrated in Fig. \ref{FIG5}, we note that positive values of $V_{SC}$, as well as negative gate voltages $V_g$, result in the electrons being pushed toward the SM-SC interface, where the screening by the superconductor reduces interaction effects.  We emphasize that, even in this situation, the energies of the eigenstates are   significantly renormalized by interactions, but the profiles of the wave functions are barely affected, as demonstrated by the low values of $\Delta E_n$ in Fig. \ref{FIG5} corresponding to this regime. Applying a positive gate potential moves the charge distribution toward the center of the wire, where the interaction effects are stronger. In addition, choosing a negative reference energy $E_0$ [see panel (b)] corresponds to the isolated nanowire being electron-doped, i.e. having more charge carriers. Increasing the charge density enhances the strength of interaction effects, including the interaction-induced change of the wave function profiles. A second factor that contributes to the enhancement of $\Delta E_n$ in panel (b) is a lower value of $V_{SC}$ [as compared to that used in panel (a)],  which diminishes the attraction of electrons toward the SM-SC interface and reduces screening. As a final comment, we note that the energy differences in Fig. \ref{FIG5} can be large on the scale relevant for Majorana physics. Thus, a perturbation theoretic treatment of Coluomb interaction may be quantitatively completely unreliable since the sub-band energy scale is large compared with the delicate energy scale associated with the near-zero-energy Majorana physics.
Also, if one is interested in Majorana devices that contain segments of the wire that are not covered by a superconductor (e.g., a tunnel barrier region), one should expect strong interaction effects, which requires a fully self-consistent treatment.  

\subsection{Electrostatic response to an applied magnetic field} \label{Applic1A}

%%%%%%%%%%%%%%%%%%%%%%%%%%%%%%%%
%%%%%%%%%%%%%%%%%%%%%%%%%%%%
\begin{figure}[t]
\begin{center}
\includegraphics[width=0.48\textwidth]{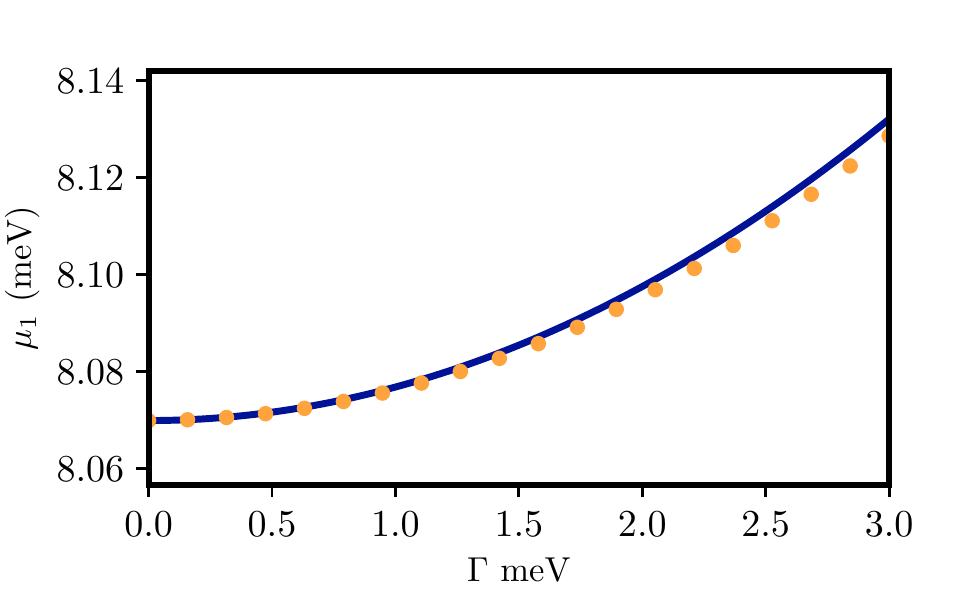}
\end{center}
\vspace{-3mm}
\caption{(Color online) Dependence of the effective chemical potential on to applied magnetic field for a system with single-band occupancy. The solid blue line corresponds to the analytic solution given by  Eq. (\ref{delmu1}), while the orange dots are the numerical results of the fully from self-consistent calculation. The parameters that control the electrostatic properties of the system are: $V_{SC}=150~$mV, $E_0=100~$meV, and $V_g=-30~$mV.}
\label{FIG6}
\vspace{-3mm}
\end{figure}
%%%%%%%%%%%%%%%%%%%%%%	
%%%%%%%%%%%%%%%%%%%%%%%%%%%%%%%%

We investigate the response of the system to an applied Zeeman field  focusing on the field dependence of the effective chemical potential. In Sec.  \ref{Methods_infinite} we have defined the  chemical potential measured relative to the bottom of a given band $n$ as $\mu_n(\Gamma) = -\frac{1}{2}[E_{n\uparrow}^\Gamma(0) + E_{n\downarrow}^\Gamma(0)]$, where $E_{n\sigma}^\Gamma(0)$ is the energy (at $k=0$) of the corresponding spin-split sub-band (see Fig. \ref{FIG2}). The dependence of $\mu_n$ on the Zeeman field $\Gamma$ has been studied in Ref. \onlinecite{Vuik2016} based on a perturbative scheme. Here, we systematically compare the perturbation theory results with the fully self-consistent calculation. This has a double purpose: on the one hand it serves as a test ground for the numerical implementation of our self-consistent scheme and, on the other hand, it provides a systematic evaluation of the accuracy of the  perturbation theory approach.  

%%%%%%%%%%%%%%%%%%%%%%%%%%%%%%%%
%%%%%%%%%%%%%%%%%%%%%%%%%%%%
\begin{figure}[t]
\begin{center}
\includegraphics[width=0.46\textwidth]{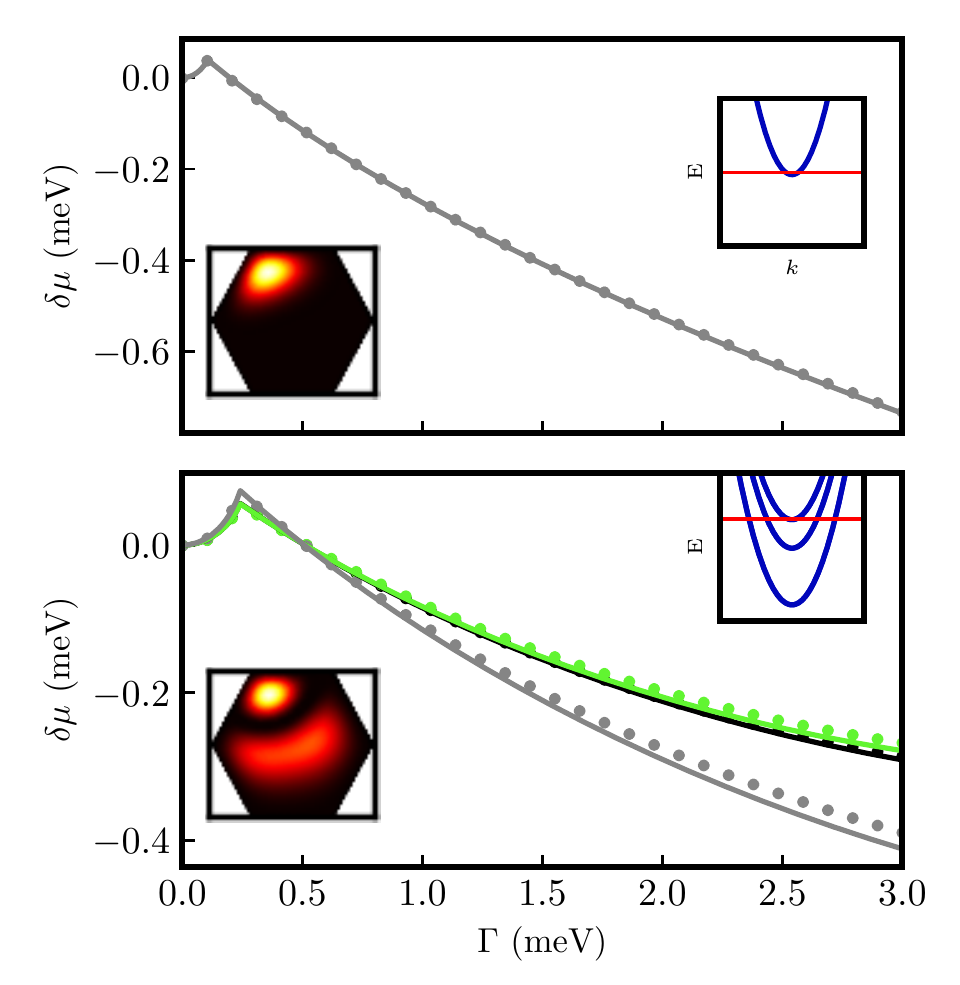}
\end{center}
\vspace{-3mm}
\caption{(Color online) Dependence of the effective chemical potential $\Gamma_n$ on the  applied Zeeman field field. The difference $\delta\mu_n$ is defined as $\delta\mu_n(\Gamma) = \mu_n(\Gamma) - \mu_n(0)$. {\em Top}: System with single band occupancy corresponding to the parameters $V_{SC} = 150~$mV, $E_0=100~$meV, and $V_g = -125~$mV.  {\em Bottom}: Same system (i.e. $V_{SC} = 150~$mV, $E_0=100~$meV), but with three occupied bands, which corresponds to applying a positive gate potential  $V_g = 75~$mV. The dotted and solid lines are obtained using the self-consistent approach and the perturbation method, respectively. The gray, green, and black lines represent the highest energy, middle, and lowest energy bands, respectively.  The corresponding spectra are shown in the upper right insets, while the wave function profiles of the highest occupied bands are shown in the lower left insets.}
\label{FIG7}
\vspace{-3mm}
\end{figure}
%%%%%%%%%%%%%%%%%%%%%%	
%%%%%%%%%%%%%%%%%%%%%%%%%%%%%%%%

We start with a comparison between the chemical potential calculated fully self-consistently for a system with single band occupancy and the low magnetic field analytical solution obtained in Appendix \ref{App3}. We note that Eq. (\ref{delmu1}) is valid in the low field (high chemical potential) regime, $\mu_1\gg \Gamma$. The results are shown in Fig. \ref{FIG6}. 
Note that the two methods are in excellent agreement, suggesting that the transverse profile of the lowest-energy band is practically independent of the applied magnetic field. In the light of the general comments made at the beginning of this section, these results are not surprising. Indeed, the relatively large (positive) $V_{SC}$ and the negative gate voltage strongly push the charge toward the SM-SC interface. Increasing the Zeeman field changes the occupation of the lowest band, but the effect is too weak to modify the transverse profile. Note, however, that the energy of the band (i.e. the effective potential $\mu_1$)  changes significantly with the applied Zeeman field. 

Next, we consider several cases characterized by different values of the  parameters that control the electrostatic properties of the system, $V_{SC}$, $E_0$, and $V_g$. In Fig. \ref{FIG7} we fix the intrinsic system parameters $V_{SC}$ and $E_0$ and tune the device from a regime characterized by single-band occupancy (top panel) to a regime characterized by three occupied bands (bottom panel)  by changing the applied gate voltage. 
The self-consistent and the perturbative results are shown as points and solid lines, respectively. The wave function profile corresponding to the 
highest-energy occupied bands are shown in the lower left insets. Note that  the effective chemical potential initially  increases  with the Zeeman field, until the highest-energy spin sub-band is completely depleted. At higher fields, the effective chemical potential decreases to reduce the amount of the charge that is added to the system as the low-energy spin sub-band ``sinks'' with increasing $\Gamma$. 

The trends revealed by Figs. \ref{FIG6} and \ref{FIG7} can be naturally interpreted as corresponding to  the intermediate regime between the constant chemical potential and constant density limits. Indeed, in the absence of electronic interactions the effective chemical potential is independent 
of the  Zeeman field. In the opposite limit, which corresponds to strong interactions, the effective chemical potential $\mu_1$ for a system with a single occupied band will decrease in a (nearly) one-to-one correspondence with the (half) Zeeman splitting $\Gamma$ to maintain a constant charge density (so as to minimize the Coulomb energy cost). The situation is slightly more complicated in the case of multiple occupied bands. Nonetheless, the results in Fig. \ref{FIG7} show clearly that the  rates of change of the effective chemical potentials with respect to $\Gamma$ are significantly  lower than the expected behavior in the constant density limit. 

An important feature in Fig. \ref{FIG7} is the good agreement between the perturbative results and the self-consistent solutions. The agreement is slightly better in the case of a single occupied band (top panel) primarily due to the wave function profile, which  is very localized near the SM-SC interface, resulting in a nearly complete screening  of the electronic interactions. By contrast, in the lower panel (i.e. for a system with three occupied bands), the top band  has a significant portion of its wavefunction near the center of the wire, where screening is incomplete. As a result, the wave function profile acquires a dependence on the applied Zeeman field and the self-consistency starts to matter. 
To clearly see why a discrepancy between the two methods implies a change in the wave function profile, recall that within first order perturbation theory one simply finds the energy shift by calculating the expectation value of the perturbation using the unperturbed wave functions. In this case, the perturbation is generated by the change in the charge density of each band, $\delta n_n(\Gamma) =n_n(\Gamma) -n_n(0)$, due to the shift $\delta \mu_n(\Gamma)$ in the effective chemical potential of the band caused by the applied magnetic field, $\delta \mu_n(\Gamma) =\mu_n(\Gamma) -\mu_n(0)$. As shown in Appendix \ref{App3}, the first order perturbation theory yields
\begin{equation}
\delta \mu_{n} = -e^2\sum_{n^\prime}^{occ.} P_{nn^\prime} ~\delta n_{n^\prime},
\end{equation}
where the matrix elements of the  reciprocal capacitance, $P_{nn^\prime} = \langle \psi_{n k \sigma}^0 |P_{n^\prime}|\psi_{n k \sigma}^0\rangle$ are calculated using the {\em fully self-consistent} wave functions  $|\psi_{n k \sigma}^\Gamma\rangle$  (for arbitrary $k$ and $\sigma$) at $\Gamma = 0$. Hence,  good agreement between the two methods implies that  $|\psi_{n k \sigma}^\Gamma\rangle$ is, practically, $\Gamma$-independent, while discrepancies reveal the change of the wave function profile with the Zeeman field.

%%%%%%%%%%%%%%%%%%%%%%%%%%%%%%%%
%%%%%%%%%%%%%%%%%%%%%%%%%%%%
\begin{figure}[t]
\begin{center}
\includegraphics[width=0.48\textwidth]{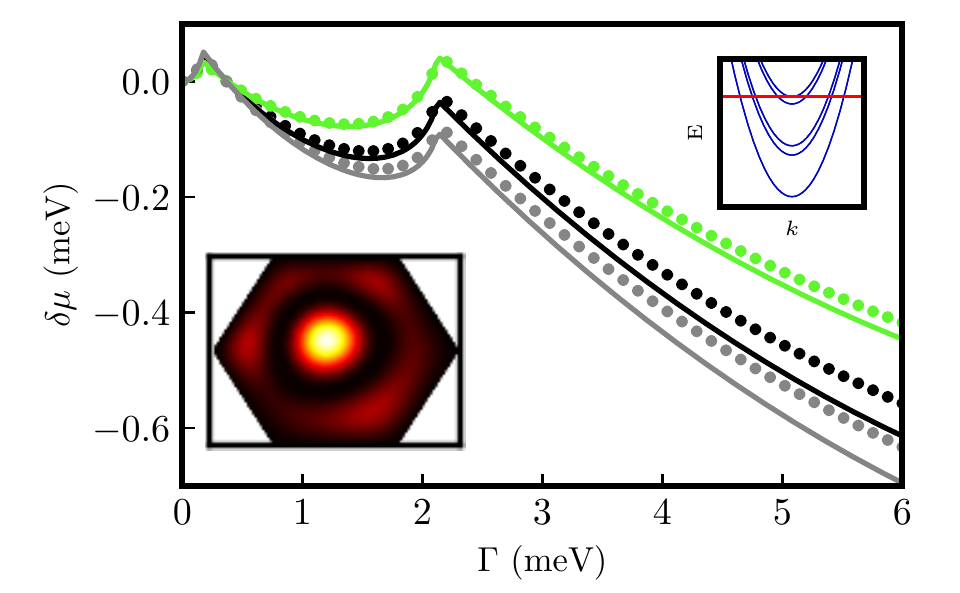}
\end{center}
\vspace{-3mm}
\caption{(Color online)  Dependence of the effective chemical potential on the  applied Zeeman field field, $\delta\mu_n(\Gamma) = \mu_n(\Gamma) - \mu_n(0)$, for a doped wire with  $V_{SC} = 50~$mV, $E_0=10~$meV, and $V_g = 12.5~$mV. The dotted and solid lines are obtained using the self-consistent approach and the perturbation method, respectively. The green, black, and gray  lines represent the third, fourth, and fifth energy bands, respectively.The wave function profile of the highest energy band (lower left inset) shows that most of the charge is localized away from the interfaces with the SC and the back gate. Consequently, the agreement between the fully self-consistent calculation and the perturbative result is significantly weaker than in Fig. \ref{FIG7}.}
\label{FIG8}
\vspace{-3mm}
\end{figure}
%%%%%%%%%%%%%%%%%%%%%%	
%%%%%%%%%%%%%%%%%%%%%%%%%%%%%%%%

To further test these findings, we consider a doped wire (i.e. $E_0<0$) with five occupied bands and a wave function profile heavily peaked  in the middle of the wire. The results are shown  in Fig.  \ref{FIG8}. The self-consistency is clearly more important in this case, although for low Zeeman fields  the perturbation theory still provides a reasonably good approximation.  In addition, we note that having two nearly degenerate top bands results in a second increase of $\delta\mu_n$ with $\Gamma$ (in the low-field regime) associated with the depletion of a spin-split sub-band. We conclude that using perturbation theory with  a reciprocal capacitance matrix $P_{nn^\prime}$ calculated (fully self-consistently) at reference field (e.g., $\Gamma=0$) provides a very good approximation over a wide regime of parameters. This result can be understood in the light of our discussion of the results shown in Fig. \ref{FIG5}. Indeed, the typical values of the Zeeman splitting are small on the energy scale corresponding to the variation of the gate voltage in Fig. \ref{FIG5}. Hence the wave function profiles are largely determined by the electrostatic parameters $V_{SC}$, $E_0$, and $V_g$ and have a very weak dependence on $\Gamma$. However, we emphasize again that the perturbative approach itself starts with a self-consistent calculation of the wave function profiles at a reference field, e.g., $\Gamma=0$, and then treats the field dependence perturbatively.

\subsection{Dependence on the work function difference} \label{Applic1B}

%%%%%%%%%%%%%%%%%%%%%%%%%%%%%%%%
%%%%%%%%%%%%%%%%%%%%%%%%%%%%
\begin{figure}[t]
\begin{center}
\includegraphics[width=0.48\textwidth]{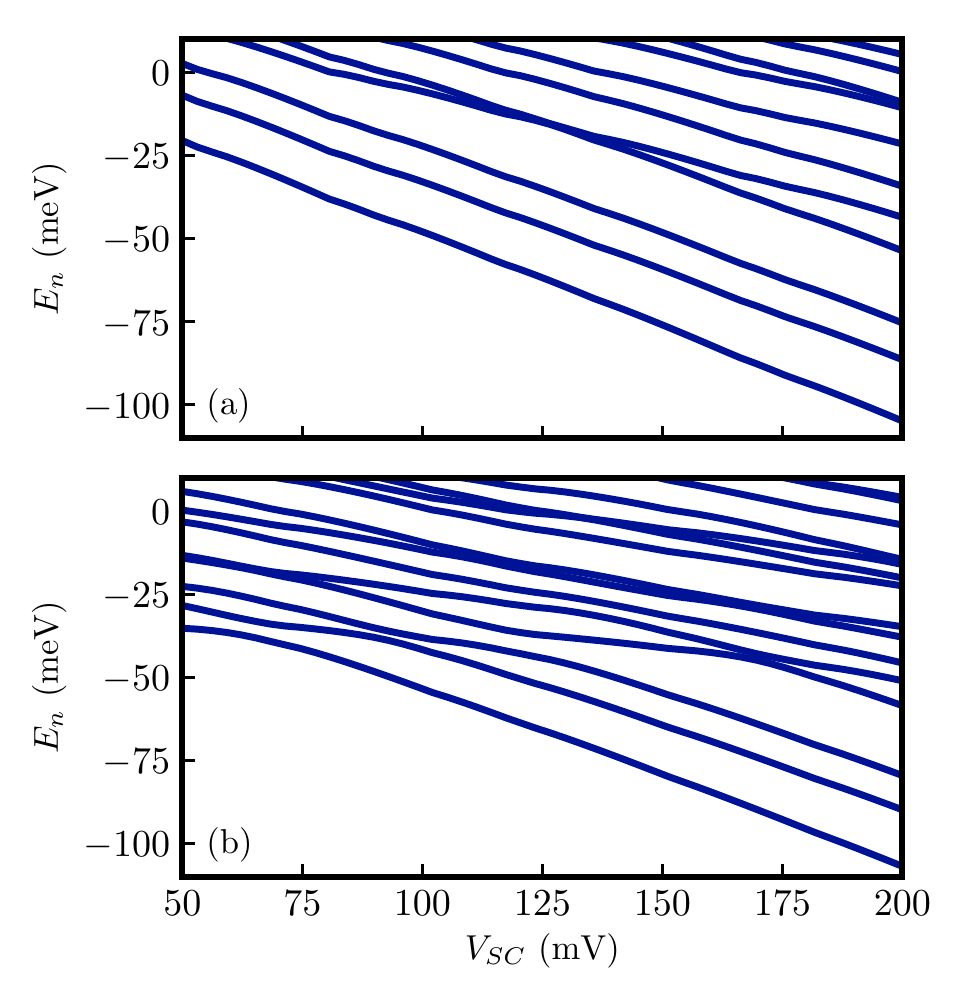}
\end{center}
\vspace{-3mm}
\caption{(Color online) Dependence of the band energies $E_n(k)$ (for arbitrary $k$) on  the work function difference $V_{SC}$ for a doped system with $E_0=-10~$meV and two different gate potentials: (a) $V_g = -75~$meV and (b) $V_g = 75~$mV. Note that the slope $|\Delta E_n/\Delta V_{SC}|$  is of the order one (more specifically, approximately $0.5$) over a wide range of parameters. When $V_{SC} < V_g$, most of the charge is located away from the SM-SC interface and the bands depend weakly on $V_{SC}$ (b). }     
\label{FIG9}
\vspace{-3mm}
%%%% Will redo a perturbation calculation will a smaller V_SC and negative E_0
\end{figure}
%%%%%%%%%%%%%%%%%%%%%%	
%%%%%%%%%%%%%%%%%%%%%%%%%%%%%%%%

A key parameter that controls the electrostatic properties of the system is the work function difference $V_{SC}$. Unfortunately, this parameter is not uniquely determined by the materials of the heterostructure (i.e. the SM and the SC),  as it depends on certain details of the SM-SC interface that, in turn, are determined by the fabrication procedure,  e.g., the exact procedure used for treating the SM wire surface before depositing the superconductor \cite{Gul2017}. In fact, it is rather difficult to obtain the interface work function difference experimentally, particularly at the level of accuracy (better than 1$\%$) relevant for Majorana physics in nanowires.  In particular, there could very easily be sample-to-sample work function differences for the same type of SM-SC hybrid structures depending on the fabrication details. In fact, even within a single sample, there could be local position dependent variations in $V_{SC}$ along the nanowire length.
Here, we treat  $V_{SC}$ as an unknown  phenomenological parameter and determine the dependence of the low-energy spectrum on this parameter by solving the Schr\"{o}dinger-Poisson problem self-consistently. 

The results for a doped nanowire with $E_0=-10~$meV and two different values of the gate potential are shown in Fig. \ref{FIG9}. First, we note that in the presence of a negative gate voltage [panel (a)] the slope $|\Delta E_n/\Delta V_{SC}|$ is approximately $0.5$ (i.e. of order one) for all low-energy bands and for a wide range of $V_{SC}$ values. This behavior can be understood in terms of the charge being pushed toward the $SM-SC$ interface, i.e. being localized in a region where the effective potential is of the order of  $V_{SC}$. Changing the sign of the applied voltage [panel (b)] results in wave functions that are more spread over the cross section of the wire. However, quite remarkably, for $V_{SC} > V_g$ most of the low-energy modes still exhibit a strong dependence on $V_{SC}$. This dependence becomes weaker when $V_{SC} < V_g$. Nonetheless, for positive gate potentials and arbitrary values of $V_{SC}$, there are many modes that are predominantly localized near the SM-SC interface and show a strong dependence on  the work function difference. In addition, there are some modes that are localized away from the interface, which exhibit a significantly weaker dependence on $V_{SC}$. These modes are also expected to have weaker proximity-induced superconductivity, as discussed below. A major consequence of the strong dependence on $V_{SC}$  illustrated in Fig. \ref{FIG9}
 is that weak inhomogeneities  in the work function difference (e.g., due to the surface treatment of the SM wire) could result in significant inhomogeneities of the effective potential along the wire. 
For example, considering the system from Fig. \ref{FIG9}, a 2$\%$ variation of $V_{SC}$ may result in a variation of the effective potential of the order of $1~$meV, which is large (typically, by a factor of $4-10$) when compared with the induced gap. We emphasize that the screening by the superconductor plays no role in reducing these potential variations. On the other hand, screening by the charge inside the wire may suppress the inhomogeneity. We will address this problem in Sec. \ref{Applic2} in the context of finite wires.

%%%%%%%%%%%%%%%%%%%%%%%%%%%%%%%%
%%%%%%%%%%%%%%%%%%%%%%%%%%%%
\begin{figure}[t]
\begin{center}
\includegraphics[width=0.48\textwidth]{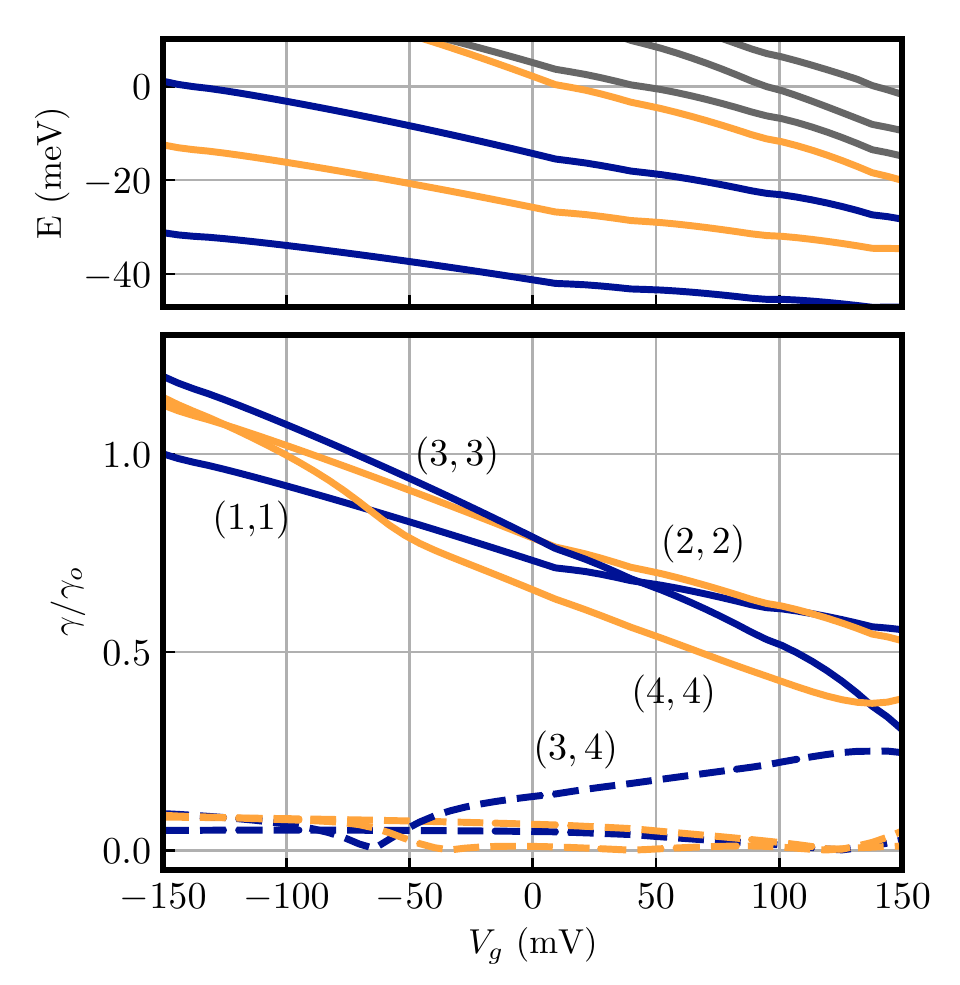}
\end{center}
\vspace{-3mm}
\caption{(Color online) {\em Top}: Energy eigenvalues as a function of the applied gate potential for a system with $V_{SC} = 200~$mV and $E_{0} = 100~$meV.  {\em Bottom}: Dependence of the normalized effective SM-SC coupling matrix,  $\gamma_{mn}(V_g)/ \gamma_{11}(-150)$, on the gate voltage, $V_g$, for the four lowest energy bands ($n,m=1, \dots, 4$). The solid and dashed lines represent diagonal matrix elements and off-diagonal elements of the form $m= n-1$, respectively. Note that all diagonal elements have  similar magnitudes, while the off-diagonal elements are negligible,  until $V_{g}$ becomes comparable to $V_{SC}$.}     
\label{FIG10}
\vspace{-3mm}
%%%% Will redo a perturbation calculation will a smaller V_SC and negative E_0
\end{figure}
%%%%%%%%%%%%%%%%%%%%%%	
%%%%%%%%%%%%%%%%%%%%%%%%%%%%%%%%

\subsection{Effective semiconductor-superconductor coupling} \label{effcoupl}

Another property that we investigate in the context of infinite wires is the dependence of the effective SM-SC coupling on the applied gate potential. 
This parameter is very important as it determines the strength of the proximity effect, including  the magnitude of the induced gap.
In general, we can define the effective SM-SC coupling as\cite{Stanescu2010,Stanescu2013}
\begin{equation}
\widetilde{\gamma}_{ij} = t_{i i^\prime} \frac{-1}{\pi}{\rm Im}\left[G_{i^\prime j^\prime}^{(SC)}(0)\right] t_{j j^\prime}^*, \label{tilgamm}
\end{equation}
where $t_{i i^\prime}$ are matrix elements for hopping across the SM-SC interface and $G^{(SC)}(\omega)$ is the surface Green's function of the parent superconductor. Working within a local approximation,\cite{Stanescu2010,Stanescu2013} we have $\widetilde{\gamma}_{ij} = \widetilde{\gamma}_i \delta_{ij}$, with $\widetilde{\gamma}_i$ being nonzero if $i$ labels a site at the SM-SC interface and zero otherwise. As evident from Eq. (\ref{tilgamm}), the effective coupling $\widetilde{\gamma}_i$ is determined by the hopping across the SM-SC interface and by the surface density of states of the parent SC. Note that the position-dependent quantity $\widetilde{\gamma}_i$ is only defined at the interface and does not contain all the information necessary for evaluating the strength of the superconducting proximity effect. Indeed, quantities such as the induced gap or phenomena such as the proximity-induced low-energy renormalization\cite{Stanescu2017a} are controlled by the band-dependent effective coupling 
\begin{equation}
\gamma_{mn} = \left< \psi_{m} \left| \widetilde{\gamma} \right| \psi_{n} \right>,  \label{gmn}
\end{equation}
where $\left|\psi_{n}\right>$ is an eigenstate of the system associated with the n$^{\rm th}$ confinement-induced band. 
For example, in the weak coupling limit, $\gamma_{nn}\ll\Delta_0$, a non-degenerate band\cite{Stanescu2017a} is characterized by an induced gap $\Delta_n = \gamma_{nn}\Delta_0/(\gamma_{nn}+\Delta_0)\approx\gamma_{nn}$.
In general, one would expect the coupling matrix $\gamma_{mn}$ to be non-diagonal and the diagonal terms $\gamma_{nn}$ to be strongly band-dependent. Note that the key ingredient in Eq. (\ref{gmn}) is the amplitude of the wave function at the SM-SC interface. In turn, this amplitude is determined by the electrostatic properties of the system, in particular by the parameters $V_{SC}$, $E_0$, and $V_g$. Consequently, the strength of the superconducting proximity effect is expected to be strongly affected by these parameters, in particular by the applied gate voltage.

%%%%%%%%%%%%%%%%%%%%%%%%%%%%%%%%
%%%%%%%%%%%%%%%%%%%%%%%%%%%%
\begin{figure}[t]
\begin{center}
\includegraphics[width=0.48\textwidth]{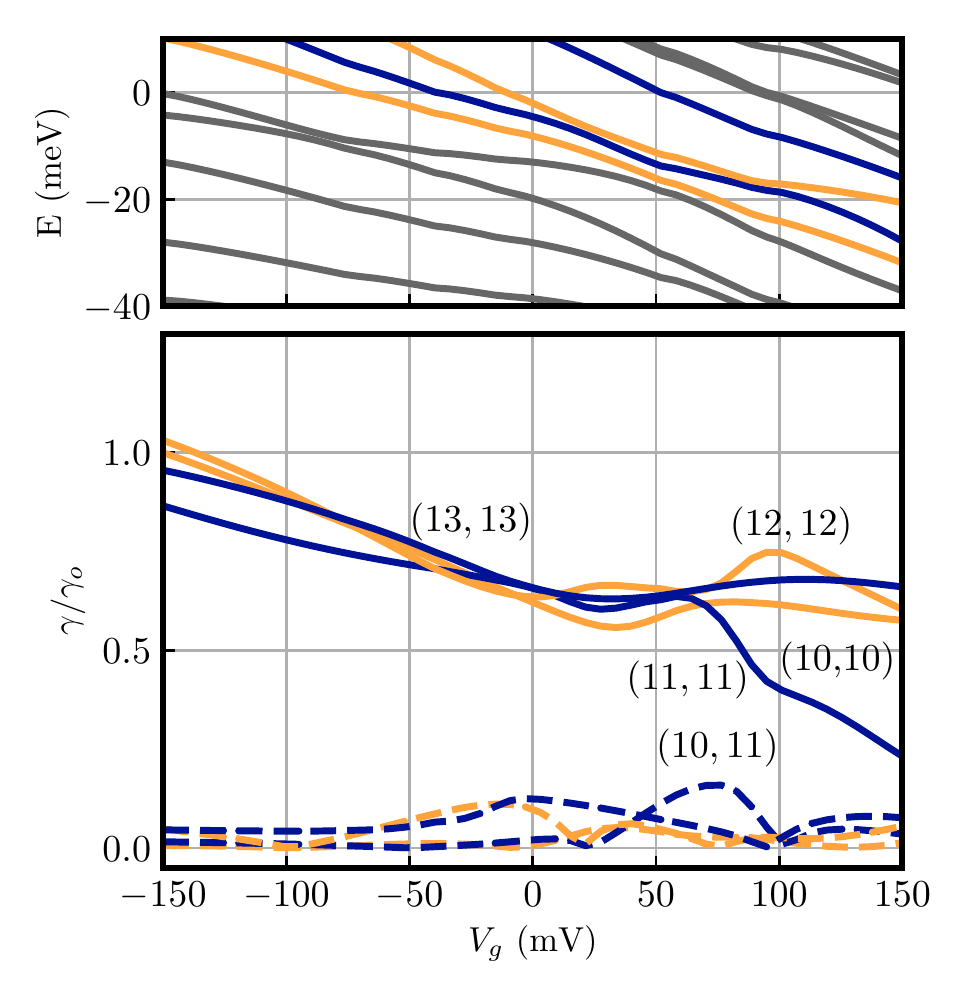}
\end{center}
\vspace{-3mm}
\caption{(Color online) {\em Top}: Energy eigenvalues as a function of the applied gate potential for a system with $V_{SC} = 200~$mV and $E_{0} = -10~$meV.  {\em Bottom}: Dependence of the normalized effective SM-SC coupling matrix,  $\gamma_{mn}(V_g)/ \gamma_{10~\!10}(-150)$, on the gate voltage, $V_g$, for the energy bands closest to the Fermi energy ($n,m=10, \dots, 13$). The solid and dashed lines represent diagonal matrix elements and off-diagonal elements of the form $m= n-1$, respectively. Note that all diagonal elements have  similar magnitudes, while the off-diagonal elements are negligible,  until $V_{g}$ becomes comparable to $V_{\!SC}$.}
\label{FIG11}
\vspace{-3mm}
\end{figure}
%%%%%%%%%%%%%%%%%%%%%%	
%%%%%%%%%%%%%%%%%%%%%%%%%%%%%%%%

To evaluate the dependence of the effective SM-SC coupling on the applied gate potential we consider two systems characterized by the same position-dependent coupling,  $\widetilde{\gamma}_i=\widetilde{\gamma}$ (i.e. independent of $i$) if $i$ is at the SM-SC interface and $\widetilde{\gamma}_i=0$ otherwise, same work function difference, $V_{SC} = 200~$mV, and different reference energies, $E_{0} = 100~$meV and $E_{0} = -10~$meV, respectively.
The dependence of the corresponding effective coupling $\gamma_{mn}$ on the applied gate potential is shown in Fig. \ref{FIG10} and \ref{FIG11}, respectively. Note that the wire  with $E_{0} = 100~$meV has $3\!-\!7$ occupied bands  (top panel of Fig. \ref{FIG10}), while the doped wire with $E_{0} = -10~$meV is characterized by a significantly higher occupancy ($9\!-\!15$ occupied bands).  Only the values of $\gamma_{mn}$ corresponding to the bands closest to the Fermi energy are shown. 

The results in Figs. \ref{FIG10} and \ref{FIG11} reveal three important features. First, we note that the off-diagonal components of $\gamma_{mn}$ are significantly smaller than the diagonal components, except in the regime characterized by large positive values of $V_g$. Second, there is a clear trend: the strength of the effective SM-SC coupling decreases with increasing $V_g$, i.e. as the electrons are attracted toward the back gate and away from the SM-SC interface. The trend is less clear in the system characterized by high occupancy (see Fig. \ref{FIG11}) for $V_g>0$. This is due to the presence of different types of modes, some localized predominantly near the SM-SC interface and some away from the interface, as discussed in the context of Fig. \ref{FIG9}. Finally, we note that the magnitude of $\gamma_{nn}$ is about the same for several low-energy bands within a significant range of parameters. This result is somehow unexpected, considering predictions based on simple noninteracting models, and has direct experimental implications. Specifically, in a system with multi-band occupancy (a class which probably includes  all experimental nanowires), the existence of (significantly) different values of the band-dependent coupling $\gamma_{nn}$ should lead to the observation of different proximity-induced gaps. By contrast, similar band-dependent couplings will lead to the observation of a single proximity-induced gap, unless a high-resolution measurement of the induced-gap features is possible. The results in Figs. \ref{FIG10} and \ref{FIG11} are consistent with the second (i.e. single-gap) scenario. Of course, a more detailed study of the specific experimental setup is necessary in order to gain complete understanding of any given device. In particular, the possibility of distinct band-dependent proximity gaps in SM-SC hybrid systems cannot be ruled out \textit{a priori}.

\subsection{Proximity-induced gap in the intermediate coupling regime} \label{indgap}

%%%%%%%%%%%%%%%%%%%%%%%%%%%%%%%%
%%%%%%%%%%%%%%%%%%%%%%%%%%%%
\begin{figure}[t]
\begin{center}
\includegraphics[width=0.48\textwidth]{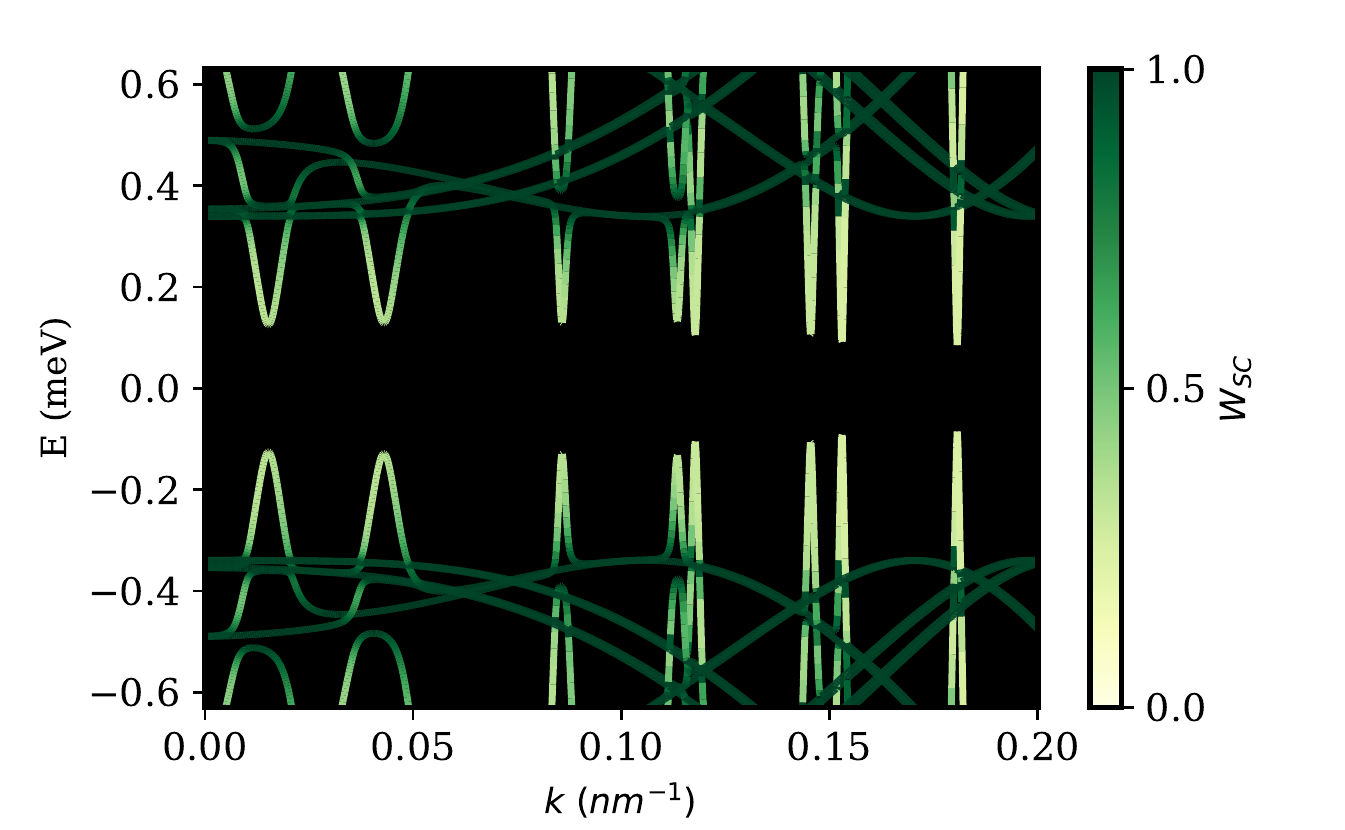}
\end{center}
\vspace{-3mm}
\caption{Low-energy spectrum of the hybrid SM-SC system in the superconducting state at zero magnetic field. The color scheme reflects the weight of a state in the superconductor: dark modes (which are almost invisible) represent SC states, while bright modes are weakly hybridized  SM states. Note that the induced gap is band-dependent, but has comparable values for different bands. The electrostatic parameters are: $V_{SC}=200~$mV, $E_0=100~$mV, and $V_g=-210~$mV.}
\label{FIG_S1}
\vspace{-3mm}
\end{figure}
%%%%%%%%%%%%%%%%%%%%%%	
%%%%%%%%%%%%%%%%%%%%%%%%%%%%%%%%

So far, our analysis has focused on the weak coupling regime characterized by $\gamma_{mn}\ll \Delta_0$. In this section, we consider the situation when the effective SM-SC coupling is comparable with the (bulk) superconducting gap, i.e. the intermediate coupling regime. As mentioned at the end of Sec. \ref{Greens_Method}, our scheme is applicable to the intermediate and strong coupling regimes, but the parent superconductor has to be included explicitly. To illustrate the implementation of our effective theory method, we consider an infinite wire proximity-coupled to a thin SC layer, as represented schematically in Fig. \ref{FIG1}, and we calculate the dependence of the induced SC gap on the applied gate potential $V_g$. The calculation is done for a wire of thickness $2R=100~$nm, with a $10~$nm superconducting shell, and a dielectric of thickness $d=30~$nm. 
The parent superconductor is described (at the mean-field level) by the BdG Hamiltonian
\begin{eqnarray}
H_{\rm sc} &=&\sum_{i, j, k,\sigma} \left[t_{sc} +\left(\frac{\hbar^2 k^2}{2m_{sc}^*}-\mu_{sc}\right)\delta_{ij}\right]a_{ik\sigma}^\dagger a_{jk\sigma} \nonumber \\
&+&\Delta_0\sum_i(a_{ik\uparrow}^\dagger a_{i~\!-\!k\downarrow}^\dagger +h.c),
\end{eqnarray}
where $a_{ik\sigma}^\dagger$ is the creation operator for an electron with spin $\sigma$ and longitudinal wave vector $k$  occupying the (transverse) site $i$ of a triangular lattice with lattice constant $a=2~$nm, $t_{sc} = 7.93~$meV is the nearest-neighbor hopping (which corresponds to an effective mass $m_{sc}^* = -0.8 m_0$),  $\mu_{sc}$ is the
chemical potential, and $\Delta_0=0.3~$meV is the SC pairing.  Note that $\mu_{sc}$ is set near the top of one of the confinement-induced bands, so that the Fermi surface of the SC represents a large hole pocket, similar to the Fermi surface of Al (in the second Brillouin zone). The parent SC is coupled to the SM wire through a coupling term of the form
\begin{equation}
H_{\rm sm-sc} = -\tilde{t}\sum_{\langle i, j\rangle, \sigma}\left(a_{ik\sigma}^\dagger c_{jk\sigma} + c_{jk\sigma}^\dagger a_{ik\sigma}\right),
\end{equation}
with $\langle i, j\rangle$ being nearest-neighbor sites located on the two sides of the SM-SC interface and $\tilde{t}=108.86~$meV is the hopping across the interface. The SM wire is described by a Hamiltonian having the noninteracting part given by Eq. \ref{HamInf} with nearest-neighbor hopping $t=-453.6~$meV (corresponding to an effective mass $m^* = 0.014 m_0$) and a spin-orbit coupling coefficient of $500~$meV$\cdot$\AA. 
The Schr\"{o}dinger-Poisson problem defined by the total (BdG) Hamiltonian $H = H_0 +H_{int}+ H_{\rm sc} + H_{\rm sm-sc}$ is solved using the generalized scheme described in \ref{Greens_Method}. In particular, the charge density in the wire is calculated in the presence of (induced) superconductivity using Eq. (\ref{rhobis}). 

%%%%%%%%%%%%%%%%%%%%%%%%%%%%%%%%
%%%%%%%%%%%%%%%%%%%%%%%%%%%%
\begin{figure}[t]
\begin{center}
\includegraphics[width=0.48\textwidth]{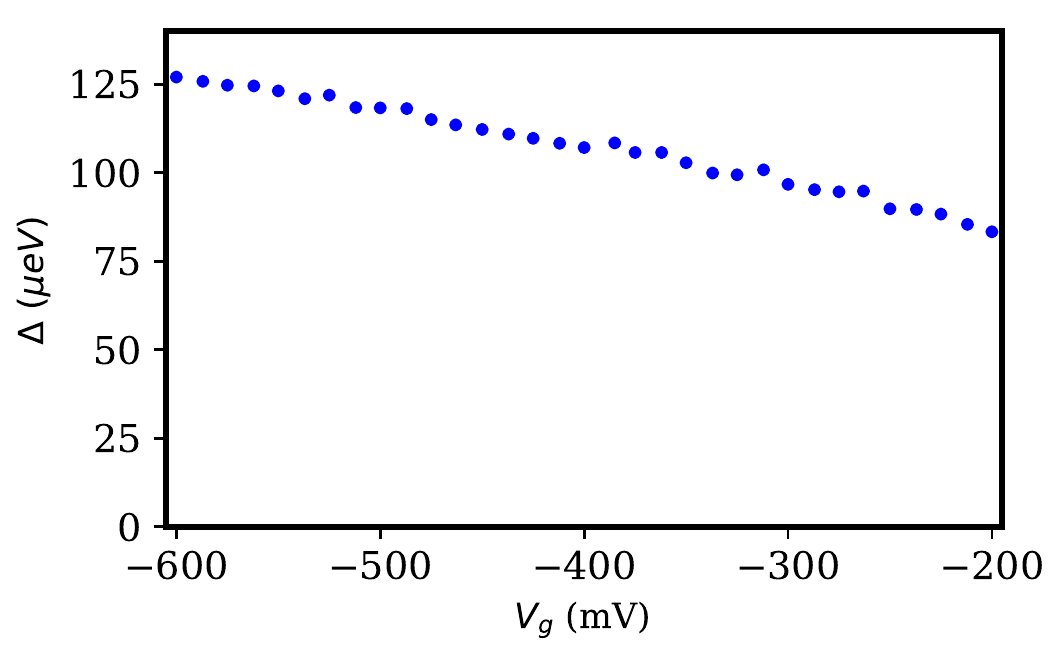}
\end{center}
\vspace{-3mm}
\caption{Dependence of the induced gap $\Delta$ on the applied gate potential. The induced gap increases relatively smoothly as the gate voltage becomes more negative, which pushes the electrons from the wire toward the SM-SC interface. We note that $\Delta$ is correlated with the weight $W_{SC}$ of the weakest hybridized band.}
\label{FIG_S2}
\vspace{-3mm}
\end{figure}
%%%%%%%%%%%%%%%%%%%%%%	
%%%%%%%%%%%%%%%%%%%%%%%%%%%%%%%%

We emphasize again that the brute-force approach is rather costly. For example, in our calculation a cross section of the hybrid system has a total of $N_\perp=2206$ sites, which implies that the BdG Hamiltonian has $4N_\perp$ degrees of freedom for each value of the wave-vector $k$. This number could increase dramatically if we consider a smaller lattice spacing or a multi-orbital tight-binding model. To efficiently address this problem, we implement our effective theory scheme by projecting onto a low-energy sub-space defined by a certain energy window $\Delta E$ (in the calculation $\Delta E \approx 200~$meV) and by constructing an effective low-energy Hamiltonian similar to that defined  by Eq. (\ref{Heff}). In our calculation the dimension of the low-energy (BdG) subspace is $550$, i.e. about $16$ times smaller than the dimension of the full Hilbert space. We note that, in principle, the low-energy sub-space is $k$-dependent. However, a low-energy basis calculated for a given value of $k$ (e.g., $k=0$) can be shown to be a good basis over a finite range of $k$ values, so that in practice the projected sub-space has to be determined only for a few different wave vectors. 

The results obtained by solving the effective BdG problem self-consistently are shown in Figs. \ref{FIG_S1} and \ref{FIG_S2}. First, we note that the hybridization between SM states and states from the superconductor is band dependent, as reflected by the color scheme in Fig. \ref{FIG_S1}. In addition, the modes with a higher weight inside the SC (i.e., larger $W_{SC}$) are characterized by larger values of the induced gap. However,  it is important to emphasize that the band-dependent induced gap $\Delta_n$ has comparable values for different bands, which is consistent with the results of Sec. \ref{effcoupl}. This property will not hold for systems with more symmetry (e.g., rectangular wires), as SM states with a given quantum number corresponding to the transverse direction parallel to the interface will only couple with SC states with the same quantum number, which may not be available at low-energy. This reveals the key importance of incorporating the details of the geometry into the model and the critical need for an efficient approach -- like the one proposed in this work -- to address the resulting numerical complexity. 
Next, we define the induced gap of the proximitized wire as $\Delta = {\rm Min}_{(n)}\left[ \Delta_n\right]$. The dependence of $\Delta$ on the applied gate potential is shown in Fig. \ref{FIG_S2}. Note that the induced gap decreases (relatively smoothly) as the potential becomes less negative, i.e. as the SM states are less confined near the SM-SC interface and hybridize less with states from the superconductor. Of course, the induced gap also depends on the SM-SC coupling $\tilde{t}$ and on the electrostatic parameters $V_{SC}$ and $E_0$, but a systematic investigation of this parameter space is beyond the scope of this proof-of-concept calculation. 

\section{Electrostatic effects in finite wires} \label{Applic2}

In this section we illustrate the implementation of our general scheme for solving the Schr\"{o}dinger-Poisson problem for finite systems by constructing the effective 1D model described in Sec. \ref{Effective_method}. As proof-of-concept examples, we consider two problems that play a major role in understanding the significance of recent experimental observations on SM-SC Majorana structures: i)  the Majorana energy splitting oscillations, and ii) the emergence of trivial low-energy states (i.e. generic low-energy non-topological Andreev bound states) in inhomogeneous Majorana wires. In addition, we investigate the convergence of our effective theory scheme and show that the low-energy projection is a well-controlled approximation.
Throughout this section we consider a finite wire of radius $R = 50~$nm (see Fig. \ref{FIG1}), length $L = 2~\mu$m, and unit cell length in the direction parallel to the wire $a_x = 10~$nm, which corresponds to dividing the wire into $N_x = 200$ layers. Each layer contains $N_\bot = 1176$ sites. The parameters are again taken to correspond to an InSb nanowire with $m^* = 0.014 m_0$ and relative permittivity $\epsilon_r = 17.7$, while  the Rashba spin-orbit coupling coefficient is set to $\alpha_{R} = 250~$meV$\cdot$\AA. 
We note that an analytical solution for the Green's functions is not possible due to the broken translation symmetry. For this reason, the Green's functions are calculated numerically using the finite element analysis software package FEniCS.\cite{Alnaes2015}

%%%%%%%%%%%%%%%%%%%%%%%%%%%%%%%%
%%%%%%%%%%%%%%%%%%%%%%%%%%%%
\begin{figure}[t]
\begin{center}
\includegraphics[width=0.48\textwidth]{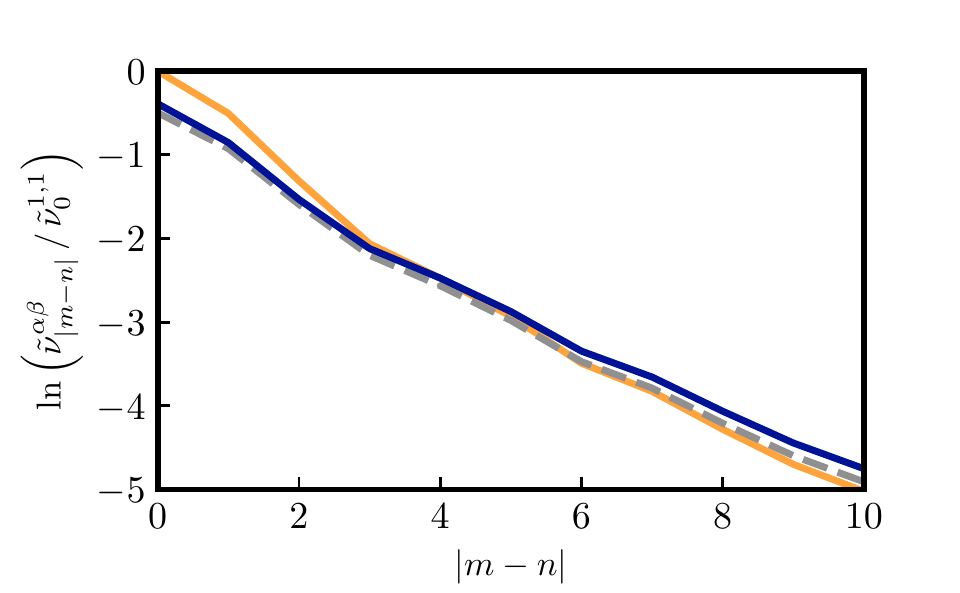}
\end{center}
\vspace{-3mm}
\caption{(Color online) Decay of the molecular orbital interaction tensor as function of distance, $\left| m-n\right|$, for a system with $V_{SC}= 200~$mV, $V_g=0$, and $E_{0}=100~$meV. The solid orange and blue lines correspond to the elements $\widetilde{\nu}_{\left|m - n\right|}^{1,1}$ and $\widetilde{\nu}_{\left|m - n\right|}^{3,3}$, respectively. The dashed gray line represents the element $\widetilde{\nu}_{\left|m - n\right|}^{1,3}$. Note that the interaction tensor decays nearly exponentially, with similar characteristic length scales for all elements. The elements that are not shown in the figure have a similar behavior.}
\label{FIG12}
\vspace{-3mm}
\end{figure}
%%%%%%%%%%%%%%%%%%%%%%	
%%%%%%%%%%%%%%%%%%%%%%%%%%%%%%%%

The results presented in the first two subsections are obtained using the method described in Sec. \ref{Effective_method} with a few additional simplifications.   First, we neglect all the terms in the effective Hamiltonian   (\ref{Heff}) that are off-diagonal in  molecular orbitals. Explicitly, we set $\widetilde{t}^\parallel_{m\alpha,n\beta} = \widetilde{\Delta U}^m_{\alpha\beta} = \alpha^{mn}_{\alpha\beta} = 0$ for $\alpha \neq \beta$. Although naively one may expect different bands to be decoupled, in general they are not, due to mixing terms introduced by the mean field fluctuations $\widetilde{\Delta U}^m_{\alpha\beta}$, by the proximity-induced band coupling $\Delta_{\alpha\beta}$, and by intrinsic inhomogeneities (e.g., the presence of a barrier region at the end of the wire), which lead to spatial variations of the transverse profiles and generate off-diagonal hopping,  $\widetilde{t}^\parallel_{m\alpha,n\beta}$ and spin-orbit coupling, $\alpha^{mn}_{\alpha\beta}$. The last two sources of band mixing can be neglected assuming weak coupling to the parent superconductor and weak inhomogeneity. Understanding the role of the off-diagonal mean-field fluctuations deserves a separate study. Neglecting band mixing leads to a simplification of Eq. (\ref{DUab}), which becomes
\begin{equation}
\widetilde{\Delta U}_{\alpha\alpha}^m = \sum \limits_\lambda^{occ.} \sum_{n,\sigma}\sum\limits_{\beta}^\bullet \widetilde{\nu}_{~m~\!n}^{\alpha ~\!\beta} ~\mathbb{A}_{\lambda, n\sigma}^{\beta\beta} -\langle\varphi_\alpha^m|U^{(m)}|\varphi_\alpha^m\rangle, \label{DUaa}
\end{equation} 
where we use the notation $\widetilde{\nu}_{~m~\!n}^{\alpha ~\!\beta} = \widetilde{\nu}_{~m~\!n}^{\alpha\alpha ~\!\beta\beta}$. 

Another simplifying assumption is that the interaction matrix between two layers  dependents only on the distance between them, $\widetilde{\nu}_{~m~\!n}^{\alpha ~\!\beta} = \widetilde{\nu}_{\left|m - n\right|}^{\alpha ~\!\beta}$. In other words, the 3D Green's function for a given site $i$  inside layer $m$ is assumed to be the same as the Green's function of corresponding site in layer $n$, up to an overall shift by $(n-m)a_x$. This approximation neglects the edge effects, but reduces the number of Green's functions that need to be calculated by a factor of $N_x$. However, we expect the edge effects to be small because of the strong screening provided by the SC and back gate. Indeed, the tensor elements $\widetilde{\nu}_{\left|m - n\right|}^{\alpha ~\!\beta}$ decay rapidly as  a function of $\left|m-n\right|$, as shown in Fig. (\ref{FIG12}). Notice the nearly exponential decay, which implies that the interaction tensor elements become negligible over distances corresponding to  a few layers (i.e. lattice sites of the effective 1D model). This demonstrates that approximation $\widetilde{\nu}_{~m~\!n}^{\alpha ~\!\beta} = \widetilde{\nu}_{~m~\!n}^{\alpha\alpha ~\!\beta\beta}$ is accurate everywhere, except the very edge of the wire.  Also note that the diagonal and off-diagonal elements in band space are of the same order. Consequently, charge fluctuations in one band will have a large effect on the other bands through the mean-field interaction term. We note that the simplifying assumption $\widetilde{\nu}_{~m~\!n}^{\alpha ~\!\beta} = \widetilde{\nu}_{\left|m - n\right|}^{\alpha ~\!\beta}$ will manifestly break down if we are interested in the electrostatic properties of a tunnel  barrier region (or any other type of strong inhomogeneity). In this case the full interaction tensor has to be calculated for the barrier region; the simplifying assumption is still valid  inside the (homogeneous) proximitized segment of the wire. 

The final simplification involves the construction of the auxiliary Hamiltonian (\ref{Haux}) for systems with  inhomogeneous electrostatic boundary conditions along the wire. We assume that the inhomogeneity is weak (in practice we consider a $1\%$ variation of $V_{SC}$) and we construct the auxiliary Hamiltonian using the local {\em boundary conditions}, instead of the local {\em electrostatic potential}. More specifically, 
we construct $V_i^{(m)}$ as the potential of an infinite wire problem with boundary conditions given by the local boundary conditions, i.e. $V_{SC}(m~\! a_x)$ and  $V_g$, of the full 3D device. Note that, ideally, one has  to solve the Laplace equation for the whole 3D devices and obtain the electrostatic potential $V_{im}$, then construct the auxiliary Hamiltonians for each layer using $V_i^{(m)} = V_{im}$. However, if the variations in the boundary conditions are small, we expect the two constructions to produce similar results, the only significant difference being the presence of spurious discontinuities in the approximate construction in regions where the boundary conditions change abruptly. Taking into account all the simplifications discussed above, the effective 1D Hamiltonian (\ref{Heff}) becomes
\begin{equation}
\begin{gathered}
H_{\rm eff} = \sum_{m,n,\sigma}\sum_{\alpha} ^\bullet t^{\parallel}_{m n} ~ {c}_{m\alpha\sigma}^{\dagger}{c}_{n\alpha\sigma}  \\
+\sum_{m,n,\sigma\sigma^\prime}\sum_{\alpha} ^\bullet \left[i \alpha_{R}^{~}(\sigma_y)_{\sigma\sigma^\prime} + {\Gamma}\left(\sigma_{x}\right)_{\sigma\sigma^{\prime}}\delta_{m n}\right] c_{m\alpha\sigma}^{\dagger}  c_{n \alpha \sigma^{\prime}}  \\  
+\sum_{m,\sigma}\sum_{\alpha} ^\bullet \left[\epsilon_{\alpha}^m + \widetilde{\Delta U}_{\alpha \alpha}^{m}\right] ~{n}_{m\alpha\sigma},
 \label{Heff1}
\end{gathered}
\end{equation}
where $t^{\parallel}_{n n} = 2t^\parallel$, while the off-diagonal elements are $t^{\parallel}_{m n} =-t^\parallel$ if $m$ and $n$ are nearest neighbors and zero otherwise. Note that the last term in Eq. (\ref{Heff1}) can be viewed as an orbital- and position-dependent effective potential,
\begin{equation}
V_{eff}^\alpha(m) = \epsilon_{\alpha}^m + \widetilde{\Delta U}_{\alpha \alpha}^{m}. \label{Veff}
\end{equation}
We emphasize that the study of the convergence of our low-energy effective scheme in Sec.  \ref{coverg} does not involve any additional approximation. 

\subsection{Majorana energy splitting oscillations} \label{Moscill}

In finite wires, the Majorana modes localized at the opposite ends of the system will, in general, acquire finite energy as a result of the hybridization generated by the (exponentially small) wave function overlap \cite{Cheng2009,Prada2012,Rainis2013}.  This energy splitting is  characterized  by  an  oscillatory  behavior  determined by the Fermi wave vector of the top occupied band \cite{Cheng2009}. Detecting correlated energy splitting oscillations at the opposite ends of the wire was proposed as a smoking gun test for the experimental confirmation of the Majorana modes \cite{DSarma2012}. This feature has never been observed in experimental systems in spite of concerted efforts. An important question concerns the electrostatic environment associated with such a measurement. Consider, for example, that our control parameter is the Zeeman field, which is varied within a certain range. Assuming a constant chemical potential results in a clear oscillatory behavior, while constant density will strongly suppress the splitting oscillations \cite{DSarma2012}. In Sec. \ref{Applic1A}, we have shown that the actual response of the system to an applied Zeeman field corresponds to the intermediate regime between the constant chemical potential and constant density limits. 

\begin{figure}[t]
\begin{center}
\includegraphics[width=0.48\textwidth]{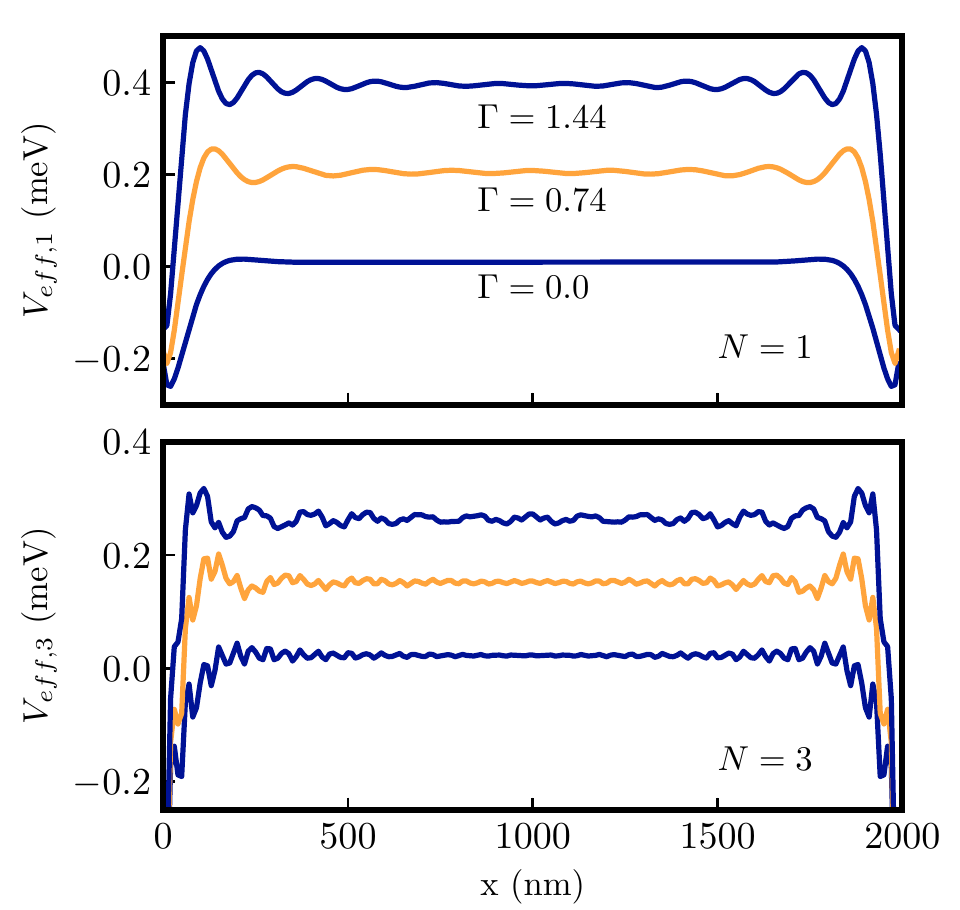}
\end{center}
\vspace{-3mm}
\caption{(Color online) Position dependence of the effective potential (\ref{Veff}) corresponding to the top occupied band (N) for different values of the model parameters corresponding to one (top) or three (bottom) occupied bands. {\em Top}:  $V_{SC}= 150~$mV, $V_g=-125~$mV, and $E_{0}=100~$meV. {\em Bottom}: $V_{SC}= 200~$mV, $V_g=-135~$mV, and $E_{0}=100~$meV. The Zeeman fields in the bottom panel match those indicated in  the top panel. The large period oscillations (see both panels) are associated with the Fermi wavelength of the top occupied band, while  the small period oscillations (bottom) are associated with the Fermi wavelength of the lower occupied bands.}
\label{FIG13}
\vspace{-3mm}
\end{figure}

Here, we go one step beyond the analysis done in Sec. \ref{Applic1A}, in the sense that we do not simply consider the dependence of the (effective) chemical potential on the applied Zeeman field, but calculate the {\em local} effective potential that is generated by the, generally non-uniform, charge distribution along the wire.  In other words, we explicitly take into account the fact that the mean-field interaction $\widetilde{\Delta U}_{\alpha\alpha}^m$ is position-dependent and evaluate the effect of this position-dependent mean-field contribution on the Majorana energy splitting oscillations. The position dependence of the effective potential defined by Eq. (\ref{Veff}) corresponding to the top occupied band of a $2~\mu$m long wire is shown in Fig. \ref{FIG13} for two different sets of parameters. The overall increase of the effective potential with the applied Zeeman field can be easily understood based on the results for infinite wires discussed in Sec. \ref{Applic1A}. Indeed, defining the variation of the effective potential with $\Gamma$ as $\delta V_{eff}^\alpha(m) = \left.V_{eff}^\alpha(m)\right|_\Gamma - \left.V_{eff}^\alpha(m)\right|_0$, we have $\delta V_{eff}^\alpha(m) \approx -\delta\mu_\alpha(\Gamma)$ for all sites $m$ that are sufficiently far away from the ends of the wire. Since $\delta\mu_\alpha(\Gamma)$ decreases with the applied Zeeman field (as shown, for example, in Fig. \ref{FIG7}) the effective potential increases with $\Gamma$. A more subtle feature, which cannot be captured by the infinite wire result, are the oscillations of the effective potential that can be clearly seen in Fig. \ref{FIG13}. These oscillations are related (through the mean field interaction term) to the Friedel oscillations of the charge density generated by the presence of the wire ends. Consequently, the periods of these oscillations are determined by the Fermi wavelengths of the occupied bands. 

%%%%%%%%%%%%%%%%%%%%%%%%%%%%%%%%
%%%%%%%%%%%%%%%%%%%%%%%%%%%%
%%%%%%%%%%%%%%%%%%%%%%	
%%%%%%%%%%%%%%%%%%%%%%%%%%%%%%%%

%%%%%%%%%%%%%%%%%%%%%%%%%%%%%%%%
%%%%%%%%%%%%%%%%%%%%%%%%%%%%
\begin{figure}[t]
\begin{center}
\includegraphics[width=0.48\textwidth]{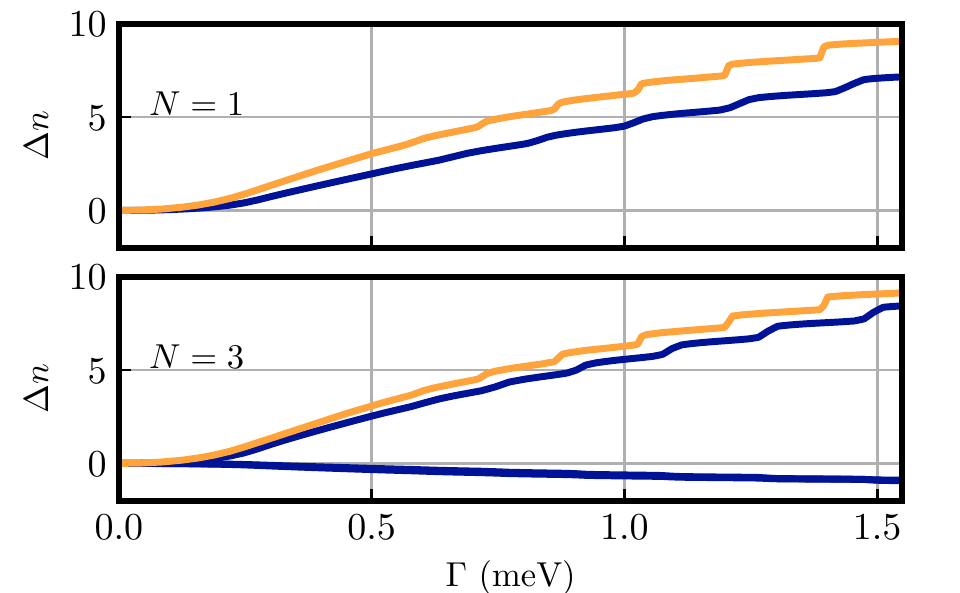}
\end{center}
\vspace{-3mm}
\caption{(Color online) Band occupation number $\Delta n_\alpha$  as function of the applied Zeeman field $\Gamma$ for a system with the same parameters as in Fig. \ref{FIG13}. The orange (light gray) lines correspond to the top occupied bands of a non-interacting  system (i.e. a system modeled by an effective Hamiltonian with $\widetilde{\Delta U}_{\alpha \alpha}^{m}=0$) with $N=1$ (top) and $N=3$ (bottom) occupied bands. The occupation numbers of the interacting system are shown in blue (dark gray). Note that the increase of $\Delta n_\alpha$ with $\Gamma$ (for the top band) is moderated by interactions. The lines with a negative slope in the bottom panel show the change in occupation of the  two lowest energy occupied bands for $N=3$. This allows the top band to accommodate more charge than in the singly occupied system (but still less than the non-interacting wire). }
\label{FIG14}
\vspace{-3mm}
\end{figure}
%%%%%%%%%%%%%%%%%%%%%%	
%%%%%%%%%%%%%%%%%%%%%%%%%%%%%%%%

\begin{figure}[t]
\begin{center}
\includegraphics[width=0.48\textwidth]{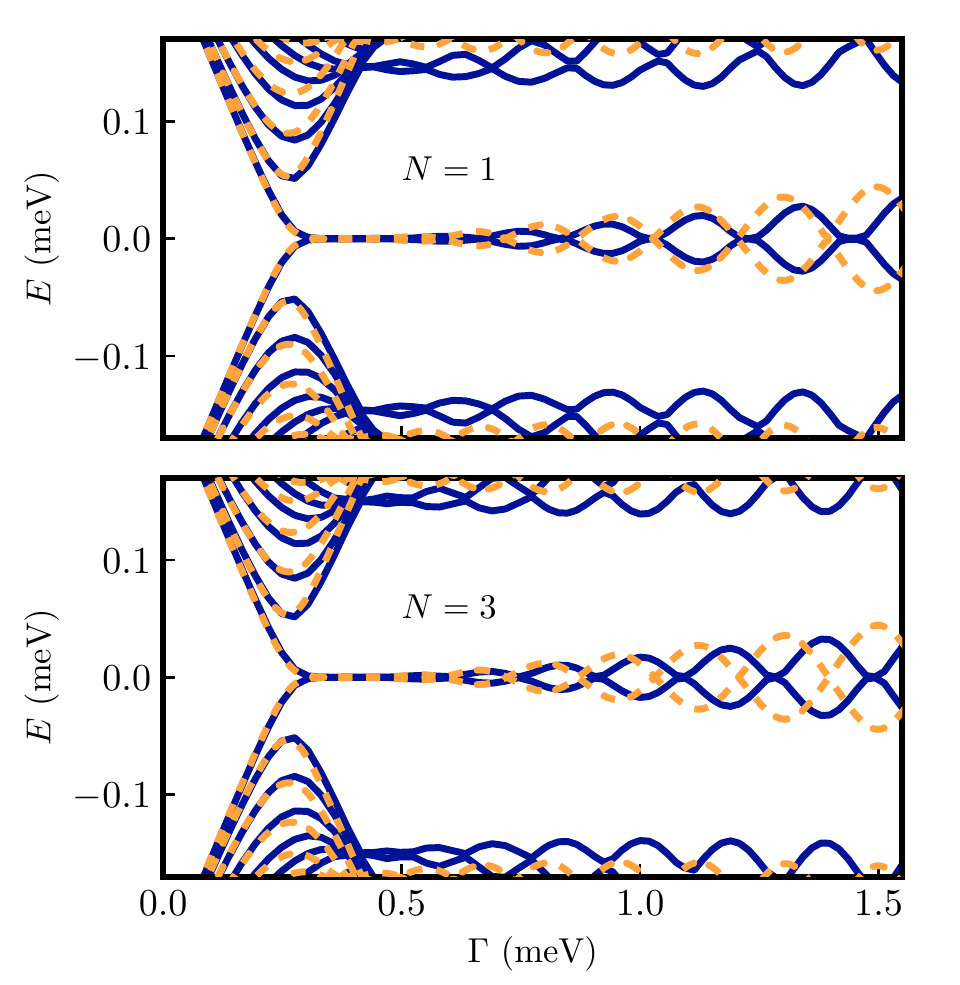}
\end{center}
\vspace{-3mm}
\caption{(Color online) Low-energy spectrum as function of the applied Zeeman field for an interacting system (solid blue lines) and a non-interacting system (dashed orange lines). Both systems are described by the same effective Hamiltonian (\ref{Heff1}), but the non-interacting system has $\widetilde{\Delta U}_{\alpha \alpha}^{m}$.  The electrostatic parameters are the same as in Fig. \ref{FIG13}, i.e.  $V_{SC}= 150~$mV, $V_g=-125~$mV, $E_{0}=100~$meV for the top panel and  $V_{SC}= 200~$mV, $V_g=-135~$mV,  $E_{0}=100~$meV for the bottom panel. The presence of interactions has two effects: i) it enhances the period of the oscillations and ii) it expands the zero-energy crossing points into  finite zero-energy segments.}
\label{FIG15}
\vspace{-3mm}
\end{figure}

As discussed in Sec. \ref{Applic1A}, upon increasing the Zeeman field the occupancy of the low-energy spin-sub-band corresponding to the top occupied band increases, which results in an overall increase of the number of electrons in the system. Interactions tend to moderate this increase by lowering the effective chemical potential (or, equivalently, increasing the effective potential $V_{eff}$). This is illustrated in Fig. \ref{FIG14}, which shows a comparison between the dependence of the (band) occupation number $\Delta n_\alpha = n_\alpha(\Gamma)-n_\alpha(0)$ on the applied Zeeman field for an interacting system and the dependence of $\Delta n_\alpha$ on $\Gamma$ in the absence of the mean-field term $\widetilde{\Delta U}$. Note that the occupation of the top band is higher in the non-interacting system (orange lines) as compared to the interacting system (blue lines). The step like features correspond to occupying the top band with an additional electron; going from one step to the next changes the parity of the top (Majorana) band and is associated with a node in the low-energy spectrum, as shown in Fig. \ref{FIG15}. 

%%%%%%%%%%%%%%%%%%%%%%%%%%%%%%%%
%%%%%%%%%%%%%%%%%%%%%%%%%%%%
%%%%%%%%%%%%%%%%%%%%%%	
%%%%%%%%%%%%%%%%%%%%%%%%%%%%%%%%

The effect of interactions on the Majorana splitting oscillations is twofold, as evident from Fig. \ref{FIG15}. On the one hand, in the interacting system the period of the splitting oscillations is larger than the oscillation period of the noninteracting wire. This is a direct consequence of the fact that, in the presence of interactions, the effective chemical potential decreases with $\Gamma$ (i.e. $V_{eff}$ increases), instead of being constant (as it is the case in the non-interacting system). This effect was also discussed in Ref. \onlinecite{Vuik2016}.  On the other hand, the presence of interactions pins the Majorana mode to zero energy over finite intervals of Zeeman fields. This behavior, which is in contrast with the typical zero-energy crossings that characterize  noninteracting system, is similar  to the zero-energy pinning reported in Ref. \onlinecite{Dominguez2016}. However, in our calculation the effect is not due to interactions with bound charges in the dielectric surroundings, as the dielectric was neglected in this particular calculation (since it is non-essential for Majorana considerations), but it is the direct result of (mean-field) electron-electron interactions. Note, that this treatment does not incorporate exchange contributions, which could be important in the case of low-energy {\em localized} states (see below). Nonetheless, it is essential to perform a {\em position-dependent}  self-consistent calculation, rather than accounting for the interaction effects through a {\em uniform} field-dependent chemical potential (of effective electrostatic potential), which does not capture the pinning behavior. This simple example clearly illustrates the importance of being able to tackle the 3D   Schr\"{o}dinger-Poisson problem, which makes the scheme proposed in this article highly relevant. Since the predicted non-local correlated Majorana splitting oscillations have never been observed experimentally, our finding of the self-consistent Coulomb interaction effect within the wire itself leading to the possible suppression of such oscillations should be taken seriously and further investigated using realistic sample parameters.

We conclude this section with a comment on the importance of including exchange contributions in our scheme. In general, there are two types of problems that should be addressed within a self-consistent Schr\"{o}dinger-Poisson approach: i) finding the effective electrostatic potential inside the SM wire (e.g., the confinement potential in a 2D electron gas structure, or the tunnel barrier potential) and/or the parameters that depend directly on this potential (e.g., the Rashba coefficient, the $g$ factor, and the pairing potential) and ii) calculating the dependence of low-energy sub-gap states (e.g., Majorana modes and Andreev bound states) on relevant control parameters (e.g., magnetic field). The first class of problems involves energy scales of the order eV (or higher). In this case, using the Hartree approximation described in this work is expected to accurately capture the relevant physics. On the other hand, the second class of problems  involves sub-eV energy scales and a proper treatment (even at a qualitative level) requires more refined approximations. Fortunately, our scheme  can be easily generalized to incorporate exchange and correlation contributions, in the spirit of the local-density approximation (LDA). The detailed implementation of these corrections will be described elsewhere. Here, we only illustrate the main idea, focusing on the self-interaction contribution to the Hartree potential. First, we note that for finite-energy delocalized states the self-interaction contribution (i.e. the interaction energy of an electron occupying such a state with itself) is small and has about the same value for all states. Consequently, we can neglect this contribution, or include it as a state-independent correction to the interaction term. By contrast, self-interaction could represent a significant contribution in the case of low-energy localized states, such as the Majorana bound states. To eliminate it from the effective potential of the bound states, we solve two Schr\"{o}dinger equations, one for the localized state and the other for the delocalized states
\begin{eqnarray}
(H_1 + U_{\rm int}^0)\psi_0 &=& E_0 \psi_0, \nonumber \\
(H_1 + U_{\rm int}^*)\psi_n &=& E_n \psi_n,               \label{Eq35}
\end{eqnarray}
where $H_1$ is the non-interacting BdG Hamiltonian of the hybrid structure (including the interactions with external fields), while  $U_{\rm int}^0$ and $U_{\rm int}^*$ are the mean-field Coulomb potentials for the localized and delocalized states, respectively. Explicitly, we have
\begin{eqnarray}
U_{\rm int}^0({\bm r})\! &=&\! -e\sum_{n}^{occ.}\!\int\!d^3r^\prime ~\!\left[G({\bm r},{\bm r}^\prime) - \delta_{n0}G_0({\bm r},{\bm r}^\prime)\right]|u_n ({\bm r}^\prime)|^2, \nonumber \\
U_{\rm int}^*({\bm r})\! &=&\! -e\sum_{n}^{occ.}\!\int\!d^3r^\prime ~\!G({\bm r},{\bm r}^\prime)|u_n ({\bm r}^\prime)|^2 \!+\! \frac{e}{\Omega}\!\! \int \!\!d^3r^\prime ~\!G_0({\bm r},\!{\bm r}^\prime),   \nonumber
\end{eqnarray}
where $G$ is the Green's function that satisfies the boundary conditions imposed by the electrostatic environment and $G_0$ is the Green's function for free space. The terms containing $G_0$ represent the self-interaction contributions. For the delocalized states we approximated this contribution with the energy of a charge $-e$ uniformly distributed over the volume $\Omega$ of the wire. The Schr\"{o}dinger equations (\ref{Eq35}) together with the equations for the  mean-field Coulomb potentials are solved self-consistently. Note that for problems involving multiple low-energy localized states ($\nu = 0, 1, \dots$) the corresponding set of equations should be appropriately expanded. Also, note that the effective potentials $U_{int}^\nu$ can incorporate exchange-correlation contributions. The actual relevance of these corrections will have to be determined using realistic system parameters.  
Since exchange-correlation corrections account for detailed quantitative effects (which perhaps may be necessary for a quantitative comparison with the experimental data, although not all the relevant model parameters corresponding to experimental nanowires  can actually be known, making such a comparison quite challenging),  they are unlikely to affect the general formalism described here and the qualitative conclusions established in the current work. We leave the inclusion of these details to future work, as an extension of the current formalism.

\subsection{System with an inhomogeneous effective potential} \label{Applic2B}

%%%%%%%%%%%%%%%%%%%%%%%%%%%%%%%%
%%%%%%%%%%%%%%%%%%%%%%%%%%%%
\begin{figure}[t]
\begin{center}
\includegraphics[width=0.46\textwidth]{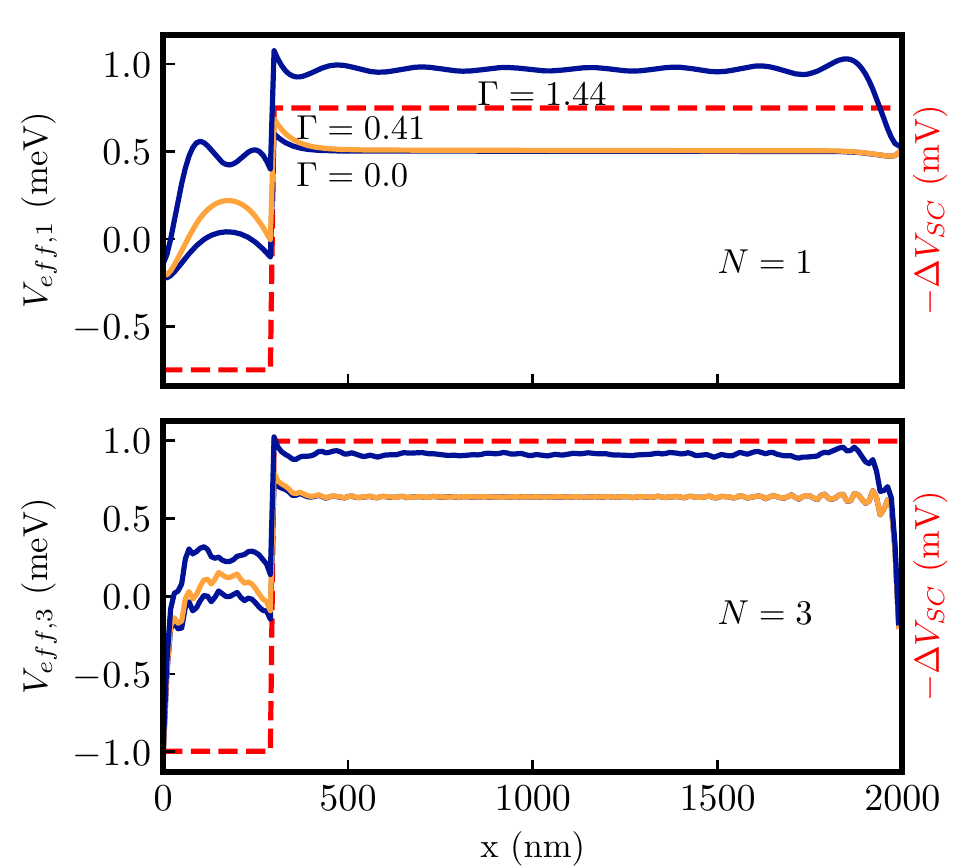}
\end{center}
\vspace{-3mm}
\caption{(Color online) Position dependence of the effective potential corresponding to the top occupied band for a system with parameters similar to Fig.  \ref{FIG13}$: V_{SC}= 150~$mV, $V_g=-125~$mV, $E_{0}=100~$meV (top panel) and  $V_{SC}= 200~$mV, $V_g=-135~$mV,  $E_{0}=100~$meV (bottom panel). A $1\%$ variation of $V_{SC}$ is assumed over a $0.3~\mu$m segment at the left end of the wire (see the red dashed line). The Zeeman energies in the bottom panel match those in the top panel.}
\label{FIG16}
\vspace{-3mm}
\end{figure}
%%%%%%%%%%%%%%%%%%%%%%	
%%%%%%%%%%%%%%%%%%%%%%%%%%%%%%%%

The emergence of trivial low-energy states in systems with non-uniform  parameters\cite{Kells2012,Prada2012,SanJose2013,Ojanen2013,Stanescu2014a,Cayao2015,Klinovaja2015} and in wires coupled to a quantum dot\cite{Liu2017a} has been discussed extensively in recent years. It was recently argued\cite{Stanescu2016,Liu2017a,Moore2018} that in certain conditions these low-energy trivial states cannot be distinguished from ``true'' Majorana zero modes localized at the ends of the wire using any type of end-of-the-wire local measurement. However, a major question concerns the very possibility of an effective potential inhomogeneity in the active (i.e. proximitized) segment of the wire that has a length scale large-enough to support stable low-energy trivial modes. After all, the strong screening due to the parent superconductor suppresses the interaction matrix elements over characteristic length scales of the order of tens of nanometers, as demonstrated by the calculations shown in Fig. \ref{FIG13}. In this section we show that such long length scale inhomogeneities are, indeed, possible in systems with non-uniform work function difference, i.e. systems with a position-dependent $V_{SC}$.

%%%%%%%%%%%%%%%%%%%%%%%%%%%%%%%%
%%%%%%%%%%%%%%%%%%%%%%%%%%%%
\begin{figure}[t]
\begin{center}
\includegraphics[width=0.46\textwidth]{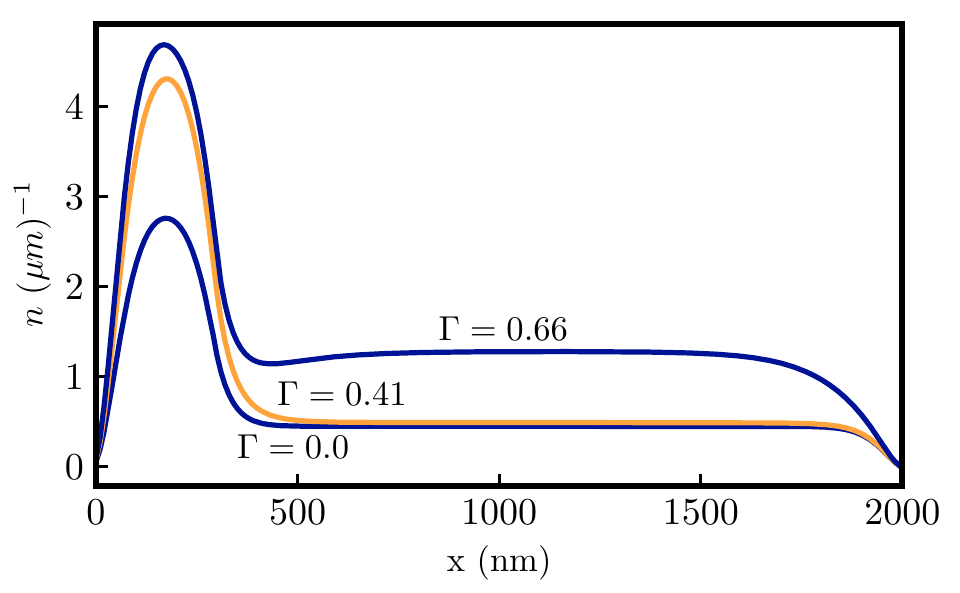}
\end{center}
\vspace{-3mm}
\caption{(Color online)  Particle density as a function of position for the system with single-band occupancy (top panel in Fig. \ref{FIG16}).  Increasing the Zeeman field adds more charge to the system; for $\Gamma <0.66~$meV the charge accumulates in region I ($x<0.3~\mu$m), while for $\Gamma >0.66~$meV the additional charge starts to leak into region II ($x>0.3~\mu$m). Note that $\Gamma=0.41~$meV corresponds to a trivial zero mode, while $\Gamma\approx0.66~$meV represents the critical field associated with the topological quantum phase transition, as shown in the top panel of Fig. \ref{FIG18}.}
\label{FIG17}
\vspace{-3mm}
\end{figure}
%%%%%%%%%%%%%%%%%%%%%%	
%%%%%%%%%%%%%%%%%%%%%%%%%%%%%%%%

Consider a $2~\mu$m long SM nawire proximity coupled to a superconductor and assume that a $0.3~\mu$m long segment at the left end of the wire has a $1\%$ variation of the work function difference $V_{SC}$, as shown in Fig. \ref{FIG16}. This variation could be the result of the procedure used for treating the SM wire surface before depositing the superconductor. Applying our self-consistent scheme, results in a position-dependent effective potential $V_{eff}^\alpha(m)$ that has significantly different (average) values inside the two segments of the wire (i.e. $x<0.3~\mu$m and $x>0.3~\mu$m, respectively), as shown in Fig. \ref{FIG16} for two sets of parameters. We emphasize that the screening by the parent superconductor plays no role in suppressing the variation of the effective potential. On the other hand, increasing the occupation of the top band (e.g., by increasing the Zeeman field) reduces the difference between the (average)  values of the effective potential inside the two segments.  This can be understood in terms of the position-dependent charge density shown in Fig. \ref{FIG17}. Indeed, as charge accumulates in region I ($x<0.3~\mu$m), the local mean field contribution given by Eq. (\ref{DUaa}) increases and the difference between the (average) effective potentials in regions I and II ($x>0.3~\mu$m) gets smaller. We note, however, that having additional occupied bands (see lower panel in Fig. \ref{FIG16}) does not affect significantly this mechanism, as the charge associated with those low-energy bands is more or less evenly distributed along the wire.  

%%%%%%%%%%%%%%%%%%%%%%%%%%%%%%%%
%%%%%%%%%%%%%%%%%%%%%%%%%%%%
\begin{figure}[t]
\begin{center}
\includegraphics[width=0.46\textwidth]{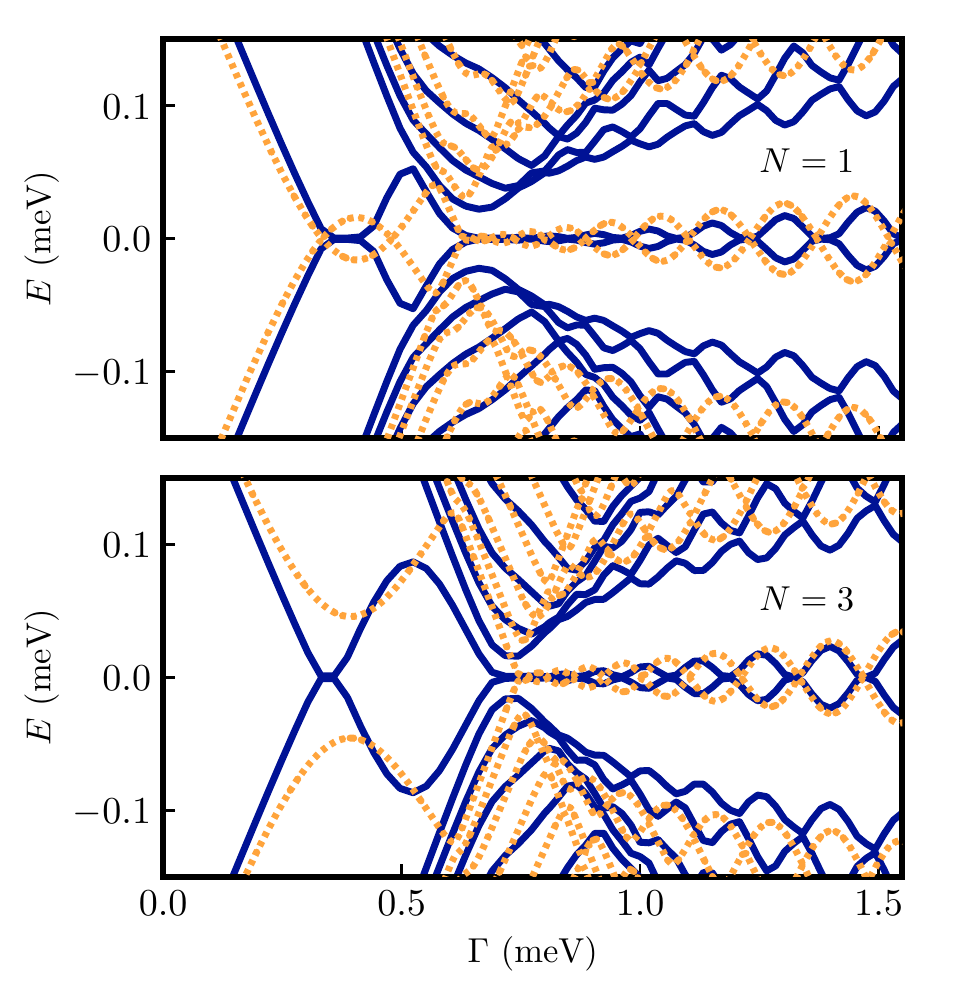}
\end{center}
\vspace{-3mm}
\caption{(Color online) Low-energy spectrum as function of the applied Zeeman field for a system with the same parameters as in Fig. \ref{FIG16}. 
The spectrum of the interacting model corresponds to the full (blue) lines; for comparison we also show the non-interacting spectrum (orange dashed lines) obtained by neglecting the mean-field term $\widetilde{\Delta U}_{\alpha \alpha}^{m}$ in the effective Hamiltonian (\ref{Heff1}). Note the topological quantum phase transition  signaled by the minimum of the bulk quasiparticle gap. The Majorana splitting oscillation associated with the topological regime have features similar to those discussed in Sec. \ref{Moscill}. A low-energy mode that sticks to zero over a finite range of $\Gamma$ emerges in the trivial regime as a result of the effective potential inhomogeneity.}
\label{FIG18}
\vspace{-3mm}
\end{figure}
%%%%%%%%%%%%%%%%%%%%%%	
%%%%%%%%%%%%%%%%%%%%%%%%%%%%%%%%

The expected consequence of having a non-uniform effective potential is the emergence of trivial low-energy states. This is illustrated in Fig. \ref{FIG18}, which shows the low-energy spectrum of a system with the same parameters as in Fig. \ref{FIG16}. For comparison, we also plot the spectrum for a non-interacting system obtained by neglecting the mean-field term $\widetilde{\Delta U}_{\alpha \alpha}^{m}$ in the effective Hamiltonian (\ref{Heff1}). First, we note that, as expected, the effective potential inhomogeneity generates trivial low-energy states. These states are already present in the non-interacting system with $N=1$ (see the dashed orange lines in the top panel of Fig. \ref{FIG18}). For $N=3$ (bottom panel) the noninteracting trivial mode has a sizable gap, but this could be reduced by fine tunning some parameters (e.g., the gate potential $V_g$). What is more important is that including the interaction effects (at the mean field level) not only does not eliminate the trivial low-energy modes, but stabilizes them.  For $N=1$ (top panel) the zero-energy crossing points are replaced by a finite field range over which the mode is pinned at zero energy. In the $N=3$ case (bottom panel) the non-interacting gap collapses and the trivial mode goes all the way to zero energy. We note that one can easily obtain more ``spectacular'' trivial modes that are pinned to zero energy over a significant range of Zeeman fields by increasing the size of region I (i.e. the length scale of the potential inhomogeneity). However, our main point is  that such long-range inhomogeneities can exist within reasonable assumptions (e.g., a $1\%$ variation of the work function difference, which almost seems inevitable in a generic experimental situation) and have to be taken seriously. A second important result is that pinning to zero-energy of trivial low-energy modes is enhanced by interaction effects. In the light of these findings, a detailed study of the low-energy physics generated by the presence of a quantum dot at the end of the wire within the framework developed in this paper is well-warranted. Earlier work on Majorana wires coupled to quantum dots\cite{Liu2017a,Moore2018} ignores self-consistent Coulomb effects.  This study suggests that the realistic situation (which is characterized by the presence of self-consistent Coulomb interaction effects) could be even worse that earlier predicted in terms of the indistinguishability between trivial Andreev bound states and topological Majorana modes.

\subsection{Convergence of the effective theory scheme} \label{coverg}

In this section we address the key question regarding the accuracy of the effective theory scheme proposed here: how large are the errors generated by the projection onto the low-energy space spanned by the molecular orbitals and how can one systematically reduce them? We start from the basic observation that including all molecular orbitals (i.e. all transverse bands) provides the exact (mean-field) solution of the original tight-binding problem. Consequently, addressing the above question implies studying the convergence of the results as the number $n_o$ of molecular orbitals included in the low-energy basis increases. Considering, for example, the energies $E_n$ of the eigenstates, we have a well-controlled scheme if i) the errors $\delta E_n$ decrease monotonically with $n_o$ and ii) the maximum error in the energy of the occupied states can be made much smaller than some relevant low-energy scale (e.g., the induced gap) by including a relatively small number of molecular orbitals, $n_o\ll N_\perp$.  To test whether or not our scheme satisfies these requirements, we consider a strongly non-uniform hybrid system consisting of a finite nanowire of length $L=0.8~\mu$m having a $400~$nm segment covered by a superconductor with $V_{SC}=150~$mV, while the other half is uncovered. A potential gate with $V_g^{SC}=-30~$mV is placed under the proximitized segment, while another gate with $V_g^{b}=300~$mV extending from $x=400~$nm to $x=550~$nm acts as a potential well at the end of the superconducting region.  Finally, a potential gate with $V_g^{u}=170~$mV is placed under the rest of the uncovered segment, $550 < x < 800~$nm.  

\begin{figure}[t]
\begin{center}
\includegraphics[width=0.5\textwidth]{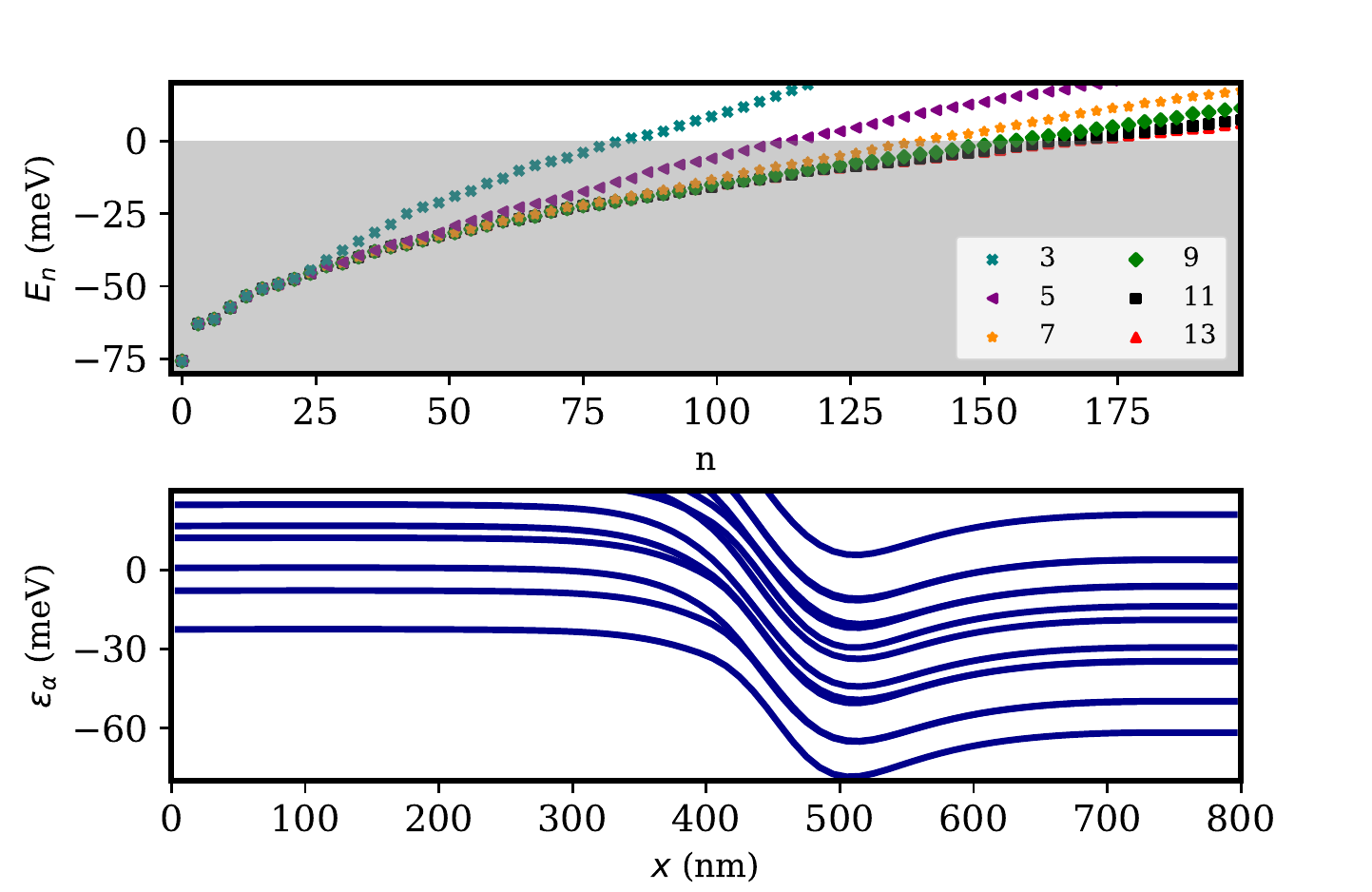}
\end{center}
\vspace{-3mm}
\caption{{\em Top}: Energy $E_n$ of the $n^{\rm th}$ low-lying state for a non-interacting effective Hamiltonian with $n_o$ molecular orbitals in the 
basis. The shaded region corresponds to the occupied states. For $n_o>13$ the result is practically indistinguishable from that corresponding to  
$n_o=13$. {\em Bottom}: Dependence of the ``effective potential'' $\epsilon_\alpha^m$ [see Eq. (\ref{Heff})] on the position along the wire for a 
noninteracting system. The segment $0 < x < 400~$nm is covered by a SC, while the rest of the wire is uncovered. Three back gates with different 
applied potentials are present, as described in the main text.}
\label{FIG_S3}
\vspace{-3mm}
\end{figure}

First, we consider a noninteracting problem and construct the effective 1D Hamiltonian (\ref{Heff}) by solving the auxiliary problem (\ref{Haux}) without including the mean-field contribution $U_i^{(m)}$, i.e. by solving the Laplace equation for the external potential $V_{im}$ with boundary conditions given by $V_{SC}$ and the gate potentials. We also neglect the mean-field contribution $\widetilde{\Delta U}$ in Eq. (\ref{Heff}). The energies $E_n$ of the low-lying states obtained by considering orbital bases of increasing dimension are shown in the top panel of Fig. \ref{FIG_S3}. The results satisfy the conditions discussed above, i.e. the errors decrease monotonically with $n_o$ and the energies of the occupied states are practically converged (i.e. $|\delta E_n| \ll 1~$meV) for $n_o > 13$, which is about two orders of magnitude smaller than the total number of bands, $N_\perp\sim 10^3$.  Note that we have done calculations for larger values of $n_o$ to verify the consistency of these conclusions. 

%%%%%%%%%%%%%%%%%%%%%%%%%%%%%%%%
%%%%%%%%%%%%%%%%%%%%%%%%%%%%
%%%%%%%%%%%%%%%%%%%%%%	
%%%%%%%%%%%%%%%%%%%%%%%%%%%%%%%%

\begin{figure}[t]
\begin{center}
\includegraphics[width=0.5\textwidth]{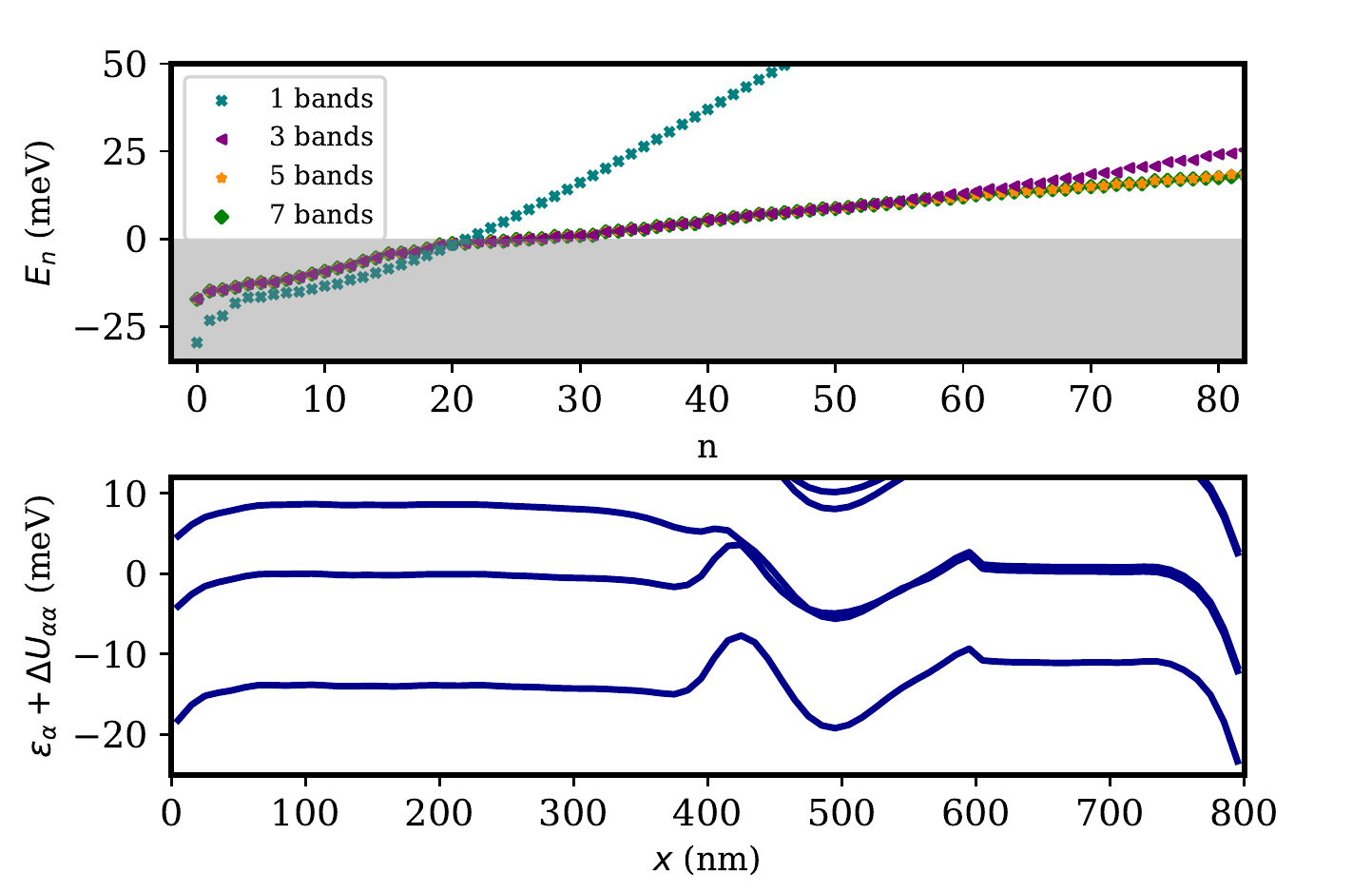}
\end{center}
\vspace{-3mm}
\caption{{\em Top}: Energy $E_n$ of the $n^{\rm th}$ low-lying state for a fully  interacting effective Hamiltonian with $n_o$ molecular orbitals in the 
basis. Both the auxiliary problem (\ref{Haux}) and the 1D effective problem (\ref{Heff}) have been solved self-consistently. {\em Bottom}: Position dependence of the ``effective potential'' $\epsilon_\alpha^m + \widetilde{\Delta U}_{\alpha \alpha}^m$, which includes (diagonal) mean-field contributions. Note that the potential profiles differ significantly from the non-interacting approximations shown in Fig. \ref{FIG_S3}.}
\label{FIG_S4}
\vspace{-3mm}
\end{figure}

 Next, we perform a fully self-consistent calculation of the same model, this time including the electron-electron interaction. The convergence of the scheme is illustrated in the top panel of Fig. \ref{FIG_S4}, which clearly supports our previous conclusions. Two remarks are warranted. First, the number $n_o$ of molecular orbitals that have to be included in the low-energy basis in order to obtain a desired value of the maximum error increases if the system becomes more non-homogeneous and if more bands become occupied. Second, the non-interacting basis corresponding to the calculation in Fig. \ref{FIG_S3} turns out to be an excellent basis for constructing the interacting effective Hamiltonian if the occupation is low. In other words, instead of solving the auxiliary problem self-consistently (for each slice), we can determine the transverse bands by simply considering the effect of the external potential. This is due to the fact that in systems with low charge density the transverse profile of the wave functions is practically determined by the external potential (and only weakly perturbed by electron-electron interaction). Of course, at high occupancy the low-energy basis has to be calculated self-consistently to ensure a fast convergence of   the scheme. In the example discussed here, while calculating the low-energy basis does not require self-consistency, including the mean-field contribution   
$\widetilde{\Delta U}$ in the effective 1D Hamiltonian is essential. The major effect of this term can be seen by comparing the lower panels in Fig.  \ref{FIG_S3} (effective potential {\em without} mean field contributions and Fig.  \ref{FIG_S4} (effective potential {\em with} mean-field contributions). The dramatic difference is due to the fact that in non-homogeneous systems electron-electron interaction leads to a redistribution of the charge {\em along} the wire. This charge redistribution is accounted for when solving the effective Hamiltonian self-consistently. The stark difference between the non-interacting and the interacting results emphasizes once more the importance of using a self-consistent Schr\"{o}dinger-Poisson scheme when studying inhomogeneous hybrid systems, e.g., proximitizedd nanowires with uncovered barrier regions or nanowires coupled to quantum dots.

%%%%%%%%%%%%%%%%%%%%%%%%%%%%%%%%
%%%%%%%%%%%%%%%%%%%%%%%%%%%%
%%%%%%%%%%%%%%%%%%%%%%	
%%%%%%%%%%%%%%%%%%%%%%%%%%%%%%%%

\section{Summary and conclusion} \label{Cnc}

We have developed a practical effective theory approach to the Schr\"{o}dinger-Poisson problem in semiconductor Majorana devices that enables one to efficiently model realistic 3D structures, including gate potentials, inhomogeneities, and multi-orbital physics.  The proposed method, which is specifically designed for lattice models and is significantly more computationally efficient than the brute-force 3D numerical methods, is based on two key ideas: i) For a given geometry, the Poisson problem is solved once using a Green's function scheme and the results are stored as an interaction tensor, which, in essence, contains information about the interaction energy between electrons occupying local orbitals. Thus, the Poisson component of the iteration loop reduces to a few straightforward summations. ii) The 3D problem is reduced to an effective multi-orbital 1D problem with molecular orbitals calculated self-consistently as the transverse modes of an infinite wire with the same electrostatic potential as the local electrostatic potential of the finite 3D device. The basic insight is that the transverse profiles of the low-energy states at a given position along the wire are similar to the profiles of the low-energy confinement-induced bands of an infinite wire under the same electrostatic conditions. Consequently, a relatively small number of molecular orbitals obtained by solving the auxiliary infinite-wire problem provide a good basis for the low-energy sub-space of the 3D problem. 

We describe in detail an implementation of our method that addresses the Schr\"{o}dinger-Poisson problem in weakly-coupled semiconductor-superconductor structures and we briefly discus a strong-coupling generalization of this scheme that  explicitly incorporates the parent superconductor, which is described at the mean-field level using a simple tight-binding Hamiltonian.  The generic strong-coupling regime necessitates additional considerations and will be discussed elsewhere. To demonstrate the capabilities of our approach, we implement it for both infinite and finite systems, which are modeled using single-orbital tight-binding Hamiltonians, and address several questions that are relevant in the context of Majorana physics in hybrid devices. More specifically, for an infinite wire we calculate the response of the system to an applied Zeeman field, the dependence of the low-energy spectrum on the work function difference at the SM-SC interface, and the dependence of the effective semiconductor-superconductor coupling as well as the induced gap (in the intermediate coupling regime) on the applied gate potential. In addition, for a finite wire we investigate the effect of interactions on the Majorana energy splitting oscillations and the emergence of effective potential inhomogeneities induced by variations of the work function difference at the SM-SC interface and we discuss the convergence of our scheme as the low-energy basis used in the construction of the effective theory is enlarged. 

The first step in the implementation of our method consists of solving the Laplace equation with non-homogeneous boundary conditions determined by the parameter $V_{SC}$, which characterizes the work-function difference at the SM-SC interface, and the gate potential(s) $V_g^n$, with $n=1,\dots, n_g$. In practice, one can take advantage of the linearity of the solution and solve the Laplace equation $n_g+1$ times, each time setting  one of the boundary conditions to unity, e.g., $V_g^1=1$, while the others are zero. The general solution can be expressed as a linear combination of these particular solutions with coefficients $V_{SC}$, $V_g^1, \dots$, which allows one to efficiently explore a large parameter space. The next step is to divide the 3D system into $N_x$ layers and solve an auxiliary infinite-wire problem for each layer. The (translation invariant) electrostatic potential for a given auxiliary problem has a transverse profile given by the solution of the Laplace equation (obtained previously) over the corresponding layer. Note that the auxiliary problem is solved self-consistently.  The Green's function used in the calculation of the interaction tensor is obtained by solving the Poisson equation with homogeneous boundary conditions, since the nontrivial boundary conditions are already incorporated into the external potential. The transverse profiles of the low-energy confinement-induced bands are then used as the basis for the effective 1D problem defined on $N_x$ lattice sites. The interaction tensor for the molecular orbitals is calculated using the Green's function for the 3D structure, which is the solution of a Poisson equation with homogeneous boundary conditions that has to be solved once (for each local orbital). Since the elements of the interaction tensor decrease rapidly with the distance between the molecular orbitals, the size of the relevant set of elements scales linearly with the length of the wire. Moreover, if the system has long homogeneous segments, the number of distinct elements can be drastically reduced. Finally, the effective 1D problem is solved self-consistently. This (second) self-consistency loop ensures that the charge is properly redistributed along the wire. The accuracy of the scheme can be tested and improved systematically by increasing the number of molecular orbitals in the basis.  We emphasize that the final solution is a solution of the original 3D problem. The full 3D spatial dependence of various quantities (e.g., effective potentials and charge distributions) can be easily reconstructed using the (known) spatial profiles of the molecular orbitals. 

The basic applications of our method discussed in this work demonstrate its potential, as well as the critical importance of electrostatic effects in the low-energy physics of semiconductor-superconductor Majorana devices. We note that all numerical calculations presented here were done on a standard laptop computer and involved typical running times of the order of minutes. By contrast, a brute-force 3D approach requires a  large parallel cluster with many nodes and huge memory. Our main results can be summarized as follows. We  demonstrate that terms in the Hamiltonian (such as, for example, the Zeeman splitting)  with energy scales much lower that the typical values of the electrostatic potentials (i.e. tens/hundreds of meV) can be treated using a perturbative scheme that involves a fully self-consistent solution calculated for a single (reference) value of the relevant parameter (e.g., zero magnetic field).  Focusing on the electrostatic response to an applied magnetic field, we evaluate the accuracy of the perturbative scheme by comparing its predictions with the fully self-consistent solution.  We also show that the strength of the effective semiconductor-superconductor coupling can be tuned by varying the applied gate potential. Rather remarkably, we find that in a wide range of parameters several low-energy bands are characterized by similar effective couplings. A direct experimental consequence of this finding is that, in certain conditions, hybrid structures with multi-band occupancy are, in fact, characterized by a single induced gap feature (rather than multiple, band-dependent induced energy scales).  In addition, we investigate the electrostatic effects in finite 3D nanowires and show that they result in a partial suppression of the Majorana energy splitting oscillations. We find that properly describing the effects of Coulomb interaction on low-energy {\em localized} states requires a careful treatment of the effective potential for the localized states, including the elimination of self-interaction and addition of exchange contributions. We note that a detailed study of the Majorana oscillations has to be done in the context of the strong-coupling implementation of our Schr\"{o}dinger-Poisson scheme, where the electrostatic effects are expected to combine with the proximity-induced energy renormalization, both acting as suppressing factors for the splitting oscillations. 

Finally, we show that the effective potential along the wire has a strong dependence on the work-function difference at the SM-SC interface. Thus, a position-dependent (inhomogeneous) work-function difference, which is a possible result of the device-fabrication process (e.g., of the procedure used for treating the SM wire surface before depositing the superconductor), results  in an inhomogeneous effective potential. In turn, the inhomogeneous potential can induce trivial low-energy states that mimic the phenomenology of Majorana zero modes.
Variations of the order of a few percent in $V_{SC}$ can induce variations of the effective potential of the order of $1-2~$meV, i.e. significantly larger than the induced gap. We emphasize that screening by the superconductor plays no role in reducing these inhomogeneities. On the other hand, the charge inside the wire can partially suppress the variations of the effective potential, but this screening mechanism is really efficient only at high occupancy.  

The ability of our scheme to efficiently solve position-dependent 3D problems suggests that it can be a valuable tool for studying realistic devices. Important types of problems that can be approached using this scheme include the study of the tunnel barrier regions at the ends of proximitized Majorana wires, the possible formation of quantum dots inside or at the ends of a wire, and the electrostatic confinement of nanowires  made lithographically in a 2D electron gas hosted by a semiconductor heterostructure. In all these problems, which are critical for the practical realization of Majorana zero modes, the electrostatic effects play a dominant role. Therefore, they can be viewed as particular aspects of the Schr\"{o}dinger-Poisson problem discussed in this work. In particular, our effective theory approach is well suited for a quantitative analysis of the spurious low-energy sub-gap states induced by the presence of junctions in Majorana devices consisting of 1D proximitized nanowires or 2D semiconductor heterostructures \cite{Stanescu2018a}. Furthermore, the method developed in this work is ideally suited for calculating the lever arms of various potential gates used in experimental devices, e.g., the change of the electrostatic potential inside the semiconductor wire corresponding to a certain variation of an applied back-gate potential. Of course, this type of analysis is relevant when considering specific devices, i.e. when taking into account details regarding the device geometry and materials properties, while a generic study is largely meaningless. However, once the details of the experimental device and gate configurations are specified, it is straightforward to calculate the lever arms using the scheme introduced here. 
We emphasize that our method is of general applicability in modeling mesoscopic hybrid devices, including different SM-SC hybrid systems, such as gatemons and quantum dot based spin qubits in 2D semiconductor heterostructures, as well as other types of hybrid devices, e.g., topological insulator-superconductor structures.
Finally, the fact that our introductory effective theory is numerically efficient and provides insight into several intriguing properties of Majorana nanowires (e.g., the possible suppression of Majorana splitting oscillations, the possible proliferation of trivial zero-energy modes, the possible band-independent single induced proximity gap) that are not accessible within the non-self-consistent minimal model used extensively in the literature implies that our self-consistent approach should be utilized systematically to understand the behavior of specific experimental SM-SC hybrid devices. 

{\em Note added}. After the completion of this work (see Woods et al., arXiv:1801.02630), we became aware of two other recent studies that address the issue of electrostatic effects in semiconductor-superconductor structures within a self-consistent Schr\"{o}dinger-Poisson approach: Antipov et al., arXiv:1801.02616 and Mikkelsen et al., arXiv:1801.03439. This activity reflects the critical importance of Coulomb interaction effects in hybrid Majorana structures and the fact that the scientific community recognizes the urgency of properly accounting for these effects. However, while Antipov et al. and Mikkelsen et al. represent applications of the Schr\"{o}dinger-Poisson scheme to systems with simplified geometries, i.e. slab geometry (which represents an effective 1D problem) in Mikkelsen et al. and infinite wire geometry (i.e. an effective 2D problem) in Antipov et al., we propose a new approach that enables the treatment of realistic models of three-dimensional (3D) hybrid devices. The major challenge in doing Schr\"{o}dinger-Poisson calculations using a realistic model is numerical complexity. Consequently, the "standard" approach is usually implemented for single-band models with simplified geometries and can be used to estimate ``bulk'' effective parameters, such as the induced superconducting gap and the Lande $g$ factor. By contrast, we have developed a Schr\"{o}dinger-Poisson approach that is applicable to realistic 3D models. Our method enables the efficient treatment of multiband models (which is essential for properly calculating the effective $g$ factor and the spin-orbit coupling) and 3D geometries (a key step towards realistic device modeling). This includes problems related to the presence of quantum dots coupled to Majorana wires, inhomogeneous gate potentials, barrier potentials, electrostatic confinement, and wire junctions, which are essential to understanding the physics of actual hybrid devices. The numerical complexity of the ``standard'' approach to these problems is extreme. This study provides a solution to this challenge. 

\begin{acknowledgments}
This  work  is  supported  by  NSF DMR-1414683 and by Laboratory for Physical Sciences and Microsoft at the University of Maryland.
\end{acknowledgments}

\appendix
\renewcommand\thefigure{\thesection.\arabic{figure}}    
\setcounter{figure}{0}

\section{Analytic Green's function for the infinite wire using conformal mapping} \label{App1}  %%%%%%%%%%%%%%%%%%
\begin{figure}[t]
\begin{center}
\includegraphics[width=0.46\textwidth]{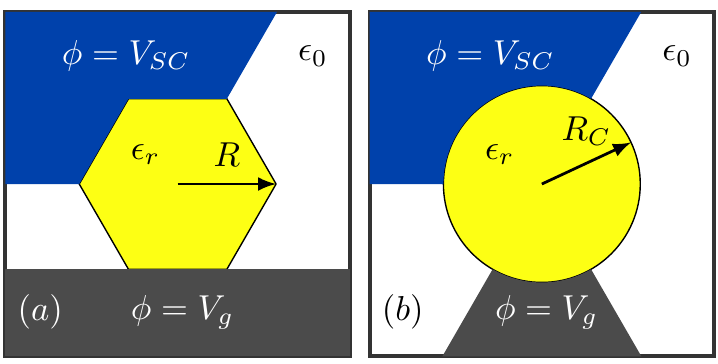}
\end{center}
\vspace{-3mm}
\caption{(Color online) (a) Schematic representation of the SM(yellow)-SC(blue) devices with simplified geometry used in the calculation of analytical Green's functions. Compared to Fig. \ref{FIG1} (see the main text), the simplified geometry neglects the dielectric layer and the finite thickness of the superconductor.  (b) Conformal mapping of the structure from panel (a) used in the calculation of the analytical Green's function.}
\label{FIGA1}
\vspace{-3mm}
\end{figure}

In this appendix we derive an analytic expression for the Green's function defined by Eq. (\ref{Geq1}) for the case of infinite nanowires (see  Sec. \ref{Methods_infinite}). To simplify the calculations, we neglect the dielectric layer and the fact that the superconductor has finite thickness. The simplified geometry is shown in Fig. \ref{FIGA1}(a). We note that changing the thickness of the SC changes very little the field lines inside the wire, hence it is expected to have a small effect.  On the other hand, the exclusion of the dielectric layer results in overestimating the screening due to the back gate. Indeed, imagine a test charge placed in the SM wire, close to the back gate.  The image charge occurring in the back gate (to satisfy the homogeneous boundary condition) is close to the SM surface, so that the potential caused by the test charge is nearly completely screened. If, on the other hand,  a dielectric layer is present, the image charge is farther away from the SM surface and the screening is reduced. However, if the test charge is localized near the SM-SC interface, the difference between the two situations is quite small. In practice, we actually expect most of the charge to be near the SM-SC interface (to ensure an effective proximity-coupling to the superconductor).  Therefore, the effect of eliminating the dielectric is expected to be small. Quantitatively, this effect can be determined using a numerical method to solve the Green's function for different values of the dielectric thickness.

\begin{figure}[t]
\begin{center}
\includegraphics[width=0.32\textwidth]{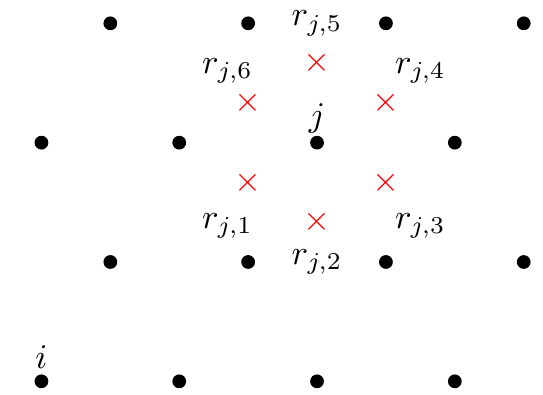}
\end{center}
\vspace{-3mm}
\caption{(Color online)  Sampling scheme used for calculating the interaction matrix when the basis states are singular, $\varphi_{i} \left(\mathbf{r}\right) = \delta^{2} \left(\mathbf{r} - \mathbf{r}_{i} \right)$. The interaction matrix element, $\nu_{ij}$ is calculated by sampling $G_i$ at the six vertices of the $j^{\rm th}$ unit cell of the hexagonal lattice [see Eq. (\ref{sampl})].}
\label{FIGA2}
\vspace{-3mm}
\end{figure}

Another simplification concerns the basis states $\varphi_i$, which are chosen to be of the form $\varphi_{i} \left(\mathbf{r}\right) = \delta^{2} \left(\mathbf{r} - \mathbf{r}_{i} \right)$, where $\mathbf{r}_{i}$ is the center of the $i^{\rm th}$ lattice site. With this choice,  $G_i$ becomes the standard Green's function used in electrostatics. Note that  due to the non-physical nature of the Dirac delta orbitals the diagonal elements $\nu_{ii}$ of the interaction matrix (\ref{Int1}) diverge. To address this issue, we calculate the interaction matrix elements using the average of the potential sampled at the vertices of the two-dimensional unit cells, as shown in Fig. (\ref{FIGA2}). Explicitly, the interaction matrix elements become
\begin{equation}
\nu_{ij} =-{e\over 6} \sum\limits_{\tau} G_i \left(\mathbf{r}_{j,\tau}\right),  \label{sampl}
\end{equation}
where $\tau$ runs over the six vertices of the $j^{\rm th}$ unit cell.

%%%%%%%%%%%%%%%%%%%%%%%%%%%%%%%%
%%%%%%%%%%%%%%%%%%%%%%%%%%%%
%%%%%%%%%%%%%%%%%%%%%%	
%%%%%%%%%%%%%%%%%%%%%%%%%%%%%%%%

The actual calculation of the Green's function is done using a cylindrical geometry. To connect it with the original geometry of the structure, we perform a conformal mapping, as shown in Fig. (\ref{FIGA1}). More specifically, the conformal mapping from the unit disk $\mathcal{D}$ in panel (b) to the hexagon $\mathcal{H}$ in panel (a) is given by 
\begin{equation}
z = wF\left({1 \over 6}, {1 \over 3}, {7\over 6}; w^6\right),
\end{equation} 
where $w \in \mathcal{D}$, $z \in \mathcal{H}$, and $F\left(a,b,c;z\right)$ is the hypergeometric function. The lattice of the hexagonal wire is mapped onto the unit disk using the inverse of this conformal mapping.  Technically, the transformation is only valid within the unit disk. Therefore, the features from the two panels that are  outside the nanowire do not perfectly map into each other. However, the solution {\em within} the nanowire should only weakly depend on the details of the outside  geometry.

%%%%%%%%%%%%%%%%%%%%%%%%%%%%%%%%
%%%%%%%%%%%%%%%%%%%%%%%%%%%%
%%%%%%%%%%%%%%%%%%%%%%	
%%%%%%%%%%%%%%%%%%%%%%%%%%%%%%%%

\begin{figure}[t]
\begin{center}
\includegraphics[width=0.4\textwidth]{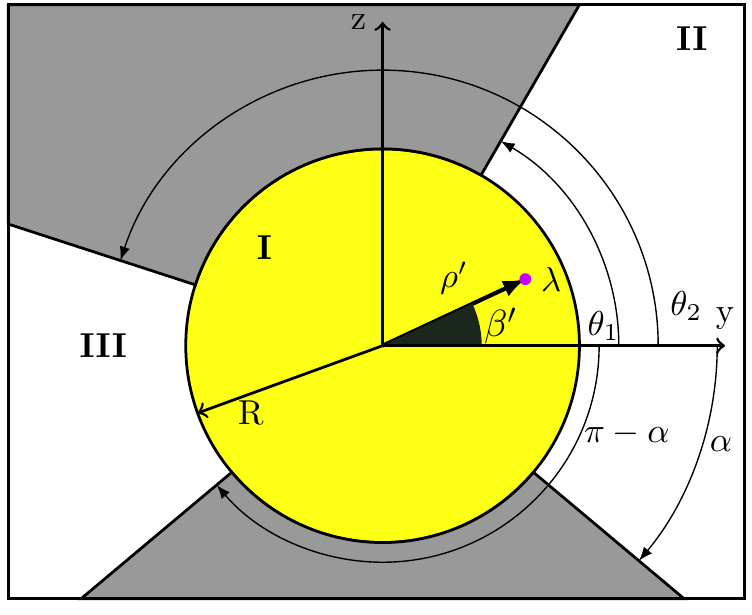}
\end{center}
\vspace{-3mm}
\caption{(Color online) Setup of the Poisson problem (to be solved analytically) for the Green's function of an infinite wire. A filament of charge (purple dot) with charge density $\lambda$ is placed at ${\bm r}^\prime$ of cylindrical coordinates  $(\rho^\prime,\beta^\prime)$ within in the nanowire (yellow). The conducting regions (dark gray) have zero potential. The problem is defined on a domain divided into three regions  (labeled I, II, and III).}
\label{FIGA3}
\vspace{-3mm}
\end{figure}

The details of the setup for the Poisson problem are shown in Fig. (\ref{FIGA3}). We divide the domain on which the problem is defined  into three regions: region I -- the nanowire --  contains a filamentary charge, while regions II and III are empty and extend to infinity. The solution of the  Poisson equation in region I can be written as $G_{I} = {G}_{P} + G_{L}$, with
\begin{equation}
\begin{aligned}
{\nabla}^{2}G_{P} &= -{\lambda\over \epsilon}~\delta^2({\bm r}-{\bm r}^\prime), \\
{\nabla}^{2}G_{L} &= 0.
\end{aligned}
\end{equation}
Note that $G_I$ should satisfy homogeneous boundary conditions at the boundaries with the back gate and the superconductor, as discussed in Sec. \ref{Greens_Method}.
A specific solution of the Poisson equation is $G_{p}(\mathbf{r}) = \frac{-\lambda}{2\pi\epsilon} \ln \left( { \frac{\left| {\mathbf{r} - \mathbf{ r}\prime}\right|}{R}} \right)$. The general solution of the Laplace equation has the form 
\begin{equation}
\begin{aligned}
G_{L}\left(\rho,\phi\right) &= G_{0} A_{0} \\
&+ G_{0}\sum\limits_{m=1}^{\infty} \left[ C_{m} \cos{\left(m\phi\right)} + D_{m} \sin{\left(m\phi\right)} \right] {\left( {\rho \over R} \right)}^{m} 
\end{aligned}
\end{equation}
with $G_{0} =\lambda/2\pi\epsilon$. Similarly, for regions II and III we have
\begin{equation}
\begin{aligned}
G_{II}\left(\rho,\phi\right) &= G_{0}\sum\limits_{m=1}^{\infty} F_{m} \sin{ \left( k_{m} \left(\phi - \alpha \right) \right)} {\left( \rho \over R \right)}^{ -k_{m}}, \\
G_{III}\left(\rho,\phi\right) &= G_{0}\sum\limits_{m=1}^{\infty} H_{m} \sin{ \left( l_{m} \left(\phi - \theta_{2} \right) \right)} {\left( \rho \over R \right)}^{-l_{m}}, \label {form2}
\end{aligned}
\end{equation}
with
\begin{equation}
k_{m} = {m\pi \over \theta_{1} - \alpha}, \quad  l_{m} = {m\pi \over \pi - \alpha - \theta_{2}}.
\end{equation}
In Eq. (\ref{form2}), we have already taken into account homogeneous boundary conditions at the boundaries with the back gate and the SC (see Sec. \ref{Greens_Method}). It is helpful to write $G_{P}$ as
\begin{equation}
\begin{aligned}
G_{P}\left(\rho,\phi\right) = G_{0} \ln{ \left( R \over \rho_{>} \right)} + G_{0} \sum\limits_{m=1}^{\infty} \biggl( {1 \over m} {\left( {\rho_{<} \over \rho_{>}} \right)} ^ {m}   \\
\cdot  \left[ \cos{\left(m\beta\right)} \cos{\left(m\phi\right)} +  \sin{\left(m\beta\right)} \sin{\left(m\phi\right)}  \right] \biggl)
\end{aligned}
\end{equation}
where $\rho_{>}$ and $\rho_{<}$ are the greater and lesser of $\rho$ and $\rho^\prime$, respectively. 

%%%%%%%%%%%%%%%%%%%%%%%%%%%%%%%%
%%%%%%%%%%%%%%%%%%%%%%%%%%%%
%%%%%%%%%%%%%%%%%%%%%%	
%%%%%%%%%%%%%%%%%%%%%%%%%%%%%%%%

Next, we impose the continuity condition for the potential and the normal component of ${\bm D}=\epsilon{\bm E}$ at the boundaries between different regions. Explicitly, we have 
\begin{eqnarray}
\left. G_I\right\vert_{\rho = R}  &=& \left. G_{II,III}\right\vert_{\rho = R}, \nonumber \\
\left. \epsilon_{r}{\partial G_{I } \over \partial \rho} \right\vert_{\rho = R}  &=& \left. {\partial G_{II,III} \over \partial \rho} \right\vert_{\rho = R}. \label {BC}
\end{eqnarray}
In addition, $G_I$ vanishes at the boundaries between region I and the conducting regions.
The equations matching the potentials at the boundaries lead to Fourier series for $A_{0}, C_{n}$, and $D_{n}$ in terms of the sets of coefficients $F_{m}$ and $H_{m}$. Similarly, Eq. (\ref{BC}) allows us to solve for $F_{n}$ and $H_{n}$ in terms of the set of coefficients $C_{m}$ and $D_{m}$. This leads to an infinite system of linear equations. Since we are not interested in the potential outside the nanowire, we eliminate the variables $F_{m}$ and $H_{m}$ and get

\begin{equation} 
\begin{aligned}
\sum \limits _{p=1}^{\infty} \left(p \Gamma_{np} ^{\gamma\gamma\delta\delta} + \delta_{n,p} \right)C_{p} + p\Gamma_{np} ^{\gamma\sigma\delta\omega}D_{p} = S_{n}, \\
\sum \limits _{p=1}^{\infty} p\Gamma_{np} ^{\sigma\gamma\omega\delta}C_{p} + \left(p \Gamma_{np} ^{\sigma\sigma\omega\omega} + \delta_{n,p} \right)D_{p}  = T_{n}, 
\end{aligned}
\end{equation}
where $\Gamma_{np} ^{abcd}, S_{n},$ and $T_{n}$ are coefficients of the form
\begin{equation}
\begin{aligned}
\Gamma_{np}^{abcd} &= {2 \epsilon_{r} \over \pi^{2}} \sum\limits_{m=1}^{\infty} {1\over m} \left( a_{mn} b_{mp} + c_{mn} d_{mp} \right), \\
S_{n}&= \Gamma_{n0}^{\gamma\gamma\Delta\Delta} - {1\over n}{\left({r \over R}\right)}^{n}\cos(n\beta) \\ &+ \sum\limits_{p=1}^{\infty} {\left( {r\over R} \right)}^{p} \left(\Gamma_{np}^{\gamma\gamma\Delta\Delta}\cos(p\beta) + \Gamma_{np}^{\gamma\sigma\Delta\omega}\sin(p\beta) \right), \\
T_{n}&= \Gamma_{n0}^{\sigma\gamma\omega\Delta} - {1\over n}{\left({r \over R}\right)}^{n}\sin(n\beta) \\ &+ \sum\limits_{p=1}^{\infty} {\left( {r\over R} \right)}^{p} \left(\Gamma_{np}^{\sigma\gamma\omega\Delta}\cos(p\beta) + \Gamma_{np}^{\sigma\sigma\omega\omega}\sin(p\beta) \right),
\end{aligned}
\end{equation}
with the matrix elements $a_{mn}$, $b_{mn}$,  $c_{mn}$, and  $d_{mn}$ being one of following matrix elements
\begin{equation}
\begin{aligned}
\mathsmaller{\gamma_{mn}} &=\mathsmaller{ { -m\pi\left(\theta_{1} - \alpha_{1} \right) \over n^{2} {\left(\theta_{1} - \alpha_{1} \right)}^{2} - {\left(m\pi\right)}^{2} } \left( \cos(n\alpha_{1}) - \cos(m\pi)\cos(n\theta_{1}) \right)}, \\
\mathsmaller{\sigma_{mn} }&= \mathsmaller{ { -m\pi\left(\theta_{1} - \alpha_{1} \right) \over n^{2} {\left(\theta_{1} - \alpha_{1} \right)}^{2} - {\left(m\pi\right)}^{2} } \left( \sin(n\alpha_{1}) - \cos(m\pi)\sin(n\theta_{1}) \right)}, \\
\mathsmaller{\Delta_{mn}} &= \mathsmaller{ { -m\pi\left(\alpha_{2} - \theta_{2} \right) \over n^{2} {\left(\alpha_{2} - \theta_{2} \right)}^{2} - {\left(m\pi\right)}^{2} } \left( \cos(n\theta_{2}) - \cos(m\pi)\cos(n\alpha_{2}) \right)}, \\
\mathsmaller{\omega_{mn} }&= \mathsmaller{ { -m\pi\left(\alpha_{2} - \theta_{2} \right) \over n^{2} {\left(\alpha_{2} - \theta_{2} \right)}^{2} - {\left(m\pi\right)}^{2} } \left( \sin(n\theta_{2}) - \cos(m\pi)\sin(n\alpha_{2}) \right)}. \label{EqA11}
\end{aligned}
\end{equation}
In Eq. (\ref{EqA11}) we have used the notation $\alpha_{2} = \pi - \alpha_{1}$.
This is an infinite system of equations. However, one can show that $S_n$ and $T_n$ decrease rapidly with increasing $n$. This is also true for the matrices $\Gamma_{np} ^{abcd}$, which decrease with increasing $n$ and $p$. Therefore, $C_n$ and $D_n$ will also decrease as $n$ increases. The solution can be approximated by taking a finite number of terms (e.g., $N=200$) and solving the corresponding (finite) matrix equation. We benchmark the solution by taking an average of $G_I$ over the conducting boundary regions, which ideally should be zero. This benchmark allows us to accurately determine when enough terms have been kept in the Fourier series.

\section{Matrix elements for the effective Hamiltonian} \label{App2}

Here we discuss the construction and structure of the various matrices in Eqs. (\ref{H3Dgen} and \ref{Heff}). Consider a generic matrix, $D$, in the 3D Hamiltoninan (\ref{H3D}). If we allow the matrix to couple two general basis states being localized on transverse sites i and j of layers m and n with spins $\sigma$ and $\sigma^\prime$, respectively, we need six indices to specify each matrix element. Explicitily, we have $D_{ijmn\sigma\sigma^\prime}$. In Sec. \ref{Effective_method} we defined $\bar{D}_{\ell \ell^\prime} = D_{ijmn\sigma\sigma^\prime}$ labeled by $\ell= \ell(m,i,\sigma)=2 (m-1) N_{\bot} +2i-1+\sigma$ and $\ell^\prime= \ell^\prime(n,j,\sigma^\prime)=2 (m-1) N_{\bot} +2i-1+\sigma$ so that we do not have to deal with the cumbersome six index notation. The six index notation is useful, however, when determining the structure of the various barred matrices in Eq. (\ref{H3Dgen}). For example, we can write the generic matrix element of $t^\bot$ as $t^\bot_{ijmn\sigma\sigma^\prime} = t^\bot_{ij} \delta_{mn} \delta_{\sigma\sigma^\prime}$. One can then clearly see that $\bar{t}^\bot$ will be a block diagonal matrix of the form
\begin{equation}
\bar{t}^\bot
=
\begin{bmatrix}
    \bar{t}^{\bot \left(1\right)} & 0 & 0 & \dots  & 0 \\
    0 & \bar{t}^{\bot \left(2\right)} & 0 & \dots  & 0 \\
    \vdots & \vdots & \vdots & \ddots & \vdots \\
    0 & 0 & 0 & \dots  & \bar{t}^{\bot \left(N_x \right)}
\end{bmatrix},
\end{equation}
where $\bar{t}^{\bot (m)}$ in the $2N_\bot  \times  2N_\bot$ hopping matrix within the $m^{th}$ layer. If the structure of the layers doesn't change along the wire then $\bar{t}^{\bot(1)} =  \bar{t}^{\bot(2)} = \dots = \bar{t}^{\bot(N_x)}$. The inter-layer hopping $t^\parallel$ has a similar structure given by 
\begin{equation}
\bar{t}^\parallel
=
\begin{bmatrix}
    -2\bar{t}^\parallel_0 & \bar{t}^{\parallel \left(1,2\right)} & 0  & \dots & 0 & 0 \\
    \bar{t}^{\parallel \left(2,1\right)} & -2\bar{t}^\parallel_0 & \bar{t}^{\parallel \left(2,3\right)}  & \dots  & 0 & 0 \\
    \vdots & \vdots & \vdots & \ddots & \vdots & \vdots \\
    0 & 0 & 0 & \dots & -2\bar{t}^\parallel_0 &  \bar{t}^{\parallel \left(N_x-1,N_x \right)} \\
    0 & 0 & 0 & \dots &  \bar{t}^{\parallel \left(N_x,N_x - 1\right)} & -2\bar{t}^\parallel_0
\end{bmatrix},
\end{equation}
where we have included only the $2N_\bot  \times  2N_\bot$ nearest neighbor hopping matrix $\bar{t}^{\parallel \left(m,n\right)}$ between the $m^{th}$ and $n^{th}$ layers and $\bar{t}^\parallel_0$ is a diagonal matrix to bring the bands to zero energy in the presence of no additional terms in the 3D Hamiltonian (\ref{H3D}). If the structure of the layers doesn't change along the wire then $\bar{t}^{\parallel(i,i+1)} = \bar{t}^{\parallel(1,2)}$ for all i. 

It is also instructive to look at the structure of the matrices in the molecular orbital basis, which are found using the relation $\tilde{D} = \bar{S} \bar{D} \bar{S}^{\dagger}$. Consider a wire with translationally invariant electrostatic conditions. This results in identical auxiliary Hamiltonians (\ref{Haux}) for each layer, which further implies identical $S$ matrices for each layer (up to a phase we take to zero). One can clearly see that the resulting $\bar{S}$ is given by 
%$\bar {S}$ =
\begin{equation}
\bar S
=
\begin{bmatrix}
    S & 0 & 0 & \dots  & 0 \\
    0 &  S & 0 & \dots  & 0 \\
    \vdots & \vdots & \vdots & \ddots & \vdots \\
    0 & 0 & 0 & \dots  &  S
\end{bmatrix},
\end{equation}
where S is a $2N_\bot  \times  2N_\bot$ matrix specifing the eigenstates of each layer's $H_{aux}^m$ (see Eq. (\ref{Smtx})). Using the form of $\bar{t}^\parallel$ above, one can show that this $\tilde{t}^\parallel_{\nu \nu^\prime} = t^\parallel ~ \delta_{\nu, \nu^\prime \pm 2N_\bot} - 2t^\parallel \delta_{\nu\nu^\prime}$, and we remind the reader that $\tilde{t}^\parallel_{\nu \nu^\prime}$ is in the molecular orbital basis with labels $\nu=\nu(m,\alpha,\sigma) = 2 (m-1) N_{\bot} +2\alpha-1+\sigma$. Writing $\widetilde{t}^\parallel$ using the (m,$\alpha$) indices gives $\widetilde{t}^\parallel_{m\alpha, n\beta} = \left(t^\parallel \delta_{m,n \pm 1} - 2t^\parallel \delta_{m,m}\right) \delta_{\alpha \beta}$.   Therefore, for the case of homogenous conditions (i.e. infinite wire), we have shown that the various molecular orbitals inter-layer hopping decouples into separate bands as expected. 

The situation is more interesting if the electrostatic potential is not translationally invariant. $\bar{S}$ is then composed of blocks that are not identical, resulting in off-diagonal hopping between the various ``molecular bands'' of the wire. Explicitly,  $\widetilde{t}^\parallel_{m\alpha, (m \pm 1)\beta} \neq 0$ for $\alpha \neq \beta$. One can then imagine that if the electrostatic potential varies a large amount over a sufficently small length scale, there will be large off-diagonal coupling between the low-energy orbitals of neighboring layers. This may lead to a ``hopping barrier'' between layers that acts much like an electrostatic barrier. We stress that this off-diagonal hopping cannot be taken account in the simplified 1D models used extensively in the literature. 

\section{Perturbation scheme for infinite nanowires} \label{App3}

To understand how large of a role the electronic interactions play in determining the transverse profiles of the eigenstates, we compared our self-consistent method with perturbation results in Sec. \ref{Applic1}. Here we present details concerning the two perturbative methods used in that comparison.

First, we describe the perturbation method used in Sec. \ref{Applic1}. In this case, we treat $H_{int}$ as a perturbation. Standard first order perturbation theory gives
\begin{equation}
E_{n}^{1} = \left< \varphi_{n}^{0} \right| H_{int} \left| \varphi_{n}^{0}  \right> \label{HintPert},
\end{equation}
where $\left| \varphi_{n}^{0} \right>$ are the eigenstates of $H_{0}$ in Eq (\ref{HamInf}). An issue arises when calculating the terms in Eq (\ref{HintPert}). Since $H_{int}$ depends on the eigen energies and states of the \textit{complete} Hamiltonian (i.e. $H_{int} = H_{int}(\{E_i\},\{\varphi_i\})$), the calculation is not straightforward. Rather, Eq (\ref{HintPert}) needs to be solved self-consistently with the energy levels assumed in the form of $H_{int}$ to coincide to with the final energy levels. We use a simply iterative method to find the self-consistent solution.

Second, we describe a first order perturbation theory devoloped by Vuik and coworkers\cite{Vuik2016} to calculate the response of the effective chemical potentials of bands given a self-consistent solution when no magnetic field is applied. This perturbation scheme is used in Sec. \ref{Applic1A}.

They define a reciprocal capacitance
\begin{equation}
P_i(y,z,\Gamma) = {\phi_i(y,z,\Gamma) \over -en_i(-\mu_i-\delta\mu_i,\Gamma,\alpha)},
\end{equation} 
where $\phi_i$ is the potential due to all the occupied states in the i$^{\rm th}$ band, and $n_i$ is the 1D electron density of the i$^{\rm th}$ band. They find the relation
\begin{equation}
\delta \mu_i=-{e}^{2} \sum\limits_{j=1}^{N} P_{ij} \delta n_j, \label{pertEq}
\end{equation}
where $\delta \mu_i (\Gamma)=\mu_i (\Gamma) - \mu_i(0) $, $\delta n_i (\Gamma)=n_i (\Gamma) - n_i(0)$, and N is the number of occupied bands. The matrix elements, $P_{ij}$, are given by
\begin{equation}
P_{ij} = \left<\psi_{i,k}\right| P_j \left|\psi_{i,k} \right>,
\end{equation}
where $\left| \psi_{i,k} \right> $ is the i$^{\rm th}$ eigenstate of the tight-binding Hamiltonian for any $k_x$ value. $P_{ij}$ is essentially the correction to the energy of the i$^{\rm th}$ band coming from the addition of an electron to the j$^{\rm th}$ band. By Green's reciprocity theorem, we have $P_{ij}=P_{ji}$.

We can see that the reciprocal capacitance, $P_i$, is the potential due to a normalized charge occupying a state in the i$^{\rm th}$ band. In other words, this quantity is independent of how many states within the band are occupied. The main approximation in the above formulation is to assume $P_i(y,z,\Gamma) \approx P_i(y,z,0)$. Thus the reciprocal capacitance becomes magnetic field-independent, and the effective chemical potentials become entirely determined by the self-consistent solution under no applied magnetic field. Since the reciprocal capacitance is independent of the occupation filling factor, the magnetic field independence is accurate over the range of Zeeman energies for which the wave function profiles of the various sub-bands remains essentially constant.
 
The 1D electron density of each band has nontrivial dependence on the effective chemical potential and magnetic field (i.e. $n_j = n_j(\mu_j,\Gamma)$). Therefore Eq. (\ref{pertEq}) must usually be solved numerically. However, for single-band occupation, zero spin-orbit coupling, and low magnetic field we find the analytical solution
\begin{equation}
\delta\mu_1 = \cfrac{\eta {\Gamma}^{2}} {4 {\left({\mu}^{0}_{1}\right)}^{3 / 2} \left(1+     \cfrac{\eta}{ {\left({\mu}^{0}_{1}\right)}^{1 / 2}  }       \right)   } + \mathcal{O}\left(   {\left(\frac{\Gamma}{ {\mu}^{0}_{1}}\right)}^{4}   \right),  \label{delmu1}
\end{equation}
for $\mu_1 \gg \Gamma$, where ${\mu}^{0}_{1}$ is the effective chemical potential at zero magnetic field and $\eta$ is given by
\begin{equation}
\eta = {e}^{2} \sqrt{ {2{m}^{*} \over {\pi}^{2}{\hbar}^{2} }} P_{11}.
\end{equation}

\bibliography{REFERENCES}

\end{document}